\documentclass[12pt,a4paper]{article}
\usepackage[dvips]{graphicx,color}
\usepackage{times}
\usepackage{xcolor}
\usepackage{cite}
\usepackage{float}
\usepackage[%
  colorlinks=true,
  urlcolor=blue,
  linkcolor=green,
  citecolor=blue]{hyperref}
\usepackage[utf8]{inputenc}
\usepackage{amsmath}
\usepackage{amsfonts}
\usepackage{makecell}
\usepackage{amssymb}
\usepackage{graphicx}
\usepackage{pgfplots}
\usepackage{subcaption}
\usepackage{caption}
\pgfplotsset{compat=1.18} % or latest version
\usepackage{multirow}
\usepackage{booktabs}
\usepackage[left=2.0cm,right=2.0cm,top=2.5cm,bottom=2.5cm]{geometry}
\begin{document}
\newcommand{\sheptitle}
{ Perturbations to $\mu-\tau$ reflection symmetry due to renormalization group running effects}
\newcommand{\shepauthor}
{Chandan Kumar Borah \footnote{ E-mail: cborah528@gmail.com}\ and\ 
 Chandan Duarah \footnote{ E-mail: chandanduarah@dibru.ac.in}}
\newcommand{\shepaddress}
   {Department of Physics, Dibrugarh University,
               Dibrugarh - 786004, India }
\newcommand{\shepabstract}
{\noindent 
 $\mu-\tau$ reflection symmetry is an attractive flavour symmetry in lepton mixing, which accommodates maximal values of atmospheric mixing angle ($\theta_{23}=\pi/4$) and Dirac CP phase ($\delta=\pi/2/3\pi/2$). Another significance of this symmetry is that it does not constrain $\theta_{13}$ to be zero. As the recent results from $T2K$ and $NO\nu A$ experiments indicate a near-maximal value of the Dirac CP phase, the role of $\mu-\tau$ reflection symmetry becomes more prominent. In this work, we study RG running effects as a perturbation to the $\mu-\tau$ reflection symmetry. Assuming the symmetry to be preserved at the seesaw scale, we study the deviations of mass eigenvalues and lepton mixing parameters at the electroweak scale due to RG running. We derive the one-loop RGEs of the mass eigenvalues and mixing parameters and solve them numerically. Numerical analysis shows that the deviations from $\mu-\tau$ reflection symmetry are consistent with $3\sigma$ range of global oscillation data.

	\bigskip
	
%---------------------------KEYWORDS (OPTIONAL)-------------------------------
	\noindent
	{\bfseries Keywords: Lepton mixing, $\mu-\tau$ reflection symmetry,  Renormalization group running, MSSM.} 
	\vspace{1.2cm}
\noindent
\pagebreak}
%888888888888888888888888888888888888888888888888888888888888888888888888
\begin{titlepage}
\begin{flushright}
%hep-ph/yymmnnn
\end{flushright}
\begin{center}
{\large{\bf\sheptitle}}
\bigskip\\
\shepauthor
\\
\mbox{}\\
{\it\shepaddress}\\
\vspace{.5in}
{\bf Abstract}
\bigskip
\end{center}
\setcounter{page}{0}
\shepabstract
\end{titlepage}

\section{Introduction}
\indent Neutrino oscillations imply that neutrinos are massive and flavours are mixed. The theory of neutrino mixing is described in terms of the lepton matrix $U$ (also known as the PMNS matrix), which connects neutrino flavor eigenstates to the mass eigenstates. In general, $U=U^{\dagger}_lU_{\nu}$, where $U_l$ and $U_\nu$ diagonalize the charged lepton mass matrix ($M_l$) and neutrino mass matrix ($M_\nu$) respectively. In the basis where $M_l$ is diagonal, $U_l$ becomes a unit matrix and $U$ coincides with $U_\nu$.
The $3\times 3$ unitary matrix $U$ is generally parametrized in terms of three mixing angles and six phases in the standard parametrization \cite{PDG} and it is given by 
\begin{equation}
    U=P_1VP_2 \label{U},
    \end{equation}  with
\begin{equation}
    V=\begin{bmatrix}
  c_{12}c_{13}&s_{12}s_{13}&s_{13}e^{-i\delta}\\-s_{12}c_{23}-c_{12}s_{23}e^{i\delta}&c_{12}c_{23}-s_{12}s_{23}e^{i\delta}&c_{13}s_{23}\\s_{12}s_{23}-c_{12}s_{13}e^{i\delta}&-c_{12}s_{23}-s_{12}c_{23}e^{i\delta}&c_{13}c_{23} \end{bmatrix},
  \label{para}
\end{equation}
 $P_1=Diag(\phi_1,\phi_2,\phi_3)$ and $P_2=Diag(e^{i{\alpha}},e^{i{\beta}},1)$. In the matrix V, $s_{ij}=\sin\theta_{ij}$ and $c_{ij}=\cos\theta_{ij}$ ($ij=12,23,13$), with $\theta_{12},\theta_{23}$ and $\theta_{13}$ being the solar mixing angle, atmospheric mixing angle and reactor mixing angle respectively and $\delta$ represents the Dirac CP-violating phase. The diagonal matrix $P_1$ contains three unphysical phases $\phi_1$, $\phi_2$ and $\phi_3$, whereas $P_2$ contains the Majorana CP phases $\alpha$ and $\beta$. If the unphysical phases are irrelevant to the treatment, they can be removed from the mixing matrix by phase redefinition of the charged lepton fields. 
 
\indent Over the past two decades, various oscillation experiments have succeeded in measuring the values of $\theta_{12}$ and $\theta_{13}$ with good precision \cite{SKT2K, SK24,SNO,MINOS, Daya,RENO,DCC, T2k1, T2k2,Nova1, Nova2,IceCube,  kamLand}. Again, the best-fit value of $\theta_{23}$ is centered at the maximal value $\pi/4$, but its exact position, whether it lies in the first or second octant, is still unresolved. This is also referred to as the problem of octant degeneracy. On the theoretical side, the overall experimental prediction of the mixing matrix elements reflects an interesting characteristic that the modulus of each $\mu$-flavour element is approximately equal to that of the corresponding $\tau$-flavour element ($|U_{\mu i}| \approx |U_{\tau i}|$). This approximate equality points towards the so-called $\mu-\tau$ flavour symmetry in neutrino mixing \cite{RVW}. It is a general feature of $\mu-\tau$ symmetry that restricts $\theta_{23}$ to be maximal. The well-known mixing patterns, such as Bi-maximal mixing \cite{PS1}, Tri-bimaximal mixing \cite{PS3}, etc. are some special types of $\mu-\tau$ symmetry (also referred to as permutation symmetry) that predict $\theta_{13}=0$. Such mixing patterns result from the invariance of the neutrino mass term under the exchange of $\nu_{\mu}$ and $\nu_{\tau}$-flavours:
$$\nu_e \rightarrow \nu_e,\  \nu_\mu \rightarrow \nu_\tau,\  \nu_\tau \rightarrow \nu_\mu.$$
Besides, there exists another type of $\mu-\tau$ symmetry, referred to as reflection symmetry in the literature, which was first introduced by Harrison and Scott \cite{RS1}. It is the symmetry such that the neutrino mass term remains invariant under the combined operation of $\mu-\tau$ exchange and charge conjugation :
$$\nu_e \rightarrow \nu_e^c,\  \nu_\mu \rightarrow \nu_\tau^c,\  \nu_\tau \rightarrow \nu_\mu^c.$$ 
This reflection symmetry additionally constrains the Dirac CP phase to be maximal ($\delta=\pi/2$ or $3\pi/2$) along with the maximal prediction of $\theta_{23}$. Since this symmetry leaves $\theta_{12}$ and $\theta_{13}$ arbitrary, it becomes more compatible with the experimental data. In recent years, we also have preliminary results on the measurement of the Dirac CP phase from $T2K$  and $NO\nu A$ experiments \cite{T2k1, T2k2, Nova1, Nova2}. These results indicate a near maximal value of the Dirac phase $\delta$ centered at $270^\circ$. The near maximal value of $\delta$ is also reflected in the recent global analysis of $3\nu$ oscillation data \cite{GlobAnal} in the inverted order(IO) scenario. In light of non-zero $\theta_{13}$ and the possible near-maximal value of $\delta$, $\mu-\tau$ reflection symmetry may serve as a fruitful candidate for the theory of lepton mixing. On the other hand, observational data generally indicate certain deviation from the exact $\mu-\tau$ symmetry. It therefore becomes important to study the deviations from $\mu-\tau$ reflection symmetry through different possible perturbation schemes. 

\begin{table}[t]
\begin{center}
\begin{tabular}{c cc cc }
\hline
 Without SK atmospheric data & & \\ \hline
\multirow{2}{*}{Parameter}& 
\multicolumn{2}{c}{Normal Ordering}&
\multicolumn{2}{c}{Inverted Ordering} \\
\cline{2-5}
 &Best-fit Value  & $3\sigma$ & best-fit value & $3\sigma$ \\ \hline
 $\theta_{12}$         & 33.68    &  31.63-35.95   & 33.68    &  31.63-35.95         \\
 $\theta_{23}$         & 48.5    & 41.0-50.5       & 48.6    & 41.4-50.6      \\
 $\theta_{13}$         & 8.52    & 8.18-8.87        & 8.58    & 8.24-8.91    \\ 
 $\delta$         & 177     & 96-422  & 285    & 201-348         \\
 $\Delta m^2_{21}(/10^{-5}eV^2)$  & 7.49   & 6.92-8.05 & 7.49   & 6.92-8.05        \\
 $\Delta m^2_{32}(/10^{-3}eV^2)$  & 2.534 & 2.463-2.606 & -2.510     & -2.584-2.438          \\ \hline
  With SK atmospheric data & & \\ \hline
 $\theta_{12}$         & 33.68    &  31.63-35.95   & 33.68    &  31.63-35.95         \\
 $\theta_{23}$         & 43.3    & 41.3-49.9       & 47.9    & 41.5-49.8      \\
 $\theta_{13}$         & 8.56    & 8.19-8.89        & 8.59    & 8.25-8.93    \\ 
 $\delta$         & 212     & 124-364  & 274    & 201-335         \\
 $\Delta m^2_{21}(/10^{-5}eV^2)$  & 7.49   & 6.92-8.05 & 7.49   & 6.92-8.05        \\
 $\Delta m^2_{32}(/10^{-3}eV^2)$  & 2.513 & 2.451-2.578 & -2.484     & -2.547-2.421          \\ \hline
\end{tabular}
\end{center}
\caption{The best-fit values and 3$\sigma$ allowed ranges of neutrino oscillation parameters in NO and IO obtained from global analysis \cite{GlobAnal}.}
\label{GA}
\end{table} 

\indent Besides the measurement of mixing angles, oscillation experiments are also sensitive to two mass squared differences- $\Delta m_{21}^2=m_2^2-m_1^2$ and $|\Delta m_{32}^2|=|m_3^2-m_2^2|$ or $|\Delta m_{31}^2|=|m_3^2-m_1^2|$; $m_1$, $m_2$ and $m_3$ being the three mass eigenvalues corresponding to three mass eigenstates. Since the oscillation experiments provide only the absolute value $|\Delta m_{32}^2|$ / $|\Delta m_{31}^2|$, this leaves two possibilities for the order of the mass eigenvalues- Normal Order (NO) with $m_1<m_2<m_3$ and inverted order (IO) with $m_3<m_1<m_2$. Although the absolute values of $m_1$, $m_2$ and $m_3$ are not known, cosmological observations provide an upper bound on the sum of mass eigenvalues given by $\sum m_i < 0.12eV$ \cite{mbound}. 

\indent Due to the attractive nature of $\mu-\tau$ reflection symmetry, compatible with experimental data, a lot of work has been devoted to $\mu-\tau$ reflection symmetry in recent years. As for example, its realization has been explored in various frameworks, including the minimal and littlest seesaw models in Refs. \cite{RRS1,RRS2,RRS3}. Modifications to the neutrino mixing from this symmetry is discussed in \cite{RRS12}. The implementation of $\mu-\tau$ reflection symmetry in trimaximal mixing is studied in \cite{RRS4,RRS5, RRS6}. Several studies have also explored its consequences for neutrino mass hierarchy and neutrinoless double beta decay \cite{RRS13,NNath}. The formulation of the $\mu-\tau$ reflection symmetry specifically for Dirac neutrinos is addressed in \cite{dirac}. Further, the general structure of reflection symmetry breaking has been discussed in \cite{RRS9}, while its breaking within discrete flavor symmetry models has been studied in \cite{VVV1,VVV2,RRS10}.  

\indent In this work, we consider Renormalization Group (RG) running effects to the $\mu-\tau$ reflection symmetry and study the consequent deviations of mixing parameters. The key idea is to assume that the symmetry to be preserved at a very high energy scale where light Majorana neutrino masses are generated effectively through some unknown physics operating at that scale. The generation of light Majorana masses can be attributed to an effective dimension-five operator, known as the Weinberg operator. The simplest realization of this operator is through the so-called type-I seesaw mechanism, where three heavy right-handed neutrinos are introduced. When these neutrinos are integrated out of the theory, light Majorana masses for the active neutrinos emerge below the right-handed neutrino mass scale $M_R$. Let us denote the high energy scale at which Majorana masses of neutrinos are generated as the flavour symmetry scale $\Lambda_{FS}$. Since $\Lambda_{FS}$ is much higher than the electroweak scale $\Lambda_{EW}$, it becomes essential to consider the RG running of neutrino parameters to connect high-scale predictions with low-energy experimental data. We assume that $\mu-\tau$ reflection symmetry is preserved at $\Lambda_{FS}$, such that the atmospheric mixing angle $\theta_{23}$ and Dirac CP phase $\delta$ are constrained at maximal values at that scale. As the energy scale runs from $\Lambda_{FS}$ down to $\Lambda_{EW}$, all the physical parameters evolve according to their RG equations and get deviated from their high-scale predictions. To account the deviations of the parameters, we consider the Minimal Supersymmetric Standard Model (MSSM) as the effective theory below $M_R$. As the energy scale runs down from $\Lambda_{FS}$, Supersymmetry is assumed to break-down at some scale $\Lambda_s$, below which the effective theory becomes the Standard Model (SM). Since there is no definite information about the exact SUSY breaking scale, following recent updates \cite{PDG}, we consider three benchmark values for $\Lambda_s\ \text{at}\ 1\ \mathrm{TeV}$, $7\ \mathrm{TeV}$, and $14\ \mathrm{TeV}$ to carry out the numerical analysis. 

\indent The study of RG running effects on neutrino parameters is rather extensive in the literature \cite{KS,RGE2,1,R1, R3, R2, poko, babu,  R4, JMei, R5,ohl,Dzhang,NNSingh,SGupta,RGERS2,YLZhou,JZhu}. To be specific, the RG running effects on the parameters involving $\mu-\tau$ reflection symmetry are studied in \cite{RRS1,RRS6,dirac, JZhu, YLZhou, RGERS2}. The renormalization group equations (RGEs) of neutrino parameters, including mass eigenvalues and mixing parameters, can be derived from the RGE of the effective coupling constant $\kappa$, which appears in the effective dimension-five operator. After spontaneous symmetry breaking, the effective operator gives rise to light Majorana neutrino masses, where the neutrino mass matrix is proportional to both $\kappa$ and the vacuum expectation value (vev) of the Higgs field. Following the standard method described in Refs.~\cite{KS,RGE2,R1, R3}, we first derive the running equations for the eigenvalues of $\kappa$ through diagonalization. As considered in \cite{v_SM, v_MSSM}, we treat in this work the vev of the Higgs field to be energy dependent and translate the RGEs of eigenvalues of $\kappa$ to corresponding RGEs of neutrino mass eigenvalues. Since the running behaviour of the vev of the Higgs field differs between the SM and MSSM, the resulting RGEs of neutrino mass eigenvalues become distinct for the two frameworks. In total, we obtain twelve RGEs corresponding to the three mass eigenvalues, three mixing angles and six CP-violating phases. All the RGEs form a set of first-order coupled differential equations which can be solved numerically. However, these RGEs also depend on the values of three gauge couplings ($g_1$, $g_2$, and $g_3$), three Yukawa couplings ($y_t$, $y_b$ and $y_{\tau}$) and the Higgs quartic coupling constant ($\lambda$). Therefore, the RGEs of these seven coupling constants must also be taken into account. In this work, we numerically solve all the nineteen coupled RGEs using a \texttt{Python} code. However, since the Higgs quartic coupling $\lambda$ is not present in the MSSM framework, in this regime we need to solve only eighteen equations. 

\indent To predict the values of neutrino parameters at the electroweak scale $\Lambda_{\mathrm{EW}}$, we adopt a top-down approach by specifying the input values at the high-energy scale $\Lambda_{\mathrm{FS}}$. Since the $\mu$–$\tau$ reflection symmetry is assumed to be exact at $\Lambda_{\mathrm{FS}}$, the input values of the atmospheric mixing angle $\theta_{23}$ and all CP-violating phases are fixed at maximal values as per the symmetry. However, the input values of the remaining two mixing angles $\theta_{12}$ and $\theta_{13}$, along with the three neutrino mass eigenvalues, remain as free parameters and are chosen such a way that the low-energy predictions match current experimental data as closely as possible. The input values of the gauge and Yukawa couplings at $\Lambda_{\mathrm{FS}}$ are obtained through bottom-up approach by solving their RGEs using low energy experimental data. 

Although RG running effects on neutrino parameters have been extensively studied in the literature, we wish to highlight certain distinctive aspects of the present work. In Ref. \cite{R1}, the RGEs of neutrino parameters are presented in compact forms in terms of the elements of the lepton mixing matrix. In the present work, we have performed detailed algebraic calculations to express the RGEs explicitly in terms of neutrino mixing parameters—namely, the mixing angles, CP-violating phases and mass eigenvalues. After obtaining the general RGEs, authors in Ref. \cite{R1} analyzed the implications of RGEs separately for the real case (without the CP phases) and complex case (with the CP phases) of lepton mixing, but the explicit running behaviours of mixing parameters were not studied in detail. An extensive treatment of running behaviour through numerical solution of simultaneous RGEs was later performed in Ref. \cite{R3}, although the unphysical phases in the PMNS matrix were not taken into account there. These unphysical phases are, however, crucial for preserving $\mu-\tau$ reflection symmetry within the standard parametrization, as will be discussed in Section 2. In this context, we have derived the complete set of RGEs for twelve neutrino parameters: three mass eigenvalues, three mixing angles, and six CP-violating phases. Together with the RGEs for the three gauge couplings, three Yukawa couplings and the Higgs quartic coupling constant, they constitute a system of nineteen coupled differential equations. In the present analysis, we have solved all the nineteen equations simultaneously to connect the high-energy symmetry-imposed boundary conditions with the experimentally constrained low-energy observables. Moreover, our analysis also includes the energy dependence of the vacuum expectation value (vev) of the Higgs field in the running equations for the mass eigenvalues. Although the effects of energy-dependent vev had been taken into account previously in Refs. \cite{v_SM, v_MSSM}, it was not incorporated in the Refs. \cite{R1,R3}.

\indent The remaining part of this paper is organized as follows: in Section~2, we present some basics of $\mu$–$\tau$ reflection symmetry and corresponding parametrization of the lepton mixing matrix. In Section~3 we derive the one-loop RGEs of neutrino mass eigenvalues and lepton mixing parameters. In Section~4, we present the numerical analysis and results. Finally, Section~5 is devoted to summary and discussion. 
\section{Basics of $\mu-\tau$ reflection symmetry and corresponding lepton mixing matrix}
The concept of $\mu$--$\tau$ reflection symmetry was first introduced by Harrison and Scott in the paper \cite{RS1}. They were motivated by the fact that the modulus of each $\mu$ flavor element and the corresponding $\tau$ flavor element in the mixing matrix are approximately equal ($|U_{\mu i}| \approx |U_{\tau i}|$) as measured by the oscillation experiments. Based on this observation, they formulated a specific parametrization of the mixing matrix given by
\begin{equation}
    V_{HS}=\begin{bmatrix}
        u_1 & u_2 &u_3 \\ v_1 &v_2 &v_3 \\ v_1^* & v_2^* & v_3^*
    \end{bmatrix},
    \label{HS}
\end{equation}
where $u_i$ are real and $v_i$ are complex parameters and the matrix is consistent with the equality $|U_{\mu i}| = |U_{\tau i}|$. It is to be noted that they have considered only the Dirac CP phase in their formulation. The mixing matrix in Eq. (\ref{HS}) remains invariant if we interchange the second and third rows of the matrix and take the complex conjugation of the entire matrix as well. This symmetry of the mixing matrix is referred to as $\mu-\tau$ reflection symmetry. The corresponding neutrino mass matrix bearing this symmetry is given by 
\begin{equation}
    M=\begin{bmatrix}
        M_{ee} & M_{e\mu} & M_{e\tau}^* \\ M_{e\mu} & M_{\mu\mu} &M_{\mu\tau} \\ M_{e\tau}^* &M_{\mu\tau} & M_{\mu\mu}^*
    \end{bmatrix},
\end{equation}
where the elements $M_{ee}$ and $M_{\mu\tau}$ are real. The mass matrix remain invariant under $\mu-\tau$ reflection operation, which can be mathematically expressed as $(A_{\mu\tau} M A_{\mu\tau})^*=M$ with
\begin{equation}
A_{\mu\tau} = 
\begin{bmatrix}
1 & 0 & 0 \\
0 & 0 & 1 \\
0 & 1 & 0
\end{bmatrix}
\end{equation}
being the $\mu-\tau$ exchange operator.

\indent The significance of $\mu$--$\tau$ reflection symmetry lies in its prediction of maximal values for the atmospheric mixing angle and the Dirac CP-violating phase, i.e., $\theta_{23} = \pi/4$ and $\delta = (\pi/2)/(3\pi/2)$. These predictions can be inferred from the Jarlskog invariants \cite{JARL} corresponding to the mixing matrix $V_{HS}$ and that of the lepton mixing matrix in standard parametrization. The Jarlskog invariant for $V_{HS}$ in Eq. (\ref{HS}) given by $J=\frac{1}{2}u_1u_2u_3$ \cite{RVW}, such that the modulus of J in terms of mixing matrix elements can be expressed as 
\begin{equation}
\left| J \right| = \frac{1}{2}\left| V_{e1} V_{e2} V_{e3} \right|.
\label{JHS}
\end{equation} 
Since the formulation of $V_{HS}$ in Eq. (\ref{HS}) concern only the Dirac phase $\delta$, in standard parametrization it is equivalent to the matrix V in Eq. (\ref{para}). Then for V, we have \cite{CL1}
\begin{equation}
\left| J \right| = \frac{1}{2} \left| V_{e1} V_{e2} V_{e3} \right| \, \left|\sin\delta \right| \, \sin 2\theta_{23}.
\label{JPARA}
\end{equation}
Now, comparison of Eqs. (\ref{JHS}) and (\ref{JPARA}) implies that $\left|\sin\delta \right| \, \sin 2\theta_{23}=1$ for non zero $\theta_{13}$. For $\theta_{23}=\pi/4$, this relation implies that $\delta=(\pi/2)/(3\pi/2)$.

\indent On the other hand, in Ref. \cite{CL1}, an issue has been addressed regarding the consistency of $\mu-\tau$ reflection symmetry in the standard parameterization of the lepton mixing matrix. It can be noted that the direct substitution of the maximal values of $\theta_{23}$ and $\delta$ predicted by reflection symmetry in the "standard" lepton mixing matrix does not reflect back the exact symmetry contained in $V_{HS}$. To be explicit, substituting $\theta_{23}=\pi/4$ and $\delta=
(\pi/2)/(3\pi/2)$ in Eq. (\ref{para}), we get the mixing matrix as 
\begin{equation}
\begin{bmatrix}
  c_{12}c_{13}&s_{12}c_{13}&\mp is_{13}\\ \frac{1}{\sqrt{2}}(-s_{12}\mp ic_{12}s_{13})&\frac{1}{\sqrt{2}}(c_{12}\mp is_{12}s_{13})&\frac{1}{\sqrt{2}}c_{13}\\\frac{1}{\sqrt{2}}(s_{12}\mp ic_{12}s_{13})&\frac{1}{\sqrt{2}}(-c_{12}\mp is_{12}s_{13})&\frac{1}{\sqrt{2}}c_{13}
  \end{bmatrix},
\end{equation}
where $-$ and $+$ sign corresponds to $\delta=\pi/2$ or $3\pi/2$ respectively. This matrix satisfy the equality $|U_{\mu i}| = |U_{\tau i}|$. However, unlike $V_{HS}$, the relation $V_{\tau i} = V_{\mu i}^*$ holds only for $i=3$, but for $i=1$, $2$, it follows that $V_{\tau i} = -V_{\mu i}^*$. More noticeably, the first row elements of the mixing matrix are not all real, which contradicts the original formulation of $V_{HS}$ in Eq. (\ref{HS}). Thus the maximal prediction of $\theta_{23}$ and $\delta$ as per reflection symmetry are not consistent with the standard parametrization of the lepton mixing matrix. In \cite{CL1}, it has been shown that the symmetry can be made consistent with the standard parametrization by assigning maximal values of Majorana phases and one of the unphysical phases. Accordingly, we consider the parametrization of the lepton mixing matrix $U$ in terms of all six phases as given in Eq. (\ref{U}) and allow specific values of the CP phases as prescribed in Ref. \cite{CL1}. Since $\mu-\tau$ reflection symmetry allows two possible maximal values of Dirac phase, we consider two separate cases:
\begin{itemize}
    \item \textbf{Case I:} $\theta_{23} = \pi/4$, $\delta = \pi/2$,
    \item \textbf{Case II:} $\theta_{23} = \pi/4$, $\delta = 3\pi/2$.
\end{itemize}
To ensure the symmetry of the lepton mixing matrix in the standard parametrization, we consider the following sets of maximal values for the Majorana and one of the unphysical phases:
\begin{equation}
    \alpha=\beta=\frac{3\pi}{2},\quad \phi_1=\frac{\pi}{2},\quad \phi_2=\phi_3=0
    \label{ch1}
\end{equation}
for Case I and 
\begin{equation}
    \alpha=\beta=\frac{\pi}{2},\quad \phi_1=\frac{3\pi}{2},\quad \phi_2=\phi_3=0
    \label{ch2}
\end{equation}
for Case-II. Substituting all the maximal values of mixing parameters considered above in Eq.(\ref{U}), we get the mixing matrix as
\begin{equation}
U_{\mu\tau}=\begin{bmatrix}
  c_{12}c_{13}&s_{12}c_{13}&s_{13}\\ \frac{1}{\sqrt{2}}(-c_{12}s_{13}\pm is_{12})&\frac{1}{\sqrt{2}}(-s_{12}s_{13}\mp ic_{12})&\frac{1}{\sqrt{2}}c_{13}\\\frac{1}{\sqrt{2}}(-c_{12}s_{13}\mp is_{12})&\frac{1}{\sqrt{2}}(-s_{12}s_{13}\pm ic_{12})&\frac{1}{\sqrt{2}}c_{13}
  \end{bmatrix},
\end{equation}
where '$+$' and '$-$' signs correspond to Case-I and Case-II respectively. The above mixing matrix is exactly similar to $V_{HS}$ in Eq. (\ref{HS}) and thus consistent with $\mu-\tau$ reflection symmetry in standard parametrization.
%%%%%%%%%%%%%%%%%%%%%%%%%%%%%%%%%%%%%%%%%%%%%%%%%%%%%%%%%%%%%%%%%% 3
\section{RG running of neutrino parameters}
While neutrinos are massless in the Standard Model (SM), massive neutrinos necessarily implies new physics beyond the SM. Various mechanisms have been proposed to generate neutrino masses through suitable extensions of the SM. Among these, the most natural approach is through an effective dimension-five operator, commonly referred to as the Weinberg operator. In the context of the SM, it is expressed as

\begin{equation}
\mathcal{L}_5^{SM} = -\frac{1}{2} \kappa(\overline{l}_L H)(H^Tl^c) + \text{h.c.},
\label{KSM}
\end{equation}
and in MSSM, it can be expressed as \cite{R3}
\begin{equation}
\mathcal{L}_5^{MSSM} =\mathcal{W}\vert_{\theta\theta}+\text{h.c.}
   = -\frac{1}{2} \kappa (\mathbb{L} \mathbb{H})(\mathbb{H}\mathbb{L})\vert_{\theta\theta} + \text{h.c.}\ .
 \label{KMSSM}  
\end{equation}
In Eq. (\ref{KSM}), $l_L$ and $H$ denote the SM lepton and Higgs doublets respectively, while, in Eq. (\ref{KMSSM}), $\mathbb{L}$ and $\mathbb{H}$ stand for lepton doublet and up-type Higgs superfield in the MSSM respectively. In both expressions $\kappa$ represents the effective coupling constant. After electroweak symmetry breaking, the dimension-five operator gives rise to effective Majorana neutrino masses as
\begin{equation}
   M_\nu = \kappa v^2 
   \label{massSM}
\end{equation}
 in SM and 
 \begin{equation}
   M_\nu = \kappa v_u^2  
   \label{massMSSM}
 \end{equation}
  in MSSM.
Here, $v =<H> \approx 174\ GeV$ is the vacuum expectation value (vev) of the SM Higgs field while $v_u$ is the vev of the up-type Higgs field in MSSM. We have $v_u=v\sin\beta$, with $tan\beta$ being the ratio of the vevs of the two Higgs doublets, viz., $ H_u $, which couples to up-type quarks and neutrinos, and $ H_d $, which couples to down-type quarks and charged leptons. \\
\indent The simplest and most elegant way to realize the dimension-five mass operator is through the type-I seesaw mechanism, where three right-handed neutrino fields \( N_{R} \) are typically introduced into the theory. These fields are singlets under the Standard Model gauge group and allow for both Dirac and Majorana mass terms. The relevant Lagrangian density is given by
\begin{equation}
\mathcal{L}_{mass} = -Y_D \overline{\nu_L}N_R v - \frac{1}{2} M_R \overline{N_R^c} N_R + \text{h.c.},
\end{equation}
where $Y_D$ is the Yukawa coupling matrix and $M_R$ is the Majorana mass matrix for the right-handed neutrinos. When the right-handed neutrinos are integrated out, we get the effective Majorana mass term  $-\frac{1}{2}M_\nu \overline{\nu_L}\nu_L^c + \text{h.c.}$ for the light neutrinos below the mass scale $M_R$. In this expression, $M_\nu$ is given by the see-saw formula: 
\begin{equation}
    M_\nu=-M_D M_R^{-1}M_D^T,
    \label{Seesawmass}
\end{equation}
where $M_D =vY_D$ is the Dirac neutrino mass matrix. From the expressions of $M_\nu$ in Eqs. (\ref{massSM}) \& (\ref{Seesawmass}), it can also be inferred that $\kappa \equiv -Y_D M_R^{-1} Y_D^T$. \\
\indent Bellow $M_R$ the energy dependence of $\kappa$ is described by its RGE. At one loop level, the RGE of $\kappa$ is given by
\begin{equation}
 16\pi^2 \frac{d\kappa}{dt}=C\left[\kappa({Y_l}^\dagger Y_l)+({Y_l}^\dagger Y_l)^T\kappa \right]+\alpha \kappa,
 \label{12}
\end{equation}
where
\begin{equation}
    C_{SM}=-\frac{3}{2}, \ \ \ \ \alpha_{SM}\approx2y_\tau^2+6y_t^2-3g_2^2+\lambda
    \label{constSM}
\end{equation}
for SM and
\begin{equation}
    C_{MSSM}=1,\  \  \  \  \alpha_{MSSM}\approx6y_t^2-\frac{6}{5}g_1^2-6g_2^2
    \label{consMSSM}
\end{equation}
for MSSM and $t\equiv ln(\mu/\mu_\circ)$, with $\mu_\circ$ being an arbitrary renormalization scale. In the above expressions, $y_{\tau}$ and $y_t$ are the $\tau$ lepton and top quark Yukuwa coupling constants respectively; $g_1$ and $g_2$ are the gauge coupling constants and $\lambda$ is the Higgs self coupling constant. $Y_l$ is the Yukawa coupling matrix for the charged leptons. We work on the basis where $Y_l$ is diagonal, i.e., $Y_l=\text{Diag}(y_e,y_\mu,y_\tau)$, such that the lepton mixing matrix $U$ is identical to the neutrino mixing matrix $U_\nu$. \\
\indent To obtain the RGEs of the neutrino parameters from Eq. (\ref{12}), we follow the standard method described in \cite{KS, RGE2, R1, R3}. We begin with the diagonalisation of $\kappa$ given by
\begin{equation}
U^T \kappa U=D(t)=Diag(\kappa_1,\kappa_2,\kappa_3),
\label{13}
\end{equation} 
where $U$ is defined in Eq. (\ref{U}). The running of $U$ is expressed in terms of an anti-hermitian matrix $T$ as
\begin{equation}
\frac{dU}{dt}=UT.
\label{Urunning}
\end{equation}
From Eqs. (\ref{12}), (\ref{13}) and (\ref{Urunning}), we arrive, through some algebric calculations, at 
\begin{equation}
\frac{dD}{dt}=\frac{1}{16\pi^2}(P'D+DP'+\alpha D)-T^*D+DT,
\label{15}
\end{equation}
where $P^\prime=U^\dagger P U$ with $P=C(Y_l^\dagger Y_l)$. Since $y_e<<y_\mu<<y_\tau$, neglecting $y_e$ and $y_\mu$ we take the approximation as: $P\approx C\ Diag(0,0,y_\tau^2)$.
Now comparing the diagonal elements of both sides of Eq. (\ref{15}), we obtain the RGE of the eigenvalues of $\kappa$ given by 
\begin{equation}
\frac{d\kappa_i}{dt}=\frac{1}{16 \pi^2}(\alpha+2P'_{ii})\kappa_i,
\label{16}
\end{equation}
where $i=1,2,3$. The RGEs of neutrino mass eigenvalues $m_1,\ m_2$ and $m_3$ can be obtained from the above RGEs of $\kappa_i$'s using the relations in Eqs. (\ref{massSM}) and (\ref{massMSSM}) for both SM and MSSM. In this work, we consider energy-dependent vev in obtaining the RGEs of $m_i$'s. The energy dependence of the SM vev $v$ is given by \cite{v_SM}  
\begin{equation}
\frac{dv}{dt}=\frac{1}{16\pi^2} \big[\frac{9}{4}(\frac{1}{5}g_1^2+g_2^2)-3y_t^2 -y_{\tau}^2\big]v,
\label{vsm}
\end{equation}   
while that of the vev $v_u$ in MSSM is given by \cite{v_MSSM}
\begin{equation}
\frac{dv_u}{dt}=\frac{1}{16\pi^2} \big[\frac{3}{20}g_1^2+\frac{3}{4}g_2^2-3y_t^2 \big]v_u.
\label{vmssm}
\end{equation}
Now using Eqs. (\ref{16}) and (\ref{vsm}) in Eq. (\ref{massSM}), we get the RGEs of the mass eigenvalues in SM- 
 \begin{equation}
      \frac{dm_1}{dt} = \frac{1} {16 \pi^2} \left[\frac{9}{10} g_1^2 + \frac{9}{2}g_2^2 +\lambda- 3y_3^2   (s_{12}^2s_{23}^2+c_{12}^2c_{23}^2s_{13}^2-\frac{1}{2}s2_{12}s2_{23}s_{13}c_\delta)\right]m_1,
      \label{m1_rgeSM}
  \end{equation} 
   \begin{equation}
      \frac{dm_2}{dt} = \frac{1} {16 \pi^2} \left[\frac{9}{10} g_1^2 + \frac{9}{2}g_2^2 +\lambda- 3y_3^2   (c_{12}^2s_{23}^2+s_{12}^2c_{23}^2s_{13}^2+\frac{1}{2}s2_{12}s2_{23}s_{13}c_\delta)\right]m_2,
  \end{equation} 
  \begin{equation}
      \frac{dm_3}{dt} = \frac{1} {16 \pi^2} \left[\frac{9}{10} g_1^2 + \frac{9}{2}g_2^2 +\lambda- 3y_3^2 c_{13}^2c_{23}^2\right]m_3,
      \label{m3_rgeSM}
  \end{equation} 
  where $s2_{12}=\sin2\theta_{12}$, $s2_{23}=\sin2\theta_{23}$ and $c_\delta=cos\delta$.
Similarly, using the Eqs. (\ref{16}) and (\ref{vmssm}) in Eq. (\ref{massMSSM}) we get the RGEs of mass eigenvalues in MSSM which are given by
  \begin{equation}
      \frac{dm_1}{dt} = \frac{1} {16 \pi^2} \left[-\frac{9}{10} g_1^2 - \frac{9}{2}g_2^2 + 2y_3^2   (s_{12}^2s_{23}^2+c_{12}^2c_{23}^2s_{13}^2-\frac{1}{2}s2_{12}s2_{23}s_{13}c_\delta)\right]m_1,
      \label{m1_regMSSM}
  \end{equation}  
  \begin{equation}
      \frac{dm_2}{dt} = \frac{1} {16 \pi^2} \left[-\frac{9}{10} g_1^2 - \frac{9}{2}g_2^2 + 2y_3^2   (c_{12}^2s_{23}^2+s_{12}^2c_{23}^2s_{13}^2+\frac{1}{2}s2_{12}s2_{23}s_{13}c_\delta) \right]m_2,
  \end{equation}
   \begin{equation}
      \frac{dm_3}{dt} = \frac{1} {16 \pi^2} \left[-\frac{9}{10} g_1^2 - \frac{9}{2}g_2^2 + 2y_3^2 c_{13}^2c_{23}^2\right]m_3.
      \label{m3_rgeMSSM}
  \end{equation}

Again, comparing the off-diagonal elements of Eq. (\ref{15}), we can obtain the elements of the anti-Hermitian matrix $T$ as
\begin{equation}
 T_{ij}=-\frac{1}{16\pi^2}[\Delta_{ij}Re(P'_{ij})+i\frac{1}{\Delta_{ij}}Im(P'_{ij})],
 \label{T}
\end{equation}
where,
\begin{equation}
\Delta_{ij}=\frac{m_i +m_j}{m_i-m_j}.
\end{equation} 
Then employing Eqs. (\ref{Urunning}) and (\ref{T}), we obtain the RGEs of the mixing angles and the CP violating phases which are given bellow:
\begin{equation}
\begin{split}
&\frac{d\theta_{23}}{dt}= \frac{Cy^2_{\tau}}{32\pi^2}  \bigg[ -s2_{23} \big(s_{12}^2 A_{31}^{\alpha+} +c_{12}^2 A_{32}^{\beta+} \big) +s2_{12}c_{23}^2s_{13}\bigg(c_{\delta}(A_{31}^{\alpha+}-A_{31}^{\beta+})+s_{\delta}(B_{32}^{\alpha-}-B_{31}^{\alpha -})\bigg)
\bigg],
\end{split}
\label{22}
\end{equation}

\begin{equation}
\begin{split}
& \frac{d\theta_{13}}{dt}= \frac{Cy^2_{\tau}}{32\pi^2} \bigg[
\frac{1}{2}s2_{12}s2_{23}c_{13}\big\{c_{\delta}(A_{31}^{\alpha+}-A_{32}^{\beta+})+s_{\delta}(-B_{31}^{\alpha -}+B_{32}^{\beta-})\big\} \\& \qquad \qquad \qquad
 -c_{23}^2s2_{23}\big\{c_{12}^2\big(c_{\delta}^2A_{31}^{\alpha+}+s_{\delta}^2 D_{31}^{\alpha+}\big)-s_{12}^2\big(c_{\delta}^2A_{32}^{\beta+}+s_{\delta}^2 D_{32}^{\beta+}\big)\big\}
\bigg],
\end{split}
\end{equation}

\begin{equation}
\begin{split}
& \frac{d\theta_{12}}{dt}= \frac{Cy^2_{\tau}c_{23}s_{13}}{32\pi^2} \bigg[ s2_{23}s_{13}\big\{(s_{12}^2A_{31}^{\alpha+}+c_{12}^2A_{32}^{\beta+}+c2_{12}A_{21}^{\gamma+})c_{\delta}  -(s_{12}^2 B_{31}^{\alpha-}+c_{12}^2 B_{32}^{\beta-}+B_{21}^{\gamma-})s_{\delta}\big\} \\& 
\qquad\quad-s2_{12}c_{23}^2s_{13}^2 \big\{c_{\delta}^2(A_{31}^{\alpha+}-A_{32}^{\beta+})+s2_{\delta}B_{31}^{\alpha -}+s_{\delta}^2(D_{31}^{\alpha+}-D_{32}^{\beta+})\big\}+A_{21}^{\gamma+}s2_{12}(c_{23}^2s_{13}^2+s_{23}^2) 
\bigg],
\end{split}
\end{equation}

\begin{equation}
\begin{split}
&\frac{d\delta}{dt}=-\frac{Cy_{\tau}^2}{16\pi^2}\bigg[
(s_{23}^2-\frac{s2_{23}}{s2_{12}}s_{13}c_{\delta}c2_{12}-c_{23}^2s_{13}^2)B_{21}^{\gamma-}+\frac{s2_{23}}{s2_{12}}s_{13}s_{\delta} D_{21}^{\gamma+} \\&
\qquad\quad+\bigg(\frac{c_{12}c2_{23}}{s_{23}}+\frac{c_{23}c_{\delta}(c_{12}^2s_{13}^2-s_{12}^2)}{s_{12}s_{13}}\bigg) \bigg(c_{12}s_{23}B_{32}^{\beta-}+s_{12}c_{23}s_{13}(c_{\delta}B_{32}^{\beta-}-s_{\delta}B_{32}^{\beta+})\bigg)\\&
\qquad\quad+\frac{c_{23}s_{\delta}(c_{12}^2s_{13}^2-s_{12}^2)}{s_{12}s_{13}}
\bigg(-c_{12}s_{23}  A_{32}^{\beta+}+s_{12}c_{23}s_{13}(-c_{\delta}A_{32}^{\beta+}+s_{\delta}B_{31}^{\alpha -})\bigg)\\&
\qquad\quad-\bigg(\frac{c_{23}c_{\delta}(c_{12}^2-s_{12}^2s_{13}^2)}{s_{13}c_{12}}-\frac{s_{12}c2_{23}}{s_{23}}\bigg) \bigg(-s_{12}s_{23}B_{31}^{\alpha -}+c_{12}c_{23}s_{13}(c_{\delta}B_{31}^{\alpha -}+s_{\delta}D_{31}^{\alpha+})\bigg)\\&
\qquad\quad-\frac{c_{23}s_{\delta}(c_{12}^2-s_{12}^2s_{13}^2)}{s_{13}c_{12}}\bigg( s_{12}c_{23}A_{31}^{\alpha+}+c_{12}c_{23}s_{13}(-c_{\delta}A_{31}^{\alpha-}+s_{\delta}B_{31}^{\alpha -})\bigg)
 \bigg],
\end{split}
\end{equation}

\begin{equation}
\begin{split}
&\frac{d\alpha}{dt}=\frac{Cy_{\tau}^2}{16\pi^2}\bigg[
(c_{12}^2s_{23}^2-\frac{c_{12}}{s_{12}}\frac{s2_{23}}{2}s_{13}c_{\delta}c2_{12}-c_{12}^2c_{23}^2s_{13}^2)B_{21}^{\gamma-}+\frac{c_{12}}{s_{12}}s_{23}c_{23}s_{13}s_{\delta}D_{21}^{\gamma+} \\&
\qquad\quad+\bigg(\frac{c_{12}c2_{23}}{s_{23}}+\frac{c_{23}s_{13}c2_{12}}{s_{12}}c_{\delta}\bigg) \bigg(c_{12}s_{23}B_{32}^{\beta-}+s_{12}c_{23}s_{13}(c_{\delta}B_{32}^{\beta-}-s_{\delta}B_{32}^{\beta+})\bigg)\\&
\qquad\quad-\frac{c_{23}s_{13}c2_{12}}{s_{12}}s_{\delta}\bigg(-c_{12}s_{23}  A_{32}^{\beta+}+s_{12}c_{23}s_{13}(-c_{\delta}A_{32}^{\beta+}+s_{\delta}B_{31}^{\alpha -})\bigg)\\&
\qquad\quad-\bigg(\frac{s_{12}c2_{23}}{s_{23}}+2c_{12}c_{23}s_{13}c_{\delta}\bigg) 
\bigg(-s_{12}s_{23}B_{31}^{\alpha -}+c_{12}c_{23}s_{13}(c_{\delta}B_{31}^{\alpha -}+s_{\delta}D_{31}^{\alpha+})\bigg)\\&
\qquad\quad+2c_{12}c_{23}s_{13}s_{\delta}\bigg( s_{12}c_{23}A_{31}^{\alpha+}+c_{12}c_{23}s_{13}(-c_{\delta}A_{31}^{\alpha-}+s_{\delta}B_{31}^{\alpha -})\bigg)
 \bigg],
\end{split}
\end{equation}

\begin{equation}
\begin{split}
&\frac{d\beta}{dt}=\frac{Cy_{\tau}^2}{16\pi^2}\bigg[
(s_{12}^2s_{23}^2-\frac{s_{12}}{c_{12}}\frac{s2_{23}}{2}s_{13}c_{\delta}c2_{12}-s_{12}^2c_{23}^2s_{13}^2)B_{21}^{\gamma-}+\frac{s_{12}}{c_{12}}s_{23}c_{23}s_{13}s_{\delta}D_{21}^{\gamma+} \\&
\qquad\quad+\bigg(\frac{c_{12}c2_{23}}{s_{23}}-2s_{12}c_{23}s_{13}c_{\delta}\bigg)
 \bigg(c_{12}s_{23}B_{32}^{\beta-}+s_{12}c_{23}s_{13}(c_{\delta}B_{32}^{\beta-}-s_{\delta}B_{32}^{\beta+}\bigg) \\&
\qquad\quad+2s_{12}c_{23}s_{13}s_{\delta}\bigg(-c_{12}s_{23}  A_{32}^{\beta+}+s_{12}c_{23}s_{13}(-c_{\delta}A_{32}^{\beta+}+s_{\delta}B_{31}^{\alpha -})\bigg)\\&
\qquad\quad-\bigg(\frac{s_{12}c_{13}c2_{23}}{s_{23}}+\frac{s_{13}c2_{12}c_{\delta}c_{23}}{c_{12}}\bigg) 
\bigg(-s_{12}s_{23}B_{31}^{\alpha -}+c_{12}c_{23}s_{13}(c_{\delta}B_{31}^{\alpha -}+s_{\delta}D_{31}^{\alpha+})\bigg)\\&
\qquad\quad +\frac{s_{12}c_{23}c2_{12}s_{\delta}}{c_{12}}\bigg( s_{12}c_{23}A_{31}^{\alpha+}+c_{12}c_{23}s_{13}(-c_{\delta}A_{31}^{\alpha-}+s_{\delta}B_{31}^{\alpha -})\bigg)
 \bigg],
\end{split}
\end{equation}

\begin{equation}
\begin{split}
&\frac{d\phi_1}{dt}=-\frac{Cy_{\tau}^2}{16\pi^2}\bigg[
(s_{23}^2-\frac{s2_{23}}{s2_{12}}s_{13}c_{\delta}c2_{12}-c_{23}^2s_{13}^2)B_{21}^{\gamma-}+\frac{s2_{23}}{s2_{12}}s_{13}s_{\delta}D_{21}^{\gamma+}  \\&
\qquad\quad+\bigg(\frac{c_{12}c2_{23}}{s_{23}}+\frac{c_{23}s_{13}c_{\delta}c2_{12}}{s_{12}}\bigg)
 \bigg(c_{12}s_{23}B_{32}^{\beta-}+s_{12}c_{23}s_{13}(c_{\delta}B_{32}^{\beta-}-s_{\delta}B_{32}^{\beta+})\bigg)\\&
\qquad\quad-\frac{c_{23}s_{13}c_{\delta}c2_{12}}{s_{12}}\bigg(-c_{12}s_{23}  A_{32}^{\beta+}+s_{12}c_{23}s_{13}(-c_{\delta}A_{32}^{\beta+}+s_{\delta}B_{31}^{\alpha -})\bigg)\\&
\qquad\quad+\bigg(\frac{s_{12}c2_{23}}{s_{23}}+\frac{c_{23}s_{13}c_{\delta}c2_{12}}{c_{12}}\bigg) 
\bigg(-s_{12}s_{23}B_{31}^{\alpha -}+c_{12}c_{23}s_{13}(c_{\delta}B_{31}^{\alpha -}+s_{\delta}D_{31}^{\alpha+})\bigg)\\&
\qquad\quad+\frac{c_{23}s_{13}c_{\delta}c2_{12}}{s_{12}}\bigg( s_{12}c_{23}A_{31}^{\alpha+}+c_{12}c_{23}s_{13}(-c_{\delta}A_{31}^{\alpha-}+s_{\delta}B_{31}^{\alpha -})\bigg)
 \bigg],
\end{split}
\end{equation}

\begin{equation}
\begin{split}
&\frac{d\phi_2}{dt}=\frac{Cy_{\tau}^2}{16\pi^2c_{13}c_{23}}\bigg[ 
\bigg(s_{12}c_{23}+c_{12}s_{23}s_{13}c_{\delta}\bigg)
\bigg(-s_{12}s_{23}B_{31}^{\alpha -}+c_{12}c_{23}s_{13}(c_{\delta}B_{31}^{\alpha -}+s_{\delta}D_{31}^{\alpha+})\bigg)  \\&
\qquad\quad -c_{12}s_{23}s_{13}s_{\delta}
\bigg(s_{12}c_{23}A_{31}^{\alpha+}+c_{12}c_{23}s_{13}(-c_{\delta}A_{31}^{\alpha-}+s_{\delta}B_{31}^{\alpha -})\bigg)\\&
\qquad\quad -\bigg(c_{12}c_{23}-s_{12}s_{23}s_{13}c_{\delta}\bigg)
\bigg(c_{12}s_{23}B_{32}^{\beta-}+s_{12}c_{23}s_{13}(c_{\delta}B_{32}^{\beta-}-s_{\delta}B_{32}^{\beta+})\bigg)\\&
\qquad\quad -s_{12}s_{23}s_{13}s_{\delta}
\bigg(-c_{12}s_{23}  A_{32}^{\beta+}+s_{12}c_{23}s_{13}(-c_{\delta}A_{32}^{\beta+}+s_{\delta}B_{31}^{\alpha -})\bigg)
  \bigg],
\end{split}
\label{30}
\end{equation}

\begin{equation}
\begin{split}
&\frac{d\phi_3}{dt}=\frac{Cy_{\tau}^2}{16\pi^2c_{13}c_{23}}\bigg[ 
-\bigg(s_{12}s_{23}-c_{12}c_{23}s_{13}c_{\delta}\bigg)
\bigg(-s_{12}s_{23}B_{31}^{\alpha -}+c_{12}c_{23}s_{13}(c_{\delta}B_{31}^{\alpha -}+s_{\delta}D_{31}^{\alpha+})\bigg)  \\&
\qquad\quad -c_{12}c_{23}s_{13}s_{\delta}
\bigg(s_{12}c_{23}A_{31}^{\alpha+}+c_{12}c_{23}s_{13}(-c_{\delta}A_{31}^{\alpha-}+s_{\delta}B_{31}^{\alpha -})\bigg)\\&
\qquad\quad +\bigg(c_{12}s_{23}+s_{12}c_{23}s_{13}c_{\delta}\bigg)
\bigg(c_{12}s_{23}B_{32}^{\beta-}+s_{12}c_{23}s_{13}(c_{\delta}B_{32}^{\beta-}-s_{\delta}B_{32}^{\beta+})\bigg)\\&
\qquad\quad -s_{12}c_{23}s_{13}s_{\delta}
\bigg(-c_{12}s_{23}  A_{32}^{\beta+}+s_{12}c_{23}s_{13}(-c_{\delta}A_{32}^{\beta+}+s_{\delta}B_{31}^{\alpha -})\bigg)
  \bigg].
\end{split}
\label{30}
\end{equation}
In the above equations, $c2_{ij}=\cos2\theta_{ij}, \quad s2_{ij}=\sin2\theta_{ij},  \quad s_\delta=\sin\delta$   and
$$A_{ij}^{x\pm}=\Delta_{ij}\cos^2x\pm\frac{1}{\Delta_{ij}}\sin^2x,$$
$$B_{ij}^{x\pm}=(\Delta_{ij}\pm\frac{1}{\Delta_{ij}})\sin x\cos x,$$
$$D_{ij}^{x\pm}=\Delta_{ij}\sin^2x\pm\frac{1}{\Delta_{ij}}\cos^2x,$$
with $x=\alpha,\beta,\gamma;\quad \ \ i,j=1,2,3$ and $\gamma=\alpha-\beta.$

%%%%%%%%%%%%%%%%%%%%%%%%%%%%%%%%%%%%%%%%%%%%% 4

\section{Numerical analysis and results}
In the previous section, we have derived the one-loop RGEs of three mass eigenvalues, three mixing angles and all six CP phases. In this section we study the running behaviour of all these parameters with the energy scale. This can be done numerically by solving all the coupled RGEs simultaneously.
However, the RGEs of neutrino parameters also involve three gauge couplings, viz., $g_1$, $g_2$ and $g_3$ corresponding to SM gauge groups $U(1)$, $SU(2)$ and $SU(3)$ respectively, three Yukawa couplings, viz., $y_t$, $y_b$ and $y_\tau$ corresponding to top quark, bottom quark and $\tau$ lepton respectively and Higgs quartic coupling $\lambda$. Therefore, we also need to account for the RGEs for these gauge couplings, Yukawa couplings and Higgs quartic coupling. 
In SM, the one-loop RGEs for the gauge and Yukawa couplings are given by \cite{v_SM,Deshpande} 
\begin{equation}
\frac{dg_i}{dt}=\frac{b_i}{16\pi^2}g_i^3,
\label{grge}
\end{equation}
\begin{equation}
\begin{split}
\frac{dy_t}{dt} &=\frac{y_t}{16\pi^2}(4.5y_t^2+1.5y_b^2+y_{\tau}^2-\sum_{i=1}^3c_ig_i^2),
                   \end{split}
                   \label{ytrge}
\end{equation}
\begin{equation}
\begin{split}
\frac{dy_b}{dt} &=\frac{y_b}{16\pi^2}(4.5y_b^2+y_{\tau}^2+1.5y_t^2-\sum_{i=1}^3c^\prime_ig_i^2),
                   \end{split}
\end{equation}
\begin{equation}
\begin{split}
\frac{dy_\tau}{dt} &=\frac{y_\tau}{16\pi^2}(2.5y_\tau^2+3y_b^2+3y_t^2-\sum_{i=1}^3c^{\prime\prime}_ig_i^2),
                   \end{split}
                   \label{y_taurge}
\end{equation}
with the values of the coefficients given by 
\begin{equation}
 b_{1,2,3}= (4.1,\  -3.167 ,\  -7.0),
\label{b_SM}
\end{equation} 
\begin{equation}
\begin{split}
& c_{1,2,3}=(0.85,  2.25, 8), \\
& c^\prime_{1,2,3}=(0.25 , 2.25 , 8),  \\
& c^{\prime\prime}_{1,2,3}=(2.25 , 2.25,  0).
\end{split}
\label{c_SM}
\end{equation}
Further, the one-loop RGE for the Higgs quartic coupling is given by \cite{v_SM, Machacek}
\begin{equation}
\begin{split}
&\frac{d\lambda}{dt} =
\frac{1}{16\pi^2}\bigg[12\lambda^2 
- \left( \frac{9}{5}g_1^2 + 9g_2^2 \right)\lambda 
+ 4(3y_t^2+3y_b^2+y_\tau^2)\lambda+ \frac{9}{4} \left( \frac{3}{25}g_1^4 + \frac{2}{5}g_1^2 g_2^2 + g_2^4 \right)
 \\
&\qquad\qquad\qquad\qquad\qquad\qquad\qquad\qquad\qquad - 4(3y_t^4+3y_b^4+y_\tau^4) \bigg].
\end{split}
\label{lambda_rge}
\end{equation}
Similarly, in the case of MSSM, the RGEs for gauge couplings are same as given in Eq. (\ref{grge}), but with different values of the coefficients given by  
\begin{equation}
 b_{1,2,3}=( 6.6, 1.0 , -3.0). 
\label{b_mssm}
\end{equation}
On the other hand, the one-loop RGEs for Yukawa couplings in MSSM are given by \cite{NSinghSSingh,22,v_SM,ParidaSingh,Barger}
\begin{equation}
\begin{split}
\frac{dy_t}{dt} &=\frac{y_t}{16\pi^2}(6y_t^2+y_b^2-\sum_{i=1}^3c_ig_i^2),
                   \end{split}
                   \label{ytrge_MSSM}
\end{equation}
\begin{equation}
\begin{split}
\frac{dy_b}{dt} &=\frac{y_b}{16\pi^2}(6y_b^2+y_{\tau}^2+y_t^2-\sum_{i=1}^3c^{\prime}_ig_i^2),
                   \end{split}
\end{equation}
\begin{equation}
\begin{split}
\frac{dy_\tau}{dt} &=\frac{y_\tau}{16\pi^2}(4y_\tau^2+3y_b^2-\sum_{i=1}^3c^{\prime\prime}_ig_i^2),
                   \end{split}
                   \label{ytaurge_mssm}
\end{equation}
with the values of the coefficients given by
 \begin{equation}
\begin{split}
& c_{1,2,3}=(13/15,  3,  16/3), \\
& c^{\prime}_{1,2,3}= (7/15,  3, 16/3),  \\
& c^{\prime\prime}_{1,2,3}=
(9/5,  3,  0).
\end{split}
\label{c_mssm}
\end{equation}
\indent To carry out the analysis, we set the high energy flavour symmetry scale at the seesaw scale ($\Lambda_{FS}=10^{14}\ GeV$), where the $\mu-\tau$ reflection symmetry is assumed to be exact. Further, we consider the top quark mass scale ($m_t=172\ GeV$) as the low-energy electroweak scale $\Lambda_{EW}$. To solve the RGEs, we employ a top-down approach and run them all from $\Lambda_{FS}$ to $\Lambda_{EW}$. Including gauge and Yukawa couplings, there are nineteen parameters involved in the RG evolution process and we need initial input values of all these parameters at $\Lambda_{FS}$. The input values of $\theta_{23}$ and all the CP phases are taken as per $\mu-\tau$ reflection symmetry. Remaining mixing parameters, $\theta_{12}$ and $\theta_{13}$ and the mass eigenvalues remain as free parameters at $\Lambda_{FS}$. The input values of $\theta_{12}$ and $\theta_{13}$ are chosen in such a way that their corresponding low-energy values become consistent with global analysis data. Similarly, we try to choose input values of mass eigenvalues such that the corresponding low-energy values 
satisfy the mass squared differences of global analysis data as well as the cosmological upper bound of $\sum m_i$. Apart from the neutrino parameters,  the input values of the gauge and Yukawa couplings at $\Lambda_{FS}$ are determined through a bottom-up approach using low-energy experimental data. The details of the calculations are described below. \\
\indent The gauge coupling constants $g_1$, $g_2$ and $g_3$ are related to the corresponding coupling constants $\alpha_1$, $\alpha_2$ and $\alpha_3$ through the relations
\begin{equation}
g_i=\sqrt{4\pi\alpha_i} ;\ \  (i=1,2,3) .
\label{39}
\end{equation}
The values of $\alpha_i$'s at the $Z$ boson mass scale $M_Z$ can be obtained from those of fine structure constant $\alpha_{em}$, strong coupling constant $\alpha_s$ and Weinberg angle $\theta_W$ at $M_Z$ scale using the matching relations-
\begin{equation}
\frac{1}{\alpha_{em}(M_Z)}=\frac{5}{3} \frac{1}{\alpha_1(M_Z)} +\frac{1}{\alpha_2(M_Z)},
\label{40}
\end{equation}
\begin{equation}
\sin^2\theta_W=\frac{\alpha_{em}(M_Z)}{\alpha_2(M_Z)},
\label{42}
\end{equation}
\begin{equation}
\alpha_3(M_Z)=\alpha_s(M_Z).
\label{alpha3}
\end{equation}
The values of $\alpha_{em}$, $\alpha_s$ and $\theta_{W}$ at $M_Z$ scale are given by
$\alpha_{em}(M_Z)=1/127.952$, $\alpha_s(M_Z)=0.1179$ and $\sin^2\theta_W(M_Z)=0.2223$ \cite{11,12,13,14}. Using these values and solving Eqs. (\ref{40}-\ref{alpha3}) we get 
\begin{equation}
\alpha^{-1}_{(1,2,3)}(M_Z)= (59.69856,28.4544,8.4817).
\end{equation}
Next the values of $\alpha_i$'s at top quark mass scale $m_t=172\ GeV$ are determined from the matching relations 
\begin{equation}
\frac{1}{\alpha_i(m_t)}=\frac{1}{\alpha_i(M_Z)}-\frac{b_i}{2\pi}ln(\frac{m_t}{M_Z}),
\label{46}
\end{equation}
where the values of $b_i$'s are given in Eq. (\ref{b_SM}) for SM framework. The calculated values are 
\begin{equation}
    \alpha^{-1}_{(1,2,3)}(m_t)= (59.06605,\ 28.064056,\ 8.88058)  .
\end{equation}
Using these values of $\alpha_i$'s at $m_t$ scale in Eq. (\ref{39}) we get 
\begin{equation}
g_{(1,2,3)}(m_t)=(0.46125,0.66239,1.18955).
\label{g_mt}
\end{equation}
Turning to Yukawa coupling constants, the values of $y_t$, $y_b$ and $y_{\tau}$ at $m_t$ scale can be obtained from the relations-
\begin{equation}
y_k(m_t)= \frac{m_k(m_t)}{v} ;\ \ k=t,b,\tau,
\label{49}
\end{equation}
where $m_t(m_t)$, $m_b(m_t)$ and $m_\tau(m_t)$ denote the running physical masses of top quark, bottom quark and $\tau$ lepton respectively at $m_t$ scale. Here $v=174\ GeV$ is the vev of the SM Higgs field. The values of $m_k(m_t)$ ($k=t,b,\tau$) are given by
\begin{equation}
m_t(m_t)=m_t(m_t),\hspace{.5cm} m_b(m_t)=\frac{m_b(m_b)}{\eta_b}, \hspace{.5cm} m_\tau(m_t)=\frac{m_\tau(m_\tau)}{\eta_\tau},
\label{52}
\end{equation}
 where $\eta_b$ and $\eta_\tau$ are the QCD-QED rescaling factors and their values are given by $\eta_b=1.017$ and $\eta_\tau=1.53$ \cite{Deshpande,NSinghSSingh}. Using the physical masses of top quark, bottom quark and tau lepton given by $m_t(m_t)=172.52\ GeV$, $m_b(m_b)=4.183\ GeV$ and $m_\tau(m_\tau)=1.777\ GeV$ \cite{PDG}, in Eqs. (\ref{49}) and (\ref{52}), the values of Yukawa couplings at $m_t$ scale are calculated as
\begin{equation}
y_{(t,b,\tau)}(m_t)=(0.9917,\ 0.01571,\ 0.01006 ).
\label{y_mt}
\end{equation}
\begin{table}[t]
\centering
\begin{tabular}{c c c c c}
\hline
\textbf{Parameter} & \textbf{Input value at $m_t$ scale} & \multicolumn{3}{c}{\textbf{Output value at $\Lambda_{FS}$ for different $\Lambda_s$}} \\ \cline{3-5}
              &            & \textbf{$1TeV$} & \textbf{$7TeV$} & \textbf{$14TeV$} \\ \hline
$g_1$         & 0.46125    & 0.633482  &  0.625793  & 0.623121         \\ 
$g_2$         &0.66239     & 0.702064  & 0.684943   & 0.679140         \\ 
$g_3$         & 1.18955    & 0.745271   & 0.721910  & 0.715024         \\ 
$y_t$         & 0.9917     & 0.763613   & 0.700936 &  0.685736         \\ 
$y_b$         & 0.01571      & 0.679651   & 0.601777  & 0.583075         \\ 
$y_{\tau}$    & 0.01006    & 0.779141  & 0.735524   & 0.725049         \\ \hline
\end{tabular}
\caption{Input values of gauge and Yukawa coupling at $m_t=172GeV$ and corresponding output values at $\Lambda_{FS}=10^{14}GeV$ for three different values of $\Lambda_s$.} 
\label{gandy}
\end{table}
Now using the values of gauge and Yukawa couplings at $m_t$ scale calculated in Eqs. (\ref{g_mt}) and (\ref{y_mt}) as inputs, we determine their high-energy values at $\Lambda_{FS}=10^{14}\ GeV$ by solving the relevant RGEs. Due to the consideration of SUSY breaking scale $\Lambda_s=1TeV/7TeV/14TeV$ intermediate between $m_t$ and $\Lambda_{FS}$, we perform the running of RGEs in two steps. In the first step, we run the RGEs given in Eqs. (\ref{grge}-\ref{y_taurge}), with values of $b_i$'s, $c_i$'s, $c_i^\prime$'s 
and $c_i^{\prime\prime}$'s given in Eqs. (\ref{b_SM}) and (\ref{c_SM}), from $m_t$ to $\Lambda_{s}$ within the SM framework. In the next step, we run the RGEs given in Eq. (\ref{grge}) (with the values of coefficients $b_i$'s given in Eq. (\ref{b_mssm})) and those in Eqs. (\ref{ytrge_MSSM}-\ref{ytaurge_mssm}),  with the values of $c_i$'s, $c^\prime_i$'s and $c^{\prime\prime}_i$'s given in Eq. (\ref{c_mssm}), from $\Lambda_s$ to $\Lambda_{FS}$ within MSSM framework.
During the transformation from the SM framework to the MSSM framework, we use the following matching relations for gauge and Yukawa couplings:
\begin{equation}
 \begin{split}
        & g_i(SUSY)=g_i(SM), \\&\ y_t(SUSY)=\frac{y_t(SM)}{\sin\beta}, \\& \ y_b(SUSY)=\frac{y_b(SM)}{\cos\beta}, \\& \ y_\tau(SUSY)=\frac{y_\tau(SM)}{\cos\beta}.
\end{split} 
\label{matching}
\end{equation}
We have chosen a higher value of $\tan\beta=58$ during the RG running. The output values of gauge and Yukawa couplings at $\Lambda_{FS}$ obtained for three different choices of $\Lambda_s=1\ TeV$, $7\ TeV$ and $14\ TeV$ are given in Table \ref{gandy}. 

\indent With all the values of gauge and Yukawa couplings determined at $\Lambda_{FS}$ (Table \ref{gandy}), we now solve the RGEs of neutrino parameters following a top-down approach. Working within the framework of MSSM, as we have considered three distinct SUSY breaking scales, namely $\Lambda_s = 1\,\mathrm{TeV},\ 7\,\mathrm{TeV}$ and $14\,\mathrm{TeV}$, lying between $\Lambda_{\mathrm{FS}}$ and $\Lambda_{\mathrm{EW}}$, this leads to a two-step RGE evolution:
\begin{itemize}
    \item \textbf{Step I}: Running the RGEs from $\Lambda_{\mathrm{FS}}$ to $\Lambda_s$ within the MSSM framework. This step involves eighteen coupled RGEs: three for mass eigenvalues, three for mixing angles, six for CP phases (including Dirac, Majorana, and unphysical phases), three for gauge couplings and three for Yukawa couplings. The relevant RGEs are Eqs.~(\ref{m1_regMSSM})--(\ref{m3_rgeMSSM}), Eqs.~(\ref{22})--(\ref{30}) with $C =1 $, Eq. (\ref{grge}) with $b_i$'s in Eq. (\ref{b_mssm}) and Eqs.~(\ref{ytrge_MSSM})--(\ref{ytaurge_mssm}), with coefficients given in Eq. (\ref{c_mssm}).

%%%%%%%%%%%%%%%%%%%%5555555555555555555555555555

\begin{table}[t]
\begin{center}
\begin{tabular}{c cc cc cc}
\hline
\multirow{2}{*}{Parameter}& 
\multicolumn{2}{c}{$\Lambda_s=1\ TeV$}&
\multicolumn{2}{c}{$\Lambda_s=7\ TeV$}&
\multicolumn{2}{c}{$\Lambda_s=14\ TeV$} \\
\cline{2-7}
 &\makecell{Input at \\ $\Lambda_{FS}$}  & \makecell{Output at \\$\Lambda_{EW}$} &\makecell{Input at \\ $\Lambda_{FS}$}&\makecell{Output at \\ $\Lambda_{EW}$ }&\makecell{Input at \\ $\Lambda_{FS}$}&\makecell{ Output at \\ $\Lambda_{EW}$} \\ \hline
 $m_1\ (eV)$& 0.01071 & 0.015496 & 0.01071 & 0.015277 & 0.01071 & 0.015172 \\
 $m_2\ (eV)$& 0.01295 & 0.017594 & 0.01295 & 0.017506 & 0.01295 & 0.017445 \\
 $m_3\ (eV)$& 0.03653 & 0.053810 & 0.03653 & 0.052693 & 0.03653 & 0.052229 \\
$\theta_{13} (/^\circ)$&8.57 & 8.5649 & 8.57 & 8.5657 & 8.57 & 8.5660\\
$\theta_{12} (/^\circ)$& 34 & 33.7228 & 34 & 33.7408 & 34 & 33.7456 \\
$\theta_{23} (/^\circ)$& 45 & 45.1032 & 45 & 45.0872 & 45 & 45.0819 \\
 $\delta (/^\circ)$& 90 & 90.0390 & 90 & 90.0220 & 90 & 90.0173 \\
 $\alpha (/^\circ)$& 270 & 270.0357 & 270 & 270.0337 & 270 & 270.0326 \\
 $\beta (/^\circ)$& 270 & 269.9427 & 270 & 269.9583 & 270 & 269.9629 \\
 $\phi_1 (/^\circ)$& 90 & 90.0097 & 90 & 89.9959 & 90 &89.9926 \\
 $\phi_2 (/^\circ)$& 0 & -0.0027 & 0 &-0.0016 & 0 & -0.0012 \\
 $\phi_3 (/^\circ)$& 0 &0.0479  & 0 &0.0374 & 0 & 0.0340\\
 $g_1$& 0.633482 & 0.461247 & 0.625793 & 0.461245 & 0.623121 & 0.461245 \\
 $g_2$& 0.702064 & 0.662409 & 0.684943 & 0.662409 & 0.679140 & 0.662409 \\
 $g_3$& 0.745271 & 1.210904 & 0.721910 & 1.193680 & 0.715024 & 1.191966\\
 $y_t$& 0.763613 & 0.988790 & 0.700936 & 0.967550 & 0.685736 & 0.963089 \\
 $y_b$& 0.679651 & 0.899016 & 0.601777 & 0.860135 & 0.583075& 0.850471\\
 $y_{\tau}$& 0.779141 & 0.569773 & 0.735524 & 0.555126 & 0.725049 & 0.550269\\
 $\Delta m^2_{21}(10^{-5}eV^2)$&- & 6.94 & - & 7.31 &- & 7.41 \\
 $\Delta m^2_{32}(10^{-3}eV^2)$&- & 2.58 & - & 2.47 &- & 2.42 \\
 $\sum_i m_i (eV)$&- & 0.08689 & - & 0.08547 &- & 0.08485 \\
 \hline
\end{tabular}
\end{center}
\caption{Input values at $\Lambda_{FS}$ and corresponding low energy values at $m_t$ scale of all the parameters for three different values of $\Lambda_s=1\ TeV,\ 7\ TeV$ and $14\ TeV$ in NO and case-I.}
\label{TNO1}
\end{table} 
\item \textbf{Step II}: Running from $\Lambda_s$ to $\Lambda_{\mathrm{EW}} = m_t$ in the Standard Model (SM) framework. This includes an additional RGE for the Higgs quartic coupling $\lambda$, with a result of total nineteen equations. The relevant RGEs are Eqs.~(\ref{m1_rgeSM})--(\ref{m3_rgeSM}), Eqs.~(\ref{22})--(\ref{30}) with $C = -\frac{3}{2}$, Eq.~(\ref{grge}) with $b_i$'s in Eq. (\ref{b_SM}) and Eqs.~(\ref{ytrge})--(\ref{y_taurge}) with coefficients given in Eq. (\ref{c_SM}).
\end{itemize}
Note that during this two-step evolution of RGEs, we use the matching relations given in Eq. (\ref{matching}) while transforming from the MSSM framework to the SM framework. Apart from the values of gauge and Yukawa couplings given in Table (\ref{gandy}), the values of $\lambda$ at the SUSY breaking scale $\Lambda_{s}$ are found to be $0.48639$, $0.46701$, and $0.46333$ corresponding to $\Lambda_s=1\,\mathrm{TeV}$, $7\,\mathrm{TeV}$, and $14\,\mathrm{TeV}$ respectively. The initial input value of $\lambda$ at the top-quark mass scale (\(m_t\)) is calculated to be $0.52$, from the relation $\lambda =m_h^2/v^2$, where the Higgs boson mass is taken as \(m_h = 125.78~\mathrm{GeV}\). \\
\indent In connection to the two possible mass ordering scenarios, we perform the numerical analysis for both the cases of NO and IO. The analysis and results are discussed in the following two subsections. In the entire analysis we have taken a higher value of $tan\beta=58$ to study the running effects. \\

\begin{figure}[!t]
  \centering
  \begin{tabular}{cc}
   \begin{subfigure}[b]{0.45\textwidth}
    \centering
    \includegraphics[height=5cm]{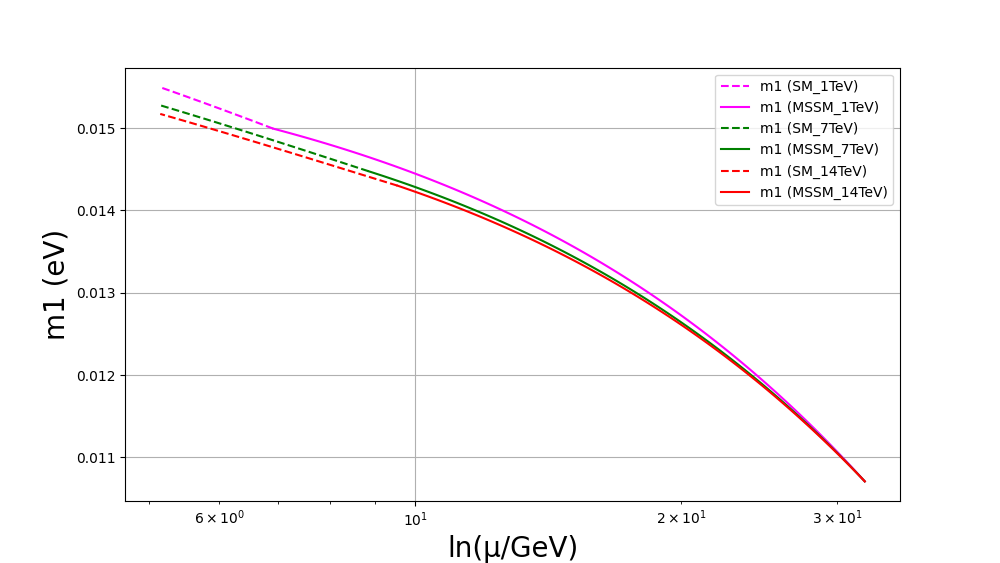}
    \caption{Evolution of $m_1$}
    \label{m1_no1}
  \end{subfigure}&
  \begin{subfigure}[b]{0.45\textwidth}
    \centering
    \includegraphics[height=5cm]{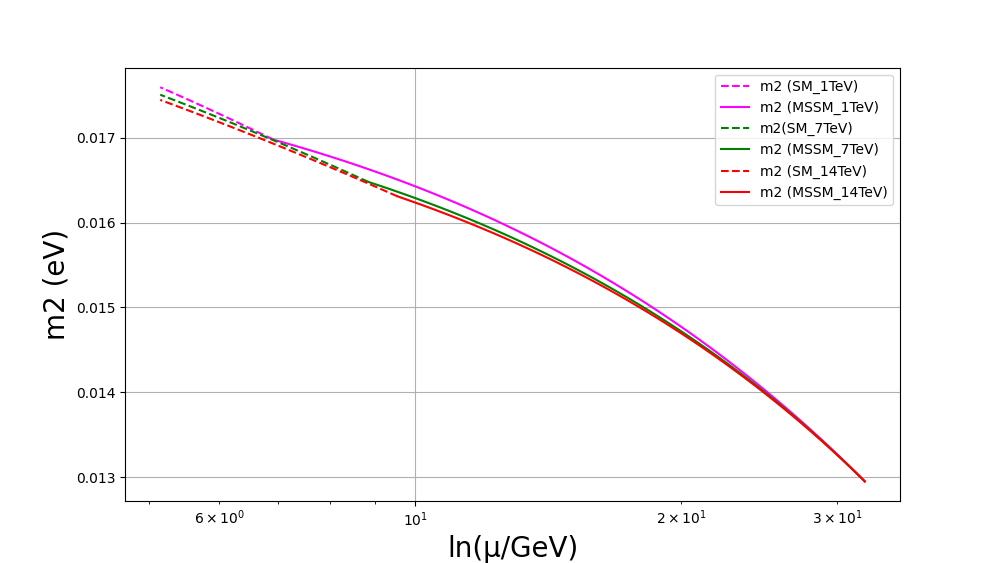}
    \caption{Evolution of $m_2$}
    \label{m2_no1}
  \end{subfigure} \\[2ex]
  \begin{subfigure}[b]{0.45\textwidth}
    \centering
    \includegraphics[height=5cm]{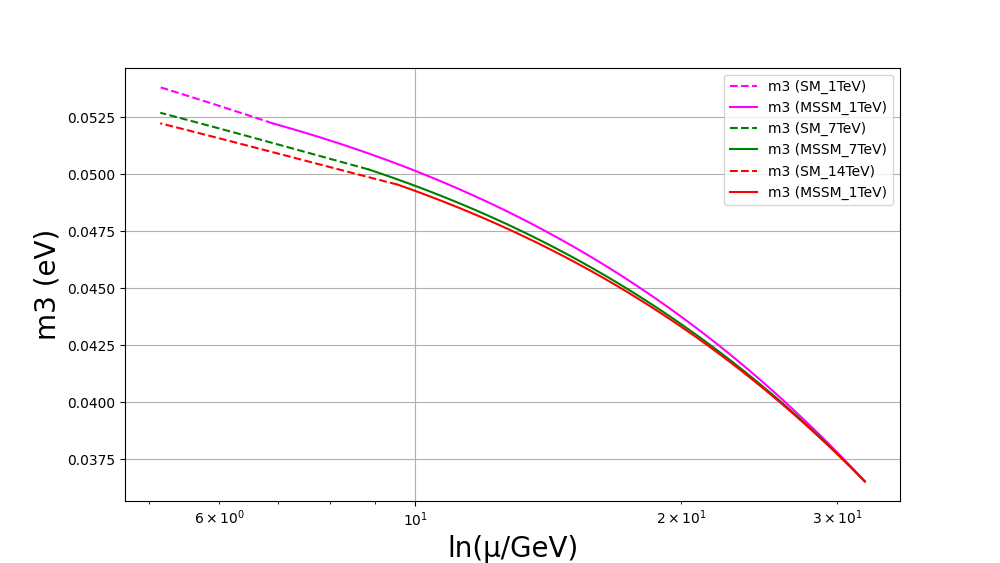}
    \caption{Evolution of $m_3$}
    \label{m3_no1}
  \end{subfigure} &
  \begin{subfigure}[b]{0.45\textwidth}
    \centering
    \includegraphics[height=5cm]{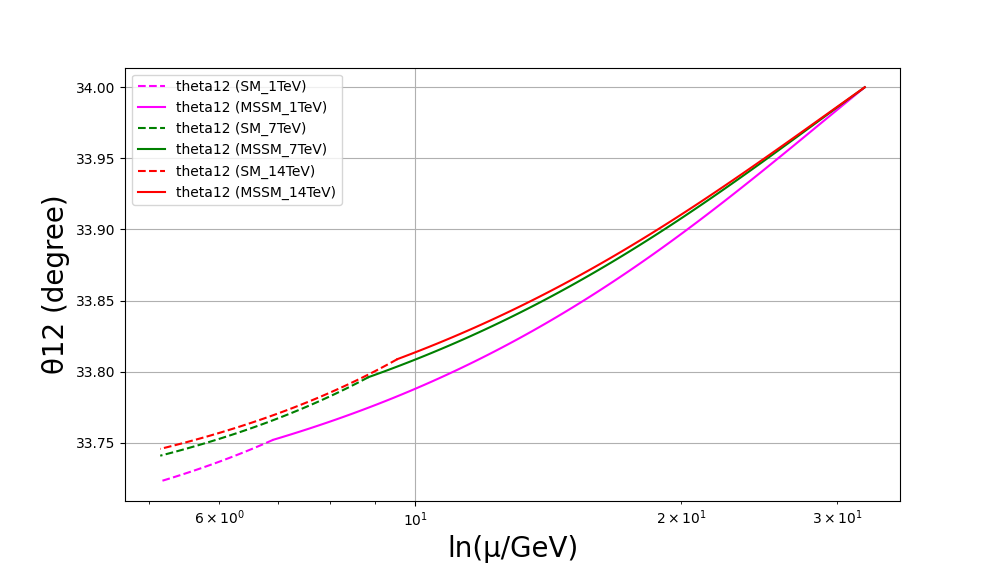}
    \caption{Evolution of $\theta_{12}$}
    \label{theta12_no1}
  \end{subfigure} \\[2ex]
  \begin{subfigure}[b]{0.45\textwidth}
    \centering
    \includegraphics[height=5cm]{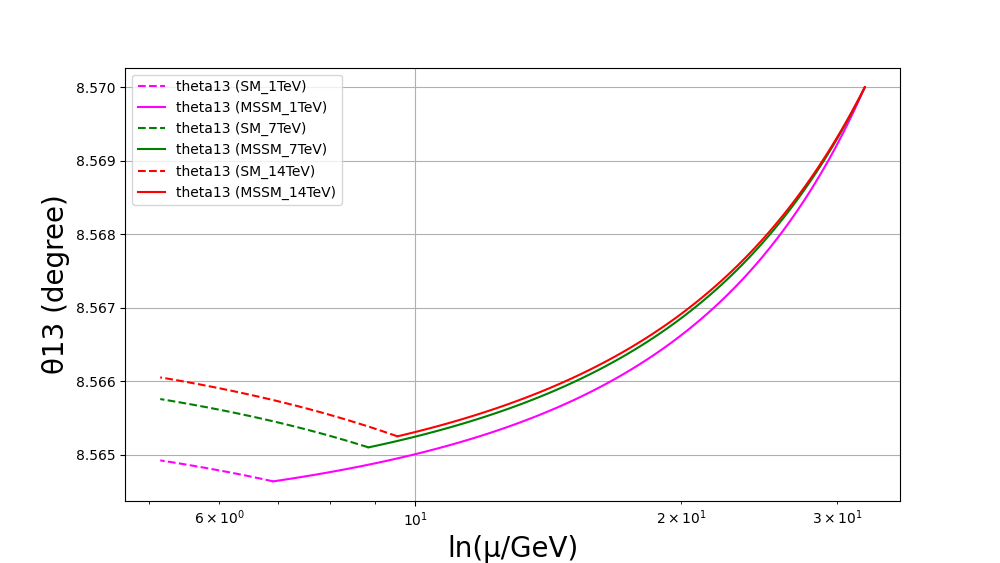}
    \caption{Evolution of $\theta_{13}$}
    \label{theta13_no1}
  \end{subfigure} &
  \begin{subfigure}[b]{0.45\textwidth}
    \centering
    \includegraphics[height=5cm]{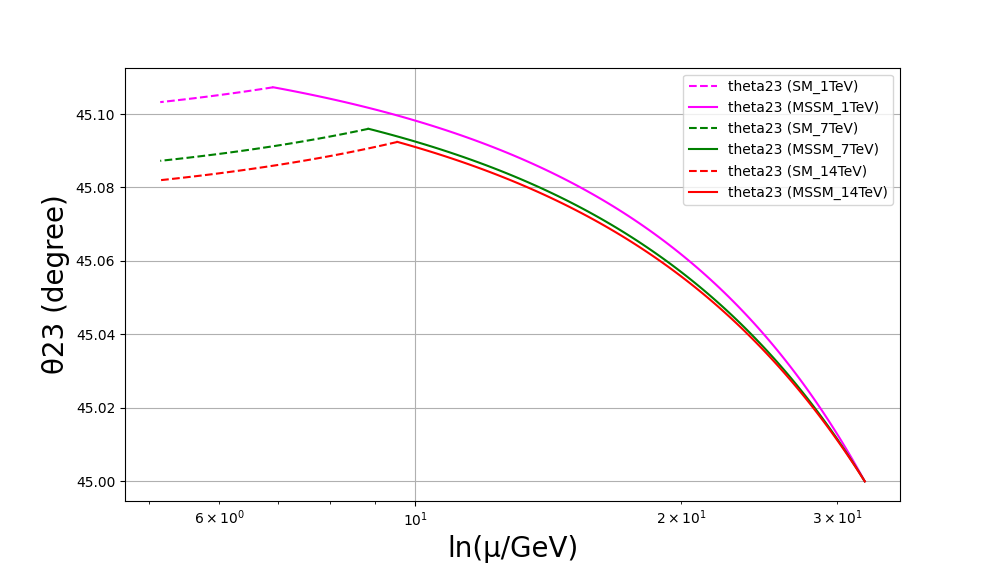}
    \caption{Evolution of $\theta_{23}$}
    \label{theta23_no1}
  \end{subfigure} \\[2ex]
    \end{tabular}
  \caption{Evolution of mass eigenvalues and mixing angles with energy scale for NO and case-I with three different values of $\Lambda_s$. Solid and dashed portions of each curve represent the evolution in the MSSM and SM regions respectively. Magenta , green and red colors respectively stands for $\Lambda_s=1\ TeV,\ 7\ TeV\ \text{and}\ 14\ TeV$.}
  \label{FNO11}
\end{figure}
\subsection{Normal Order ($m_1<m_2<m_3$)}
\indent For NO, we choose the high energy input values of mass eigenvalues at $\Lambda_{FS}$, maintaining the order $m_1<m_2<m_3$. As the reflection symmetry allows two possible values of $\delta$, we have identified two separate cases, viz., case I ($\theta_{23}=\pi/4$ and $\delta=\pi/2$) and case II ($\theta_{23}=\pi/4$ and $\delta=3\pi/2$) in section 2. Accordingly the analysis is carried out separately for the two different cases.\\
\indent For case I, the numerical results obtained by solving the full set of RGEs are presented in Table \ref{TNO1}. 
\begin{figure}[!t]
  \centering
  \begin{tabular}{cc}
  \begin{subfigure}[b]{0.45\textwidth}
    \centering
    \includegraphics[height=5cm]{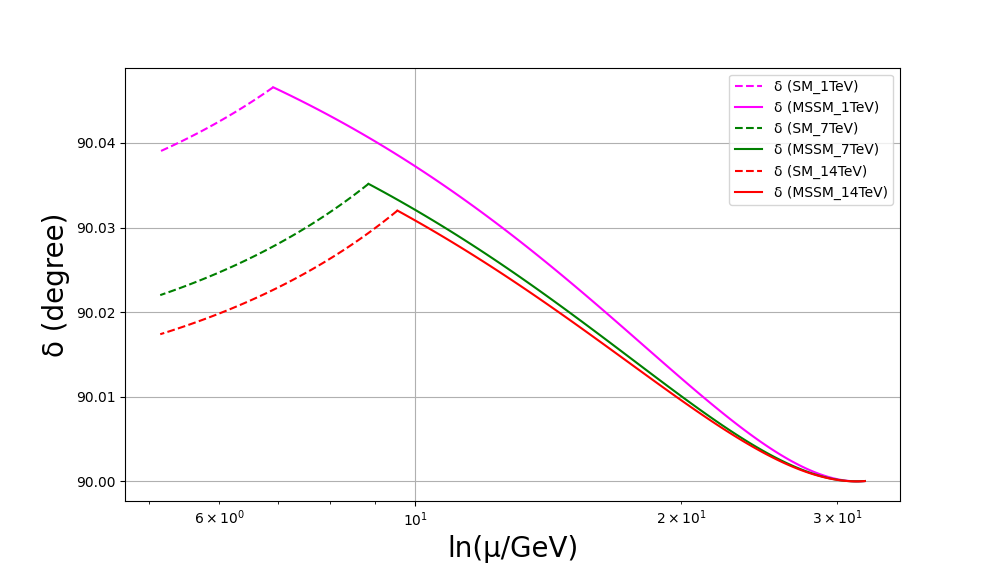}
    \caption{Evolution of $\delta$}
    \label{delta_no1}
  \end{subfigure} &
  \begin{subfigure}[b]{0.45\textwidth}
    \centering
    \includegraphics[height=5cm]{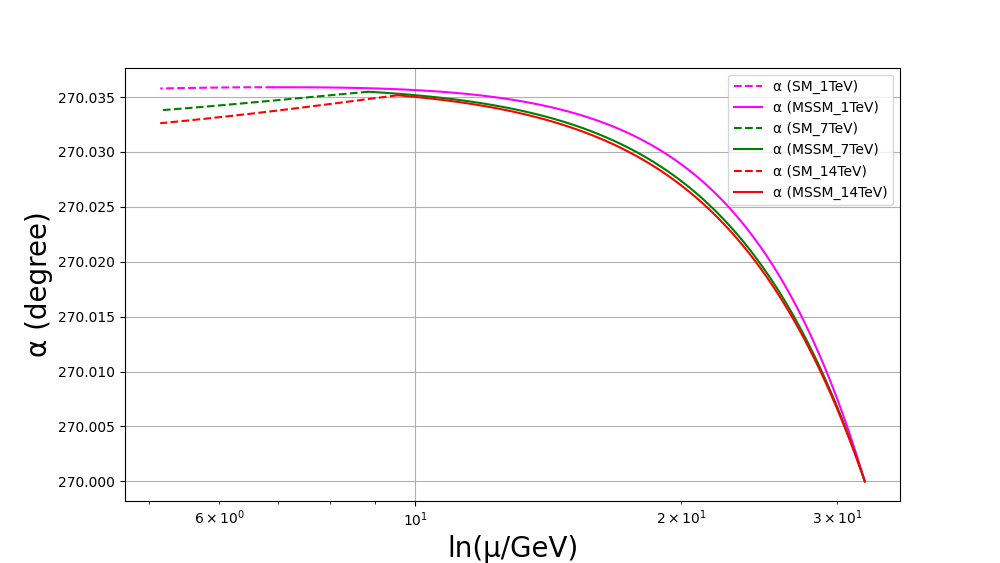}
    \caption{Evolution of $\alpha$}
  \end{subfigure} \\[2ex]
  \begin{subfigure}[b]{0.45\textwidth}
    \centering
    \includegraphics[height=5cm]{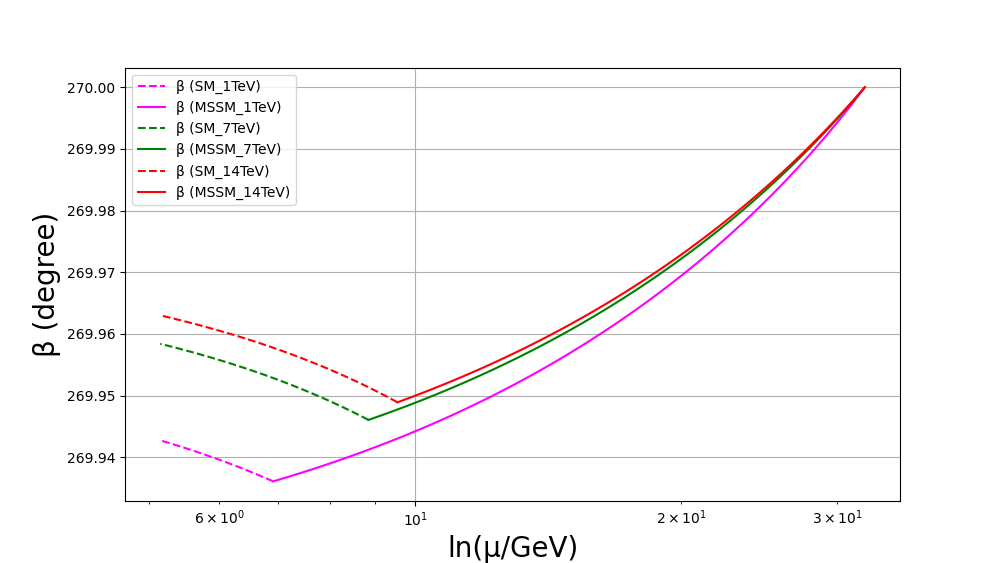}
    \caption{Evolution of $\beta$ }
  \end{subfigure} &
  \begin{subfigure}[b]{0.45\textwidth}
    \centering
    \includegraphics[height=5cm]{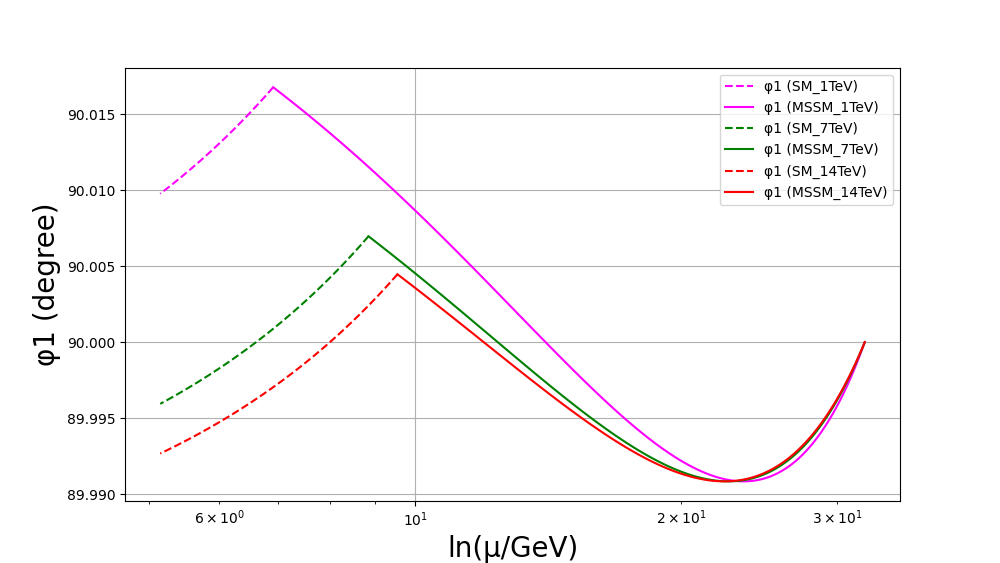}
    \caption{Evolution of $\phi_1$}
  \end{subfigure} \\[2ex]
  \begin{subfigure}[b]{0.45\textwidth}
    \centering
    \includegraphics[height=5cm]{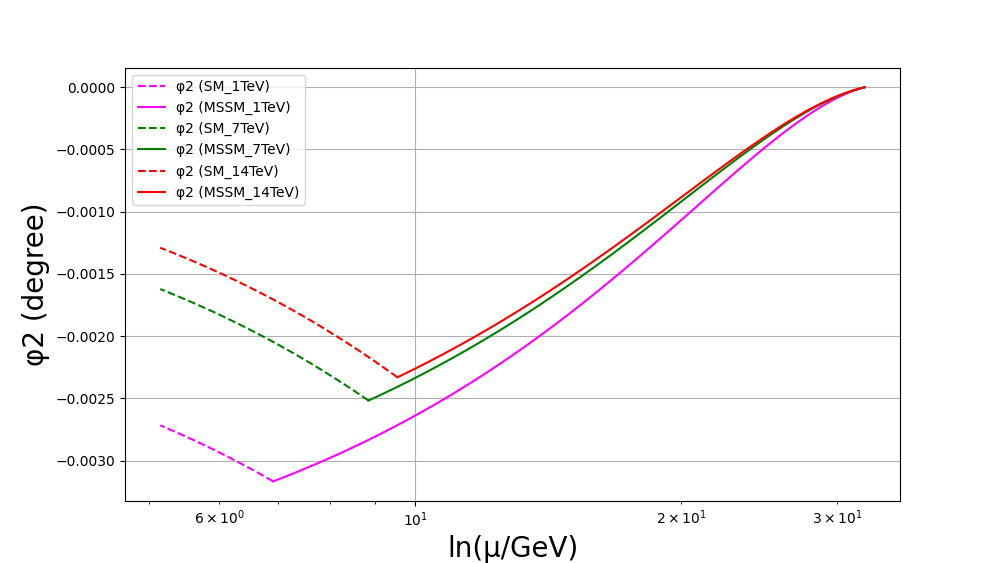}
    \caption{Evolution of $\phi_2$}
  \end{subfigure} & 
  \begin{subfigure}[b]{0.45\textwidth}
    \centering
    \includegraphics[height=5cm]{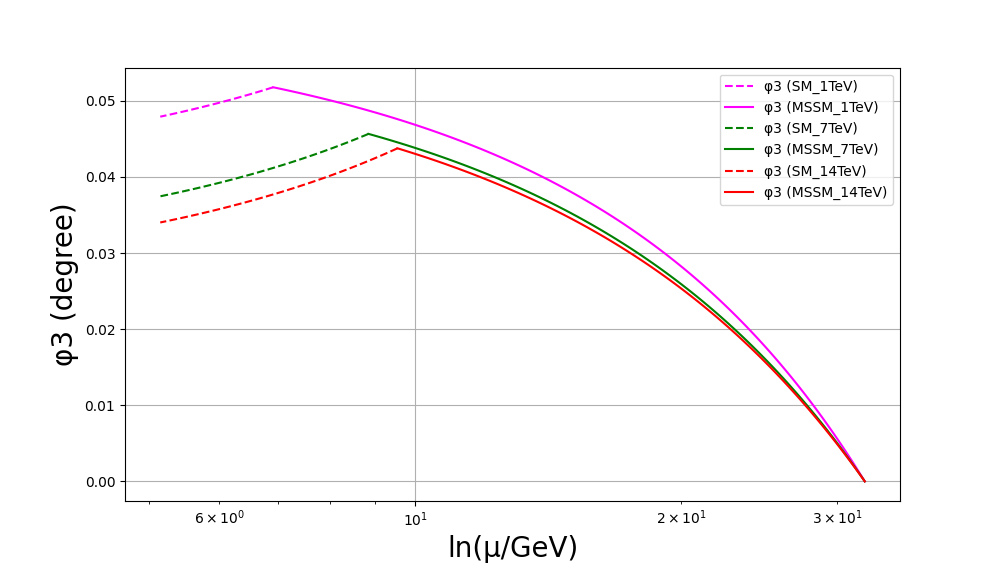}
    \caption{Evolution of $\phi_3$}
    \label{phi3_no1}
  \end{subfigure}  
  \end{tabular}
  \caption{Evolution of CP phases with energy scale for NO and case-I with three different values of $\Lambda_s$. Solid and dashed portions of each curve represent the evolution in the MSSM and SM regions respectively. Magenta , green and red colors respectively stands for $\Lambda_s=1\ TeV,\ 7\ TeV\ \text{and}\ 14\ TeV$}
  \label{FNO12}
\end{figure}
The table shows the input values of all the parameters taken at $\Lambda_{FS}$ and their corresponding output values obtained at $m_t$ scale for the case of three SUSY breaking scales $\Lambda_s=1\ TeV,\ 7\ TeV\ \text{and}\ 14\ TeV$. The evolutions of mass eigenvalues and mixing angles with energy scales are depicted in Figs. 1 (a)-(f) and those of the CP phases are presented in Figs. 2 (a)-(f). Each figure displays the running behaviour of the given parameter for the three different values of $\Lambda_s$. In all figures, the evolution curves in magenta, green and red colors represent the scenario with $\Lambda_s=1 \ TeV$, $7\ TeV$ and $14\ TeV$ respectively. Further, solid and dashed portions of a given curve represent the evolution in MSSM and SM regions respectively. \\
\indent As for example, let us illustrate the numerical analysis and results for the case of $\Lambda_s=1\ TeV$. Since the input values of mass eigenvalues are treated as free parameters at $\Lambda_{\text{FS}}$, they must be chosen carefully to satisfy low-energy experimental constraints. Even small changes in the mass eigenvalues at $\Lambda_{\text{FS}}$ can lead to substantial variations in the mass-squared differences and the sum of the mass eigenvalues. Within the allowed upper bound of $\sum m_i$, we observe that the mass-squared differences initially approach the experimental best-fit values as the input masses increase. However, beyond a certain point, they begin to deviate again. This optimal set of input mass eigenvalues yields mass-squared differences that are closest to the experimental best-fit values. A small increase or decrease from this optimal set leads to deviations from the best-fit mass-squared differences. The chosen mass eigenvalues for the case of $\Lambda_s=1\ TeV$ are $m_1=0.01071\ eV< m_2=\ 0.01295\ eV<m_3=0.03653\ eV$. Accordingly, we have obtained the low-energy mass eigenvalues at $m_t$ scale as $m_1=0.015496\ eV$, $m_2=\ 0.017594\ eV$ and $m_3=0.053810\ eV$. These low-energy mass eigenvalues yield $\sum m_i \approx 0.087\ eV$, which lies within the cosmological upper bound $0.12\ eV$ \cite{mbound}. They also lead to $\Delta m_{21}^2=6.94\times 10^{-5}\ eV^2$ and $\Delta m_{32}^2=2.58\times 10^{-3}\ eV^2$, which are close to corresponding best-fit values of global analysis data (Table \ref{GA}). From Figs. 1(a)-(c), we see that the mass eigenvalues tend to increase as we run down from high to low energy scales.\\
\indent Similar to the mass eigenvalues, the mixing angles $\theta_{13}$ and $\theta_{12}$ are also treated as free parameters.  We take their input values as $8.57^\circ$ and $34^\circ$ respectively, to ensure that their low-energy values align closely with the experimental best-fit values. We get $\theta_{12}\approx33.75^\circ$ and $\theta_{13}\approx8.56^\circ$ at $m_t$ scale, which are close to the global best-fit values (Table \ref{GA}). Due to RG running effects, their values tend to decrease with the decrease of energy scales (Figs. 1(d) and 1(e). We see that running of $\theta_{13}$ is relatively week as compared to that of $\theta_{12}$. The input value of $\theta_{23}$ is taken as $45^\circ$, as per $\mu$--$\tau$ reflection symmetry. This angle does not exhibit a significant deviation from the maximal value due to RG running; the low-energy output value is obtained to be $45.10^\circ$. Unlike $\theta_{12}$ and $\theta_{13}$, it tends to increase with the decrease of energy scale and thereby prefers to lie in the second octant due to RG running. Note that the current best-fit value of $\theta_{23}$ is $48.5^\circ$ (without SK data) / $43.3^\circ$ (with SK data) in NO scenario (Table \ref{GA}). In view of this, low energy prediction of $\theta_{23}$ due to RG running effects is consistent with the global analysis without SK data as it tends to increase over $45^\circ$ prefering the second octant.\\
\indent For case I, we take the high-energy value of Dirac CP phase $\delta$ at $\Lambda_{FS}$ as $90^\circ$ as per $\mu-\tau$ reflection symmetry and for the rest of the CP phases, the input values are taken as given in Eq. (\ref{ch1}). The deviation of the CP phases from maximal values are found to be very small. For example, the low energy value of $\delta$ is found to be only $90.03^\circ$. Low energy values of other CP phases with respect to the high energy input values can be read off Table \ref{TNO1}. Corresponding RG evolutions of the CP phases are depicted in Figs. 2(a)-(f).\\
\indent To analyse the RG running effects on neutrino parameters, for the case of SUSY breaking scale $\Lambda_s= 7\ TeV$ and $14\ TeV$ in the scenario of NO and case I, we use the same set of input values of parameters as chosen for the case of $\Lambda_s=1\ TeV$. Due to the increase of $\Lambda_s$, mild variations in low energy values of each parameter occur as can be read off Table (\ref{TNO1}). However, low-energy prediction of mass-squared differences and $\sum m_i$ are still found to be consistent with observational data. As for example, low energy value of $\Delta m^2_{21}$ changes from $6.94\times 10^{-5}\ eV^2$ (at $\Lambda_s=1\ TeV$) to $7.31\times 10^{-5}\ eV^2$ at $\Lambda_s=7\ TeV$; $\Delta m^2_{32}$ changes from $2.58\times 10^{-3}\ eV^2$ (at $\Lambda_s=1\ TeV$) to $2.47 \times 10^{-3}$ (at $\Lambda_s=7\ TeV$) and $\sum m_i$ changes from $0.087\ eV$ to $0.085\ eV$. The splitting between the different energy evolution curves in each figure reflects the effects of variation of the SUSY breaking scale on the respective parameter. From Figs. 1(a)-(f), we see that the evolution curves of a given parameter corresponding to $\Lambda_s=1\ TeV,\ 7\ TeV$ and $14\ TeV$ are close to each other. This implies that the running effects as well as the low energy output values do not vary significantly with the variation of $\Lambda_s$. However, in case of CP phases (Figs.2(a)-(f)) we can see that the evolution curves in green and red colors, corresponding to $7\ TeV$ and $14\ TeV$ respectively, are relatively closer to each other but that in magenta color is quite distinct from the former two. This reflects the running effects for $\Lambda_s=7\ TeV/14\ TeV$ are quite distinguishable from those corresponding to $\Lambda_s=1\ TeV$.
\begin{table}[t]
\begin{center}
\begin{tabular}{c cc cc cc}
\hline
\multirow{2}{*}{Parameter}& 
\multicolumn{2}{c}{$\Lambda_s=1\ TeV$}&
\multicolumn{2}{c}{$\Lambda_s=7\ TeV$}&
\multicolumn{2}{c}{$\Lambda_s=14\ TeV$} \\
\cline{2-7}
 &\makecell{Input at \\ $\Lambda_{FS}$}  & \makecell{Output at \\$\Lambda_{EW}$} &\makecell{Input at \\ $\Lambda_{FS}$}&\makecell{Output at \\ $\Lambda_{EW}$ }&\makecell{Input at \\ $\Lambda_{FS}$}&\makecell{ Output at \\ $\Lambda_{EW}$} \\ \hline
 $m_1(eV)$& 0.01235 & 0.016569 & 0.01235 & 0.016722 & 0.01235 & 0.016737 \\
 $m_2(eV)$& 0.01295 & 0.019232 & 0.01295 & 0.018915 & 0.01295 & 0.018779 \\
 $m_3(eV)$& 0.03653 & 0.054028 & 0.03653 & 0.053275 & 0.03653 & 0.052930 \\
$\theta_{13} (/^\circ)$&8.57 & 8.5646 & 8.57 & 8.5657 & 8.57 & 8.5660\\
$\theta_{12} (/^\circ)$& 34 & 33.5479 & 34 & 33.5843 & 34 & 33.5991 \\
$\theta_{23} (/^\circ)$& 45 & 45.0302 & 45 & 45.0248 & 45 & 45.0230 \\
$\delta  (/^\circ)$& 270 & 269.9684 & 270 & 269.9752 & 270 & 269.9774 \\
 $\alpha (/^\circ)$& 90 & 89.9599 & 90 & 270.0337 & 90 & 89.9692 \\
 $\beta (/^\circ)$& 90 & 90.0163 & 90 & 90.0119 & 90 & 90.0106 \\
 $\phi_1 (/^\circ)$& 270 & 269.9971 & 270 & 269.9997 & 270 &270.0005 \\
 $\phi_2 (/^\circ)$& 0 & 0.0053 & 0 &0.0049 & 0 & 0.0048 \\
 $\phi_3 (/^\circ)$& 0 &0.0286  & 0 &0.0229 & 0 & 0.0209\\
 $g_1$& 0.633482 & 0.461267 & 0.625793 & 0.461247 & 0.623121 & 0.461245 \\
 $g_2$& 0.702064 & 0.662409 & 0.684943 & 0.662409 & 0.679140 & 0.662409 \\
 $g_3$& 0.745271 & 1.210904 & 0.721910 & 1.193678 & 0.715024 & 1.191964\\
 $y_t$& 0.763613 & 0.988786 & 0.700936 & 0.967549 & 0.685736 & 0.963088 \\
 $y_b$& 0.679651 & 0.899012 & 0.601777 & 0.860134 & 0.583075& 0.850470\\
 $y_{\tau}$& 0.779141 & 0.569769 & 0.735524 & 0.555125 & 0.725049 & 0.550269\\
 $\Delta m^2_{21}(10^{-5}eV^2)$& -& 9.53 & - & 7.81 &- & 7.25 \\
 $\Delta m^2_{32}(10^{-3}eV^2)$& -& 2.55 & - & 2.48 & -& 2.45 \\
 $\sum_i m_i (eV)$&- & 0.08983 & - & 0.08891 &- & 0.08845 \\
 \hline
\end{tabular}
\end{center}
\caption{Input values at $\Lambda_{FS}$ and corresponding low energy values at $m_t$ scale of all the parameters for three different values of $\Lambda_s=1\ TeV,\ 7\ TeV$ and $14\ TeV$ in NO and case-II. }
\label{TNO2}
\end{table} 

\indent For Case II, the high-energy input value of the Dirac CP-violating phase $\delta$ at $\Lambda_{\text{FS}}$ is taken to be $270^\circ$ along with $\theta_{23}=45^\circ$. The input values of remaining CP phases are taken as specified in Eq.~(\ref{ch2}), consistent with $\mu-\tau$ reflection symmetry. Input values of the mass eigenvalues and the remaining two mixing angles are taken same as those chosen in case I, except $m_1$. To satisfy the low energy constraints on the mass-squared differences, the input value of $m_1$ is adjusted from $0.01071\ eV$ (in case I) to $0.01235\ eV$ in case II. As before, the numerical analysis is carried out for all three SUSY breaking scales $\Lambda_s=1\ TeV,\ 7\ TeV$ and $14\ TeV$ and the results are summarised in Table \ref{TNO2}. To analyse the effects of variation of $\Lambda_s$ on the low energy predictions, we use the same set of input values for all three cases of SUSY breaking scale. The evolution of the mass eigenvalues and mixing angles as functions of the energy scale are shown in Figs. 3(a)–3(f), while the corresponding running behaviors of the CP-violating phases are depicted in Figs. 4(a)-4(f). Each figure demonstrates the RG running of the respective parameter for the three chosen values of $\Lambda_s$. As before, the evolution curves of a given parameter in magenta, green and red colors stand for the scenarios with $\Lambda_s=1\ TeV,\ 7\ TeV$ and $14\ TeV$ respectively. The solid and dashed segments of each curve denote the evolution in the MSSM and SM regions respectively.\\
\begin{figure}[!t]
  \centering
  \begin{tabular}{cc}
   \begin{subfigure}[b]{0.45\textwidth}
    \centering
    \includegraphics[height=5cm]{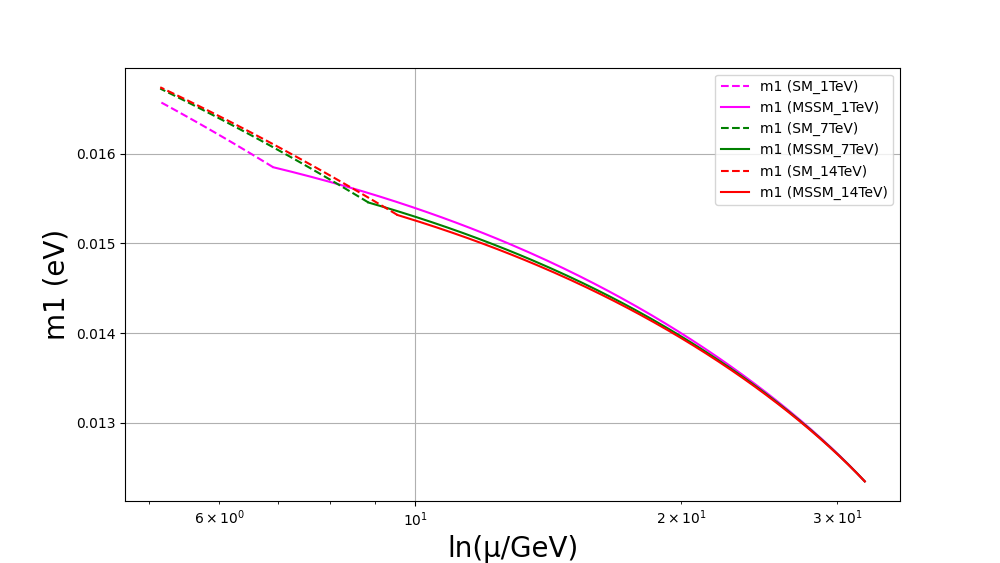}
    \caption{Evolution of $m_1$}
  \end{subfigure}&
  \begin{subfigure}[b]{0.45\textwidth}
    \centering
    \includegraphics[height=5cm]{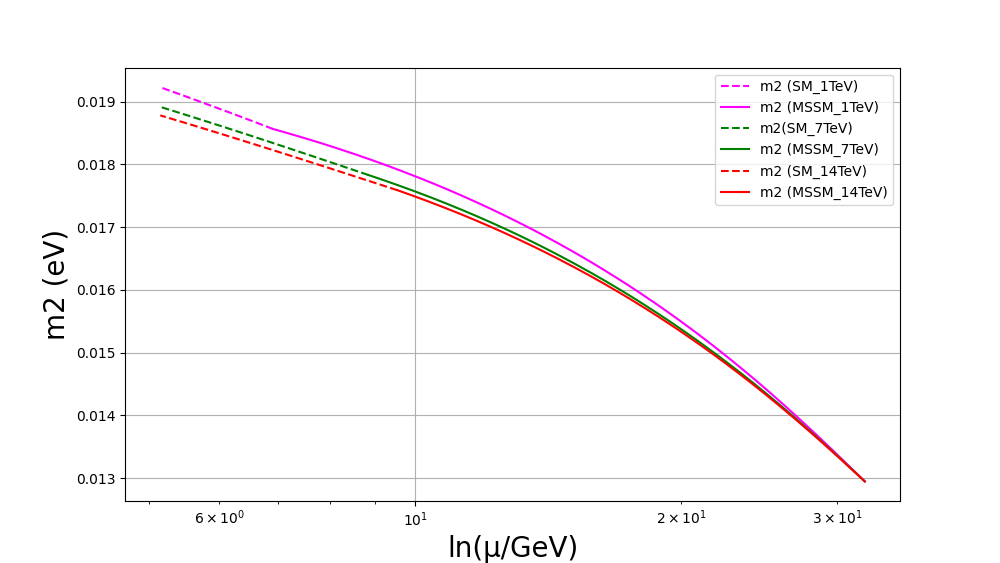}
    \caption{Evolution of $m_2$}
  \end{subfigure} \\[2ex]
  \begin{subfigure}[b]{0.45\textwidth}
    \centering
    \includegraphics[height=5cm]{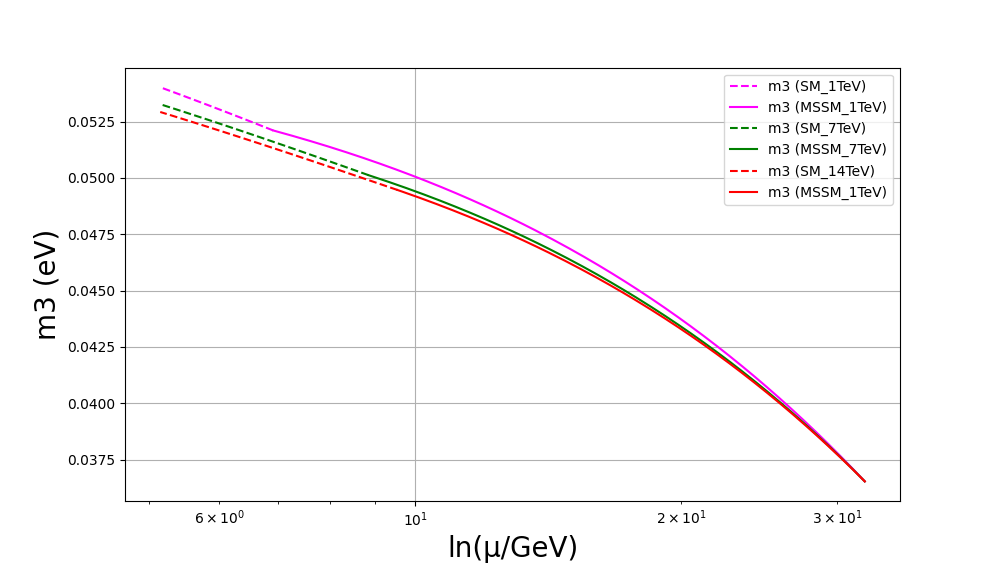}
    \caption{Evolution of $m_3$}
  \end{subfigure} &
  \begin{subfigure}[b]{0.45\textwidth}
    \centering
    \includegraphics[height=5cm]{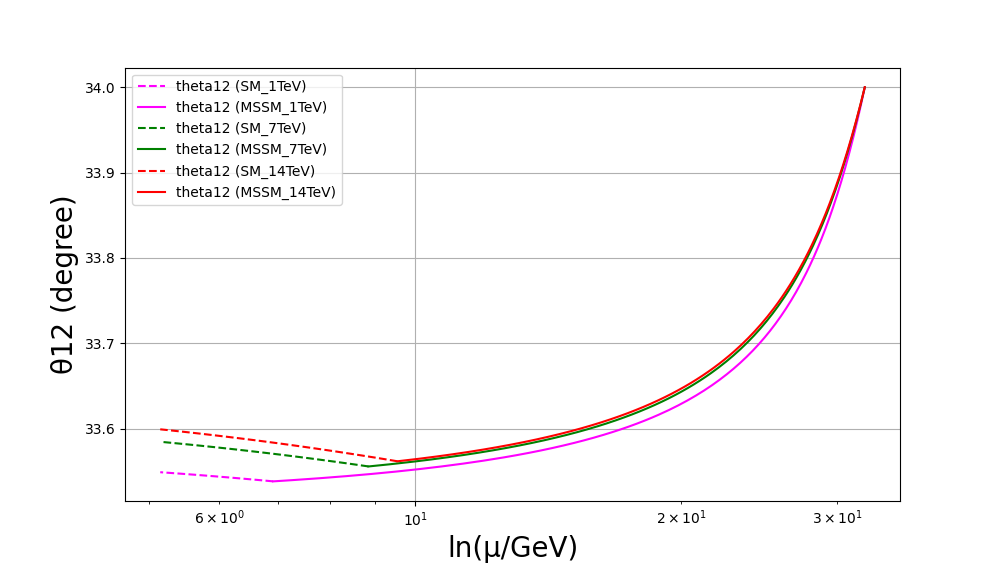}
    \caption{Evolution of $\theta_{12}$}
  \end{subfigure} \\[2ex]
  \begin{subfigure}[b]{0.45\textwidth}
    \centering
    \includegraphics[height=5cm]{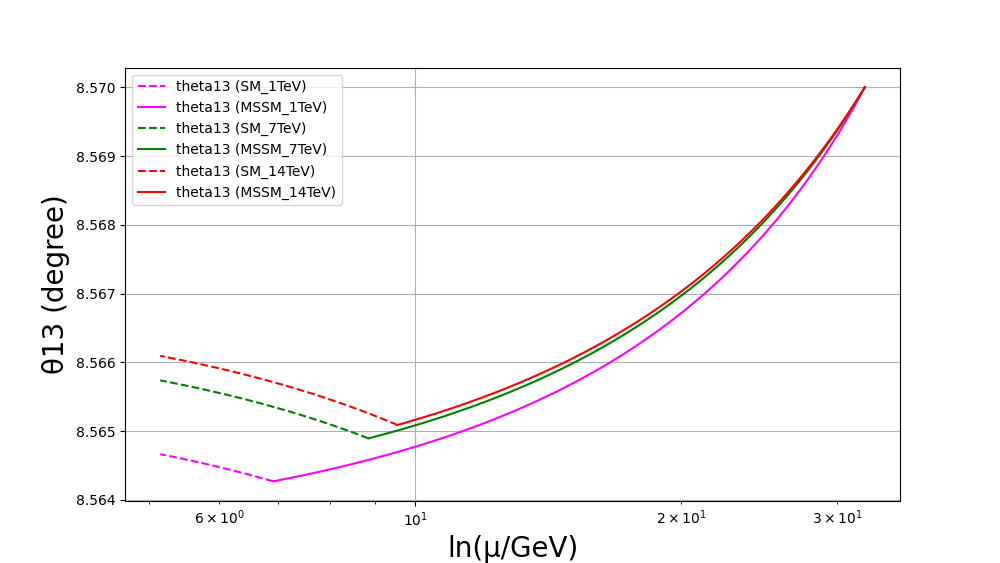}
    \caption{Evolution of $\theta_{13}$}
  \end{subfigure} &
  \begin{subfigure}[b]{0.45\textwidth}
    \centering
    \includegraphics[height=5cm]{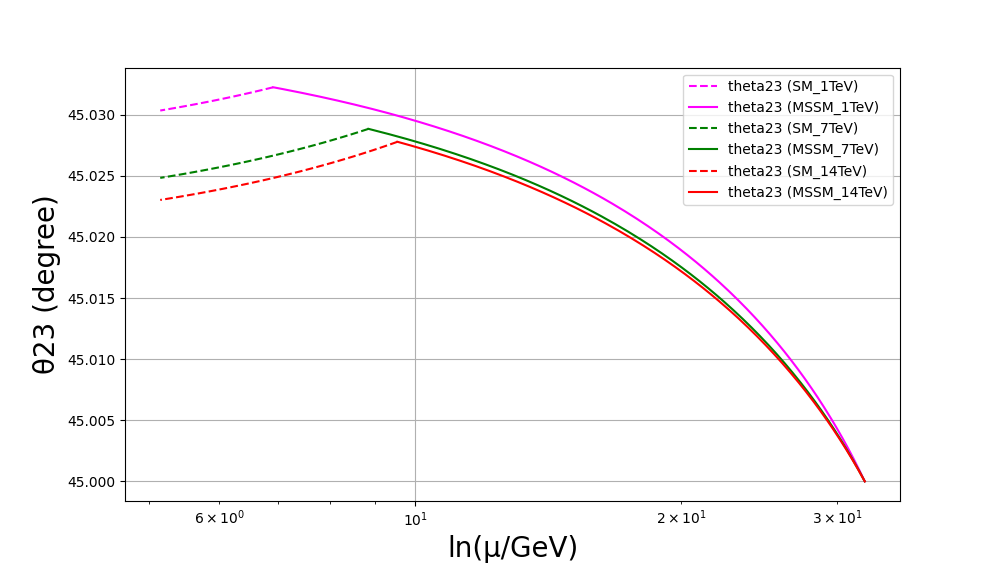}
    \caption{Evolution of $\theta_{23}$}
  \end{subfigure} \\[2ex]
    \end{tabular}
  \caption{Evolution of mass eigenvalues and mixing angles with energy scale for NO and case-II with three different values of $\Lambda_s$. Solid and dashed portions of each curve represent the evolution in the MSSM and SM regions respectively. Magenta , green and red colors respectively stands for $\Lambda_s=1\ TeV,\ 7\ TeV\ \text{and}\ 14\ TeV$.}
  \label{FNO21}
\end{figure}
\indent The observations are qualitatively similar to those in case I. All the high energy input values chosen for the mass eigenvalues and two mixing angles serve as an optimal set in predicting low energy parameters best consistent with low energy data. A slight deviation from this optimal set of input values leads to noticeable discrepancies in the prediction of mass squared differences from their global best-fit values. The chosen set of input values yield $\Delta m_{21}^2 \approx(9.53,\ 7.81,\ 7.25)\times10^{-5}\ eV^2$ corresponding to $\Lambda_s=1\ TeV,\ 7\ TeV$ and $14\ TeV$ respectively. While the predictions corresponding to $\Lambda_s=7\ TeV$ and $14\ TeV$ are close to the global best-fit value (Table \ref{GA}), it is less close to the best-fit value in the case of $\Lambda_s=1\ TeV$. This, in other words, reflects the impact of the variation of the SUSY breaking scale. However, low energy predictions of $\Delta m_{32}^2$ ($\approx 2.55 / 2.48/2.45 \times 10^{-3}\ eV^2$ corresponding to $\Lambda_s=1/7/14\ TeV$) are all close to the global best-fit value $2.513\times 10^{-3}\ eV^2$ (Table \ref{GA}). The low energy prediction of the sum of mass eigenvalues ($\sum m_i \approx0.089\ eV$) lies well within the cosmological upper bound. RG running behaviour of the mass eigenvalues (Figs.3(a)-(c)) are similar to that observed in case I; they tend to increase with the decrease of energy scale. \\
 \begin{figure}[!t]
  \centering
  \begin{tabular}{cc}
  \begin{subfigure}[b]{0.45\textwidth}
    \centering
    \includegraphics[height=5cm]{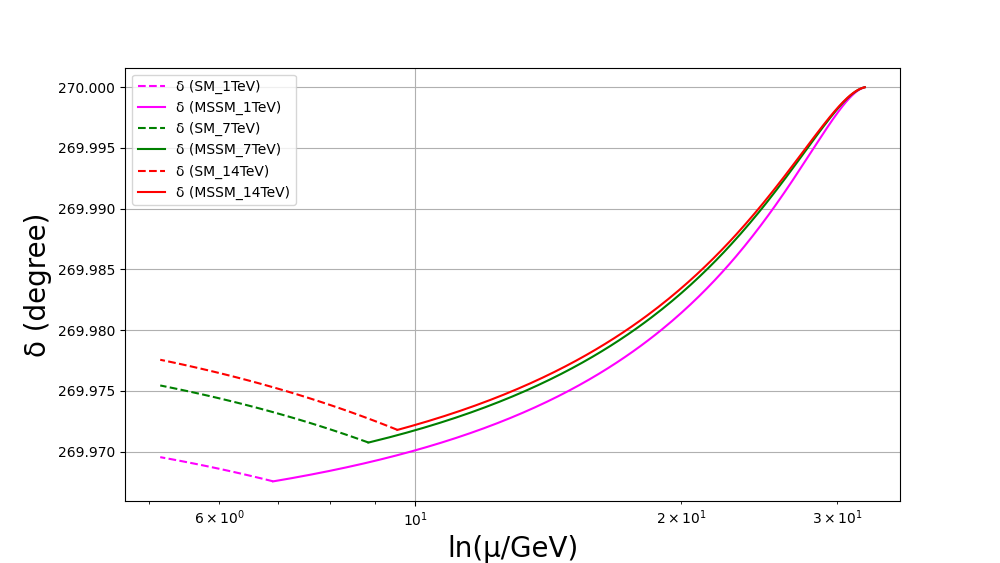}
    \caption{Evolution of $\delta$}
  \end{subfigure} &
  \begin{subfigure}[b]{0.45\textwidth}
    \centering
    \includegraphics[height=5cm]{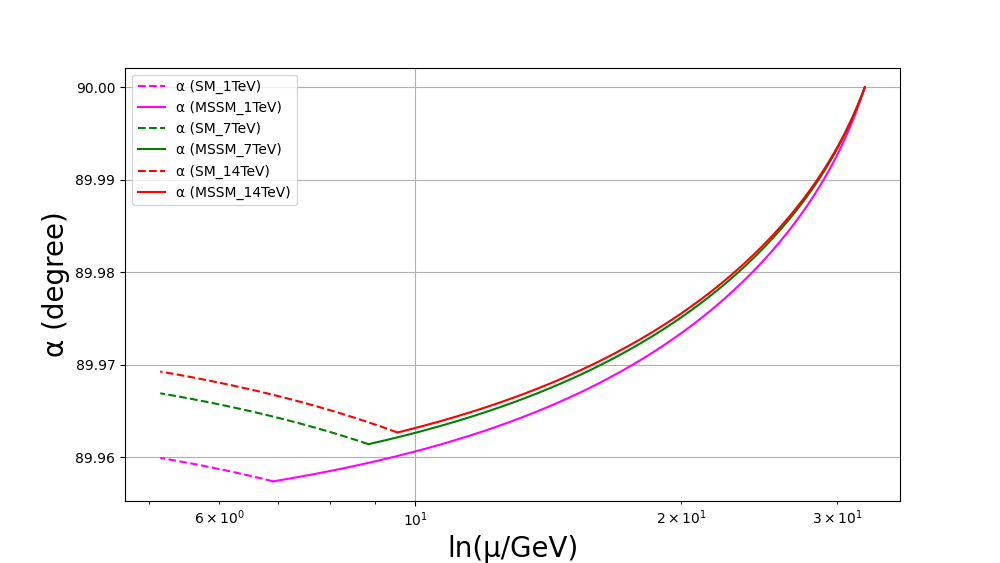}
    \caption{Evolution of $\alpha$}
  \end{subfigure} \\[2ex]
  \begin{subfigure}[b]{0.45\textwidth}
    \centering
    \includegraphics[height=5cm]{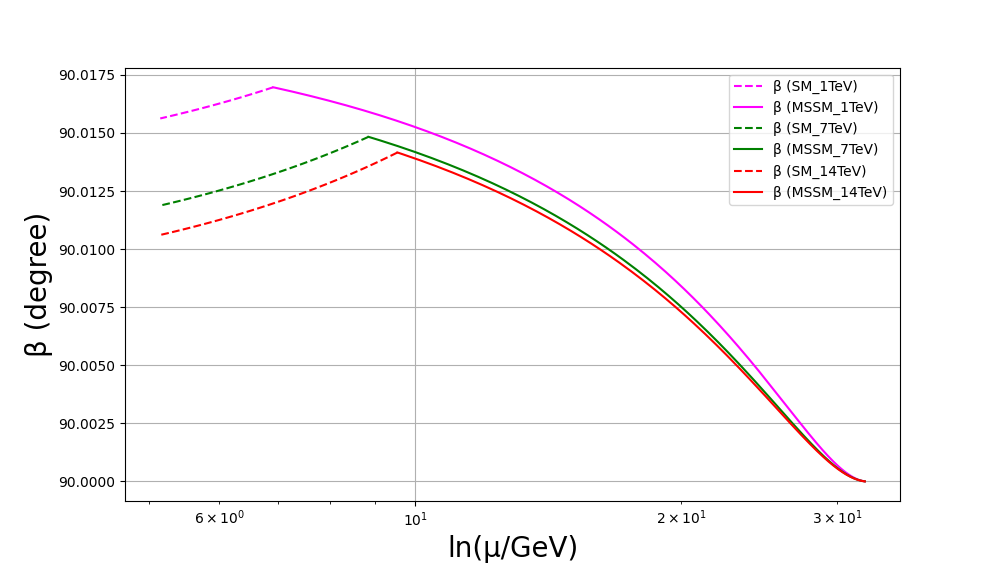}
    \caption{Evolution of $\beta$}
  \end{subfigure} &
  \begin{subfigure}[b]{0.45\textwidth}
    \centering
    \includegraphics[height=5cm]{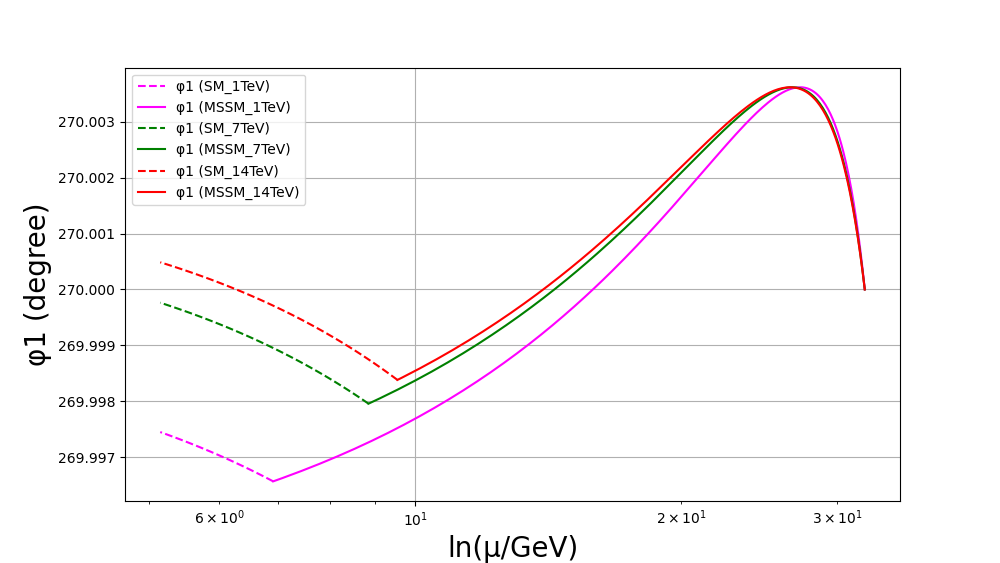}
    \caption{Evolution of $\phi_1$}
  \end{subfigure} \\[2ex]
  \begin{subfigure}[b]{0.45\textwidth}
    \centering
    \includegraphics[height=5cm]{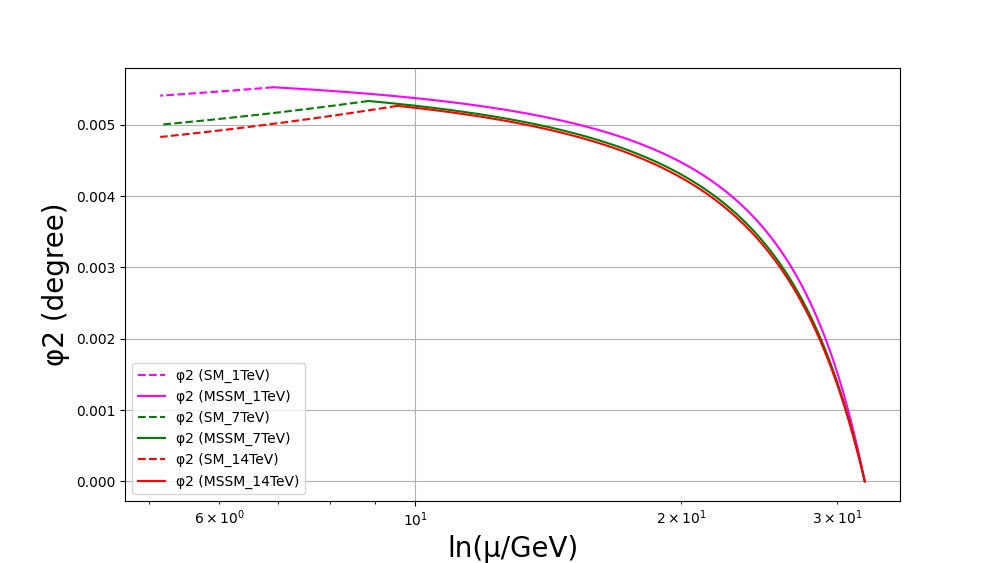}
    \caption{Evolution of $\phi_2$}
  \end{subfigure} & 
  \begin{subfigure}[b]{0.45\textwidth}
    \centering
    \includegraphics[height=5cm]{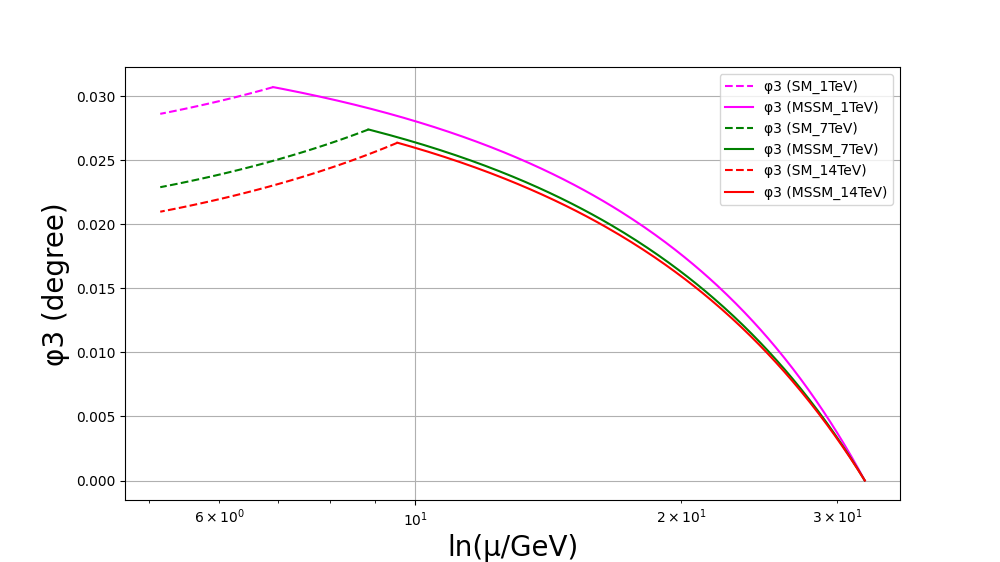}
    \caption{Evolution of $\phi_3$}
  \end{subfigure}  
  \end{tabular}
  \caption{Evolution of CP phases with energy scale for NO and case-II with three different values of $\Lambda_s$. Solid and dashed portions of each curve represent the evolution in the MSSM and SM regions respectively. Magenta , green and red colors respectively stands for $\Lambda_s=1\ TeV,\ 7\ TeV\ \text{and}\ 14\ TeV$.}
  \label{FNO22}
\end{figure}
\indent Similar to case I, the running of the mixing angles and CP phases in case II is also relatively weak, resulting in very small deviations of the parameters. The low energy output values of $\theta_{12}$ and $\theta_{13}$ at $m_t$ scale are found to be $33.54^\circ$ and $8.56^\circ$ respectively against the corresponding high energy input values $34^\circ$ and $8.57^\circ$. The predictions are in good agreement with the global best-fit values (Table \ref{GA}). The running effects on $\theta_{23}$ are found to remain same as in case I. It tends to lie in the second octant due to RG running with low energy predicted value $\sim 45.03^\circ$ for all the three choices of SUSY breaking scale. All the CP phases including $\delta$ suffer only small amount of deviations due to RG running. Low energy prediction of $\delta$ is found to be $\approx 269.97^\circ$ for all the three choices of SUSY breaking scale. Predictions of other CP phases at $m_t$ scale can be read off Table \ref{TNO2}. \\
\indent The energy evolution curves of the parameters generated for the three choices of $\Lambda_s$ in case II exhibit similar characteristics as observed in case I. For the mass eigenvalues and mixing angles (Figs. 3(a)-(f)), all three curves corresponding to $\Lambda_s=1\ TeV,\ 7\ TeV$ and $14\ TeV$ are relatively closer, implying that the impact of variation of $\Lambda_s$ is less significant. On the other hand for CP phases running effects corresponding to $\Lambda_s=7\ TeV/14\ TeV$ relatively differ from those corresponding to $\Lambda_s=1\ TeV$ (evolution curve in magenta is distinct from the other two in Figs. 4(a)-(f)). 
\begin{table}[!t]
\begin{center}
\begin{tabular}{c cc cc cc}
\hline
\multirow{2}{*}{Parameter}& 
\multicolumn{2}{c}{$\Lambda_s=1\ TeV$}&
\multicolumn{2}{c}{$\Lambda_s=7\ TeV$}&
\multicolumn{2}{c}{$\Lambda_s=14\ TeV$} \\
\cline{2-7}
 &\makecell{Input at \\ $\Lambda_{FS}$}  & \makecell{Output at \\$\Lambda_{EW}$} &\makecell{Input at \\ $\Lambda_{FS}$}&\makecell{Output at \\ $\Lambda_{EW}$ }&\makecell{Input at \\ $\Lambda_{FS}$}&\makecell{ Output at \\ $\Lambda_{EW}$} \\ \hline
 $m_1\ (eV)$& 0.03440 & 0.050051 & 0.03440 & 0.049849 & 0.03440 & 0.049708\\
 $m_2\ (eV)$& 0.03699 & 0.050841 & 0.03699 &  0.050597 & 0.03699 & 0.050399 \\
 $m_3\ (eV)$& 0.00805   &0.011863 & 0.00805 & 0.011714 & 0.00755  & 0.010920 \\
$\theta_{13} (/^\circ)$&8.57 & 8.5781 & 8.57 & 8.5761 & 8.57 &  8.5755\\
$\theta_{12} (/^\circ)$& 34.35 & 33.5174 & 34.4 & 34.1419 & 34.37 & 34.4248 \\
$\theta_{23} (/^\circ)$& 45 & 44.9129 & 45 & 44.9256 & 45 & 44.9314 \\
$\delta  (/^\circ)$& 90 & 93.1843 & 90 & 93.0997 & 90 & 92.9596 \\
 $\alpha (/^\circ)$& 270 & 266.8891 & 270 & 266.9269 & 270 & 267.0464 \\
 $\beta (/^\circ)$& 270 & 269.9454 & 270 & 269.9913 & 270 & 270.0016 \\
 $\phi_1 (/^\circ)$& 90 & 93.1847 & 90 & 93.0942 & 90 &92.9620 \\
 $\phi_2 (/^\circ)$& 0 & -0.0075 & 0 &-0.0062 & 0 & -0.0034 \\
 $\phi_3 (/^\circ)$& 0 &-0.0002  & 0 &-0.0010 & 0 &  -0.0047\\
 $g_1$& 0.633482 & 0.461240 & 0.625793 & 0.461241 & 0.623121 & 0.461241 \\
 $g_2$& 0.702064 & 0.662409 & 0.684943 & 0.461241 & 0.679140 & 0.662409 \\
 $g_3$& 0.745271 & 1.210910 & 0.721910 & 1.193682 & 0.715024 & 1.191968\\
 $y_t$& 0.763613 & 0.988794 & 0.700936 & 0.967551 & 0.685736 & 0.963090 \\
 $y_b$& 0.679651 & 0.899020 & 0.601777 & 0.860137 & 0.583075& 0.850472\\
 $y_{\tau}$& 0.779141 &  0.569774 & 0.735524 & 0.555126 & 0.725049 & 0.550269\\
 $\Delta m^2_{21}(10^{-5}eV^2)$&- & 7.97 & - &  7.51 & -& 6.91 \\
 $\Delta m^2_{32}(10^{-3}eV^2)$&- & -2.44 & - & -2.42 & -& -2.42 \\
 $\sum_i m_i (eV)$&- & 0.11275 & - & 0.1122 & -& 0.11103 \\
 \hline
\end{tabular}
\end{center}
\caption{Input values at $\Lambda_{FS}$ and corresponding low energy values at $m_t$ scale of all the parameters for three different values of $\Lambda_s=1\ TeV,\ 7\ TeV$ and $14\ TeV$ in IO and case-I }
\label{TIO1}
\end{table} 
\subsection{Inverted Order ($m_3<m_1<m_2$)}
We have performed a similar analysis for the IO scenario as was done for the case of NO. In comparison to the NO scenario, it is relatively difficult in the present scenario to choose the high energy input values of the free parameters, which can yield the low energy predictions compatible with observational data. However, through careful investigation, an optimal set of input parameters could be found such that corresponding low energy predictions become consistent with $3\sigma$ allowed ranges of global analysis data. Let us first present the numerical results for case I corresponding to the maximal prediction of $\theta_{23}=45^\circ$ and $\delta=90^\circ$ as per $\mu-\tau$ reflection symmetry. 
\begin{figure}[!t]
  \centering
  \begin{tabular}{cc}
   \begin{subfigure}[b]{0.45\textwidth}
    \centering
    \includegraphics[height=5cm]{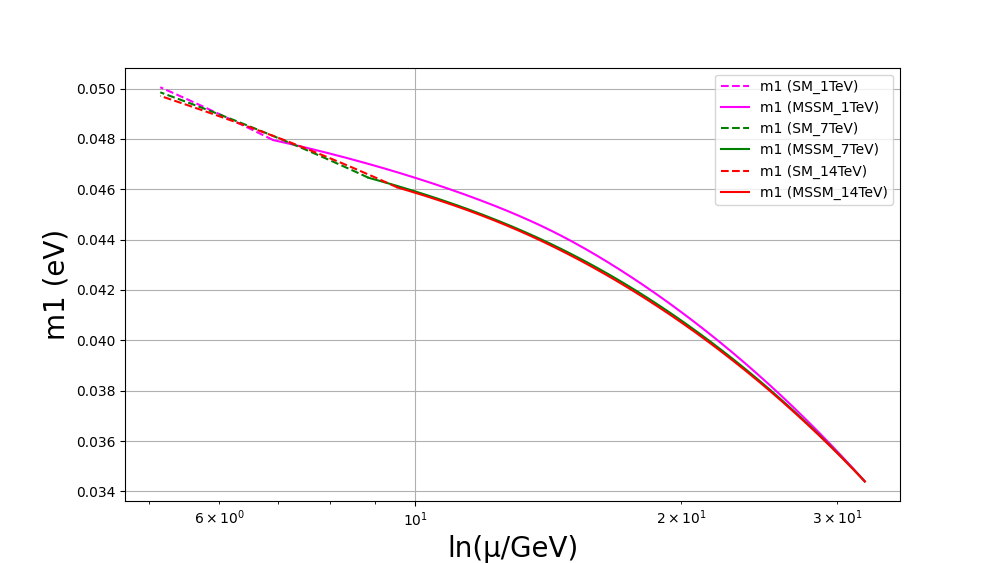}
    \caption{Evolution of $m_1$}
  \end{subfigure}&
  \begin{subfigure}[b]{0.45\textwidth}
    \centering
    \includegraphics[height=5cm]{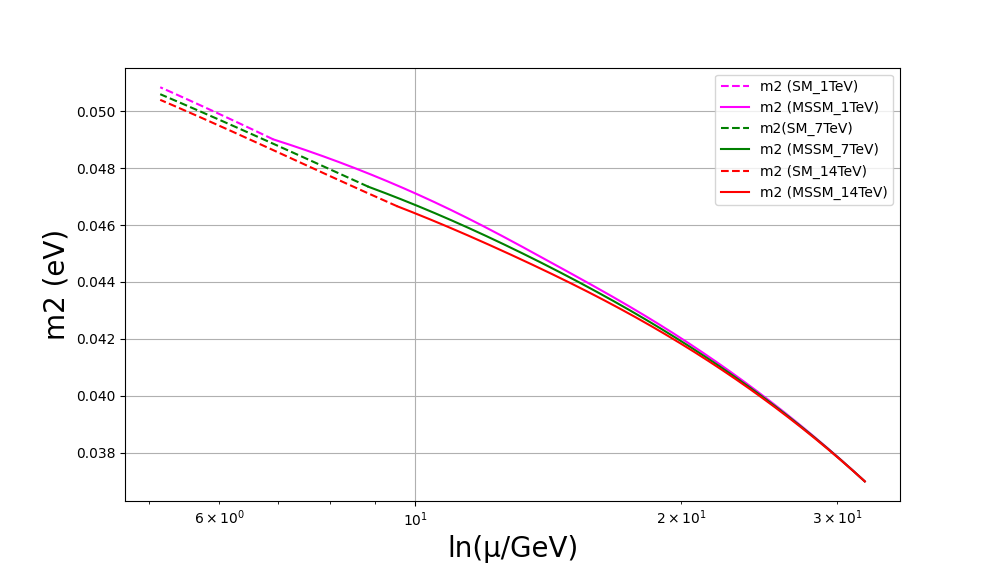}
    \caption{Evolution of $m_2$}
  \end{subfigure} \\[2ex]
  \begin{subfigure}[b]{0.45\textwidth}
    \centering
    \includegraphics[height=5cm]{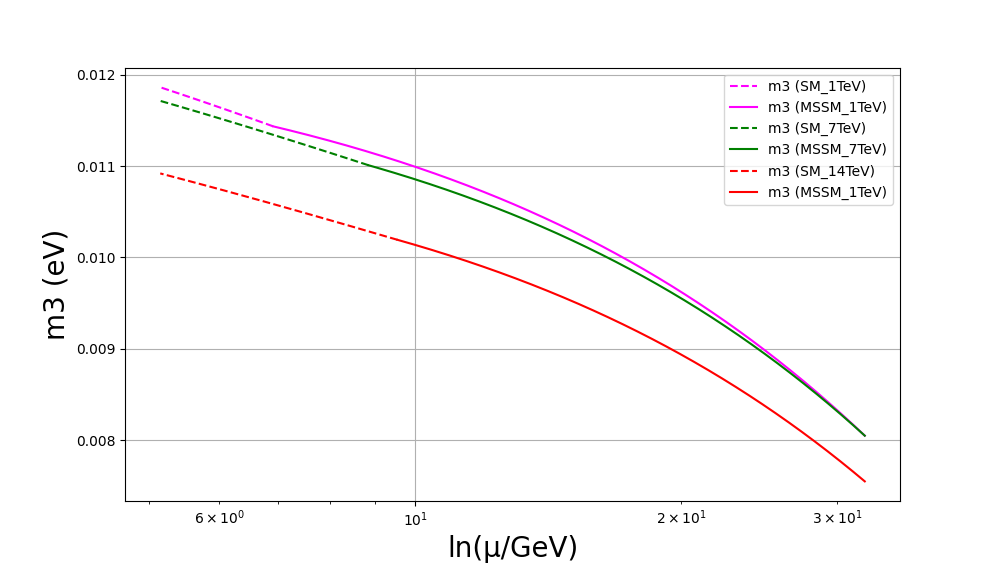}
    \caption{Evolution of $m_3$}
  \end{subfigure} &
  \begin{subfigure}[b]{0.45\textwidth}
    \centering
    \includegraphics[height=5cm]{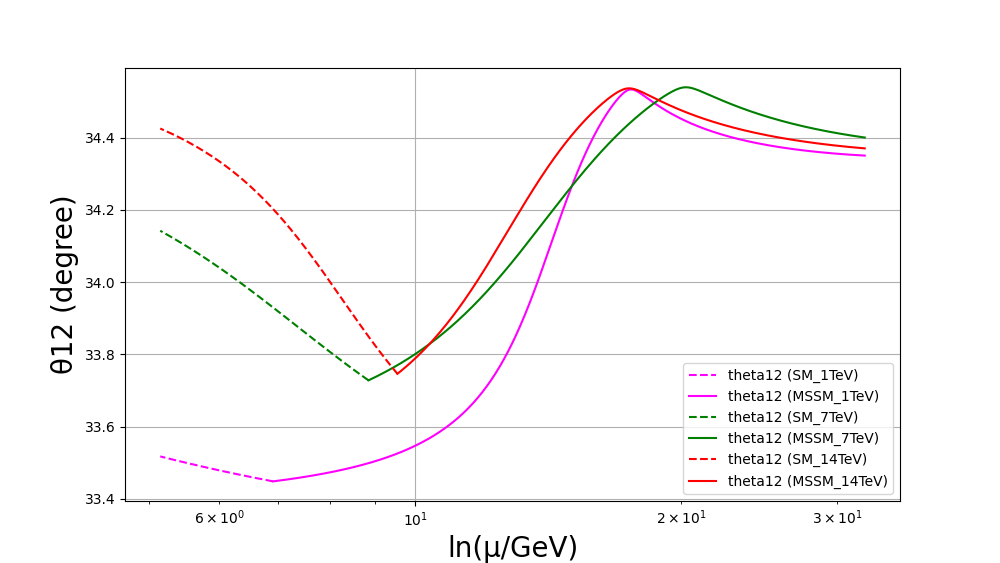}
    \caption{Evolution of $\theta_{12}$}
  \end{subfigure} \\[2ex]
  \begin{subfigure}[b]{0.45\textwidth}
    \centering
    \includegraphics[height=5cm]{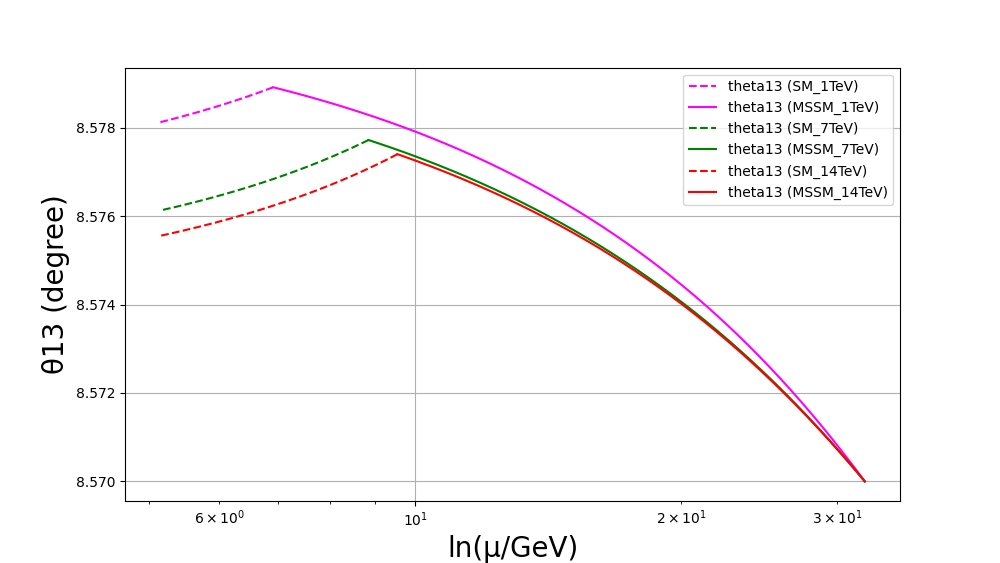}
    \caption{Evolution of $\theta_{13}$}
  \end{subfigure} &
  \begin{subfigure}[b]{0.45\textwidth}
    \centering
    \includegraphics[height=5cm]{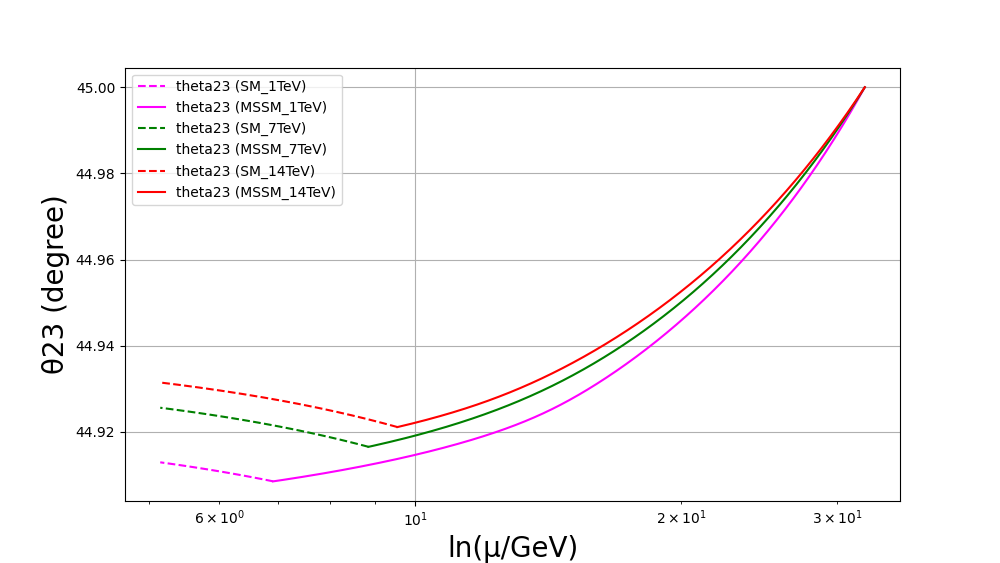}
    \caption{Evolution of $\theta_{23}$}
  \end{subfigure} \\[2ex]
    \end{tabular}
  \caption{Evolution of mass eigenvalues and mixing angles with energy scale for IO and case-I with three different values of $\Lambda_s$. Solid and dashed portions of each curve represent the evolution in the MSSM and SM regions respectively. Magenta , green and red colors respectively stands for $\Lambda_s=1\ TeV,\ 7\ TeV\ \text{and}\ 14\ TeV$.}
  \label{FIO11}
\end{figure}
We have chosen the high-energy input values of the mass eigenvalues at $\Lambda_{FS}$ as $m_3 = 0.00805~eV < m_1 = 0.03440~eV < m_2 = 0.03699~eV$, for the case of the SUSY-breaking scale $\Lambda_s = 1~TeV$. Due to RG running low energy mass eigenvalues at $m_t$ scale are found to be $m_3 = 0.011863~eV < m_1 = 0.050051~eV < m_2 = 0.050841~eV$. These low-energy mass eigenvalues yield $\sum m_i \approx 0.1127 eV$, which is consistent with the cosmological upper bound $0.12 eV$. The corresponding low-energy mass-squared differences are found to be $\Delta m^2_{21}=7.97\times 10^{-5}eV^2$ and $\Delta m^2_{32}=-2.44\times 10^{-3}eV^2$. Both these values of $\Delta m^2_{32}$ and $\Delta m^2_{21}$ are consistent with the $3\sigma$ range of global data (Table \ref{GA}). 
\begin{figure}[!t]
  \centering
  \begin{tabular}{cc}
  \begin{subfigure}[b]{0.45\textwidth}
    \centering
    \includegraphics[height=5cm]{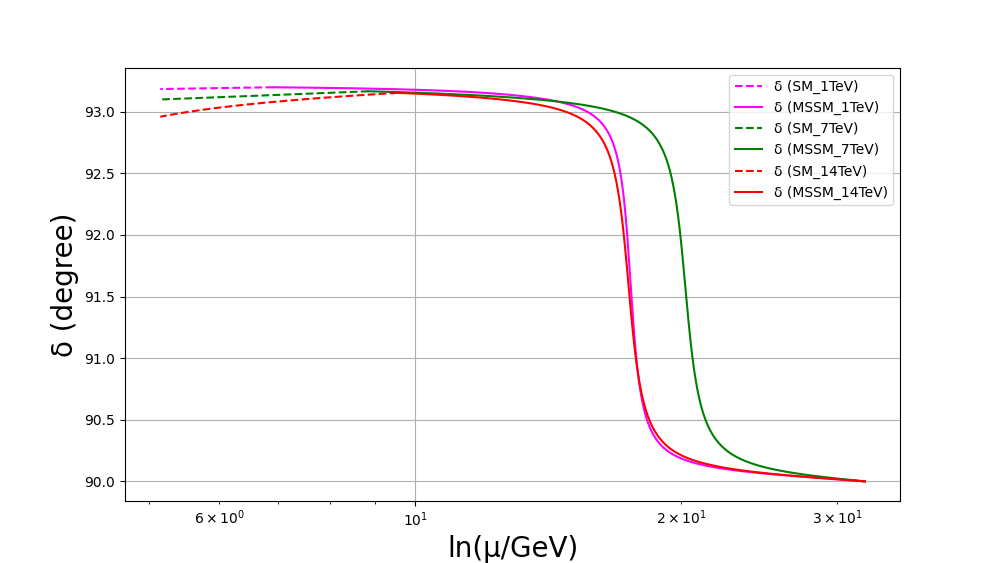}
    \caption{Evolution of $\delta$}
  \end{subfigure} &
  \begin{subfigure}[b]{0.45\textwidth}
    \centering
    \includegraphics[height=5cm]{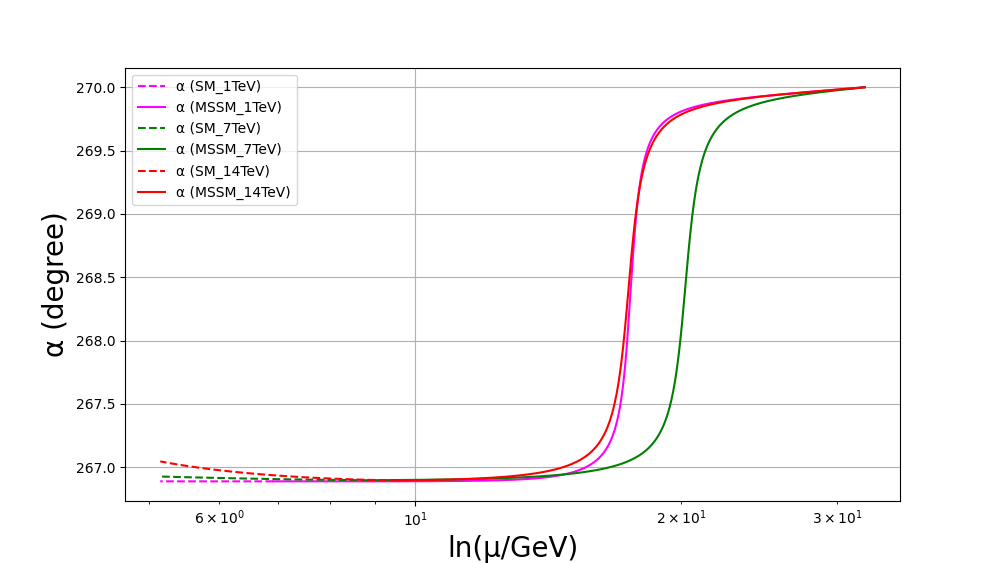}
    \caption{Evolution of $\alpha$}
  \end{subfigure} \\[2ex]
  \begin{subfigure}[b]{0.45\textwidth}
    \centering
    \includegraphics[height=5cm]{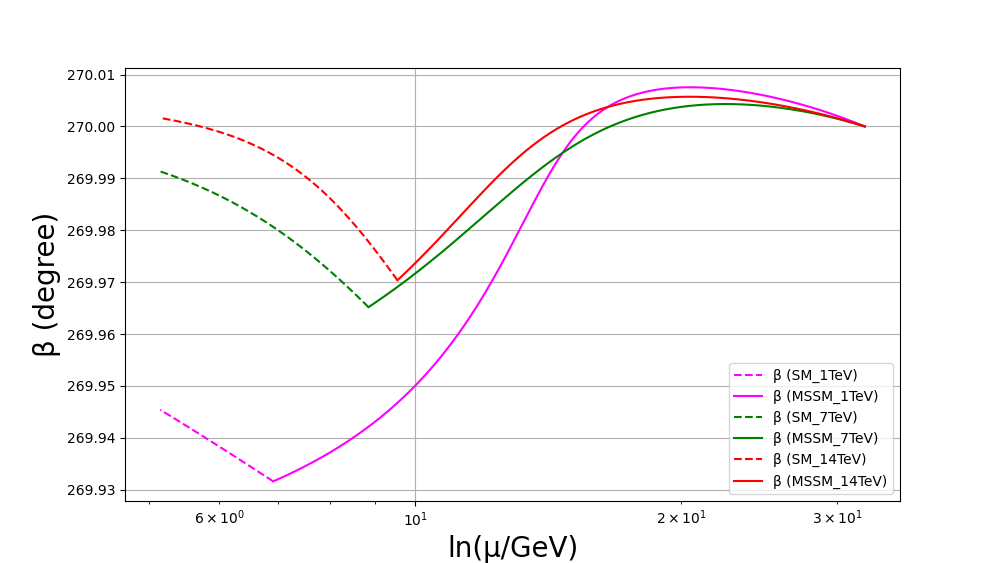}
    \caption{Evolution of $\beta$}
  \end{subfigure} &
  \begin{subfigure}[b]{0.45\textwidth}
    \centering
    \includegraphics[height=5cm]{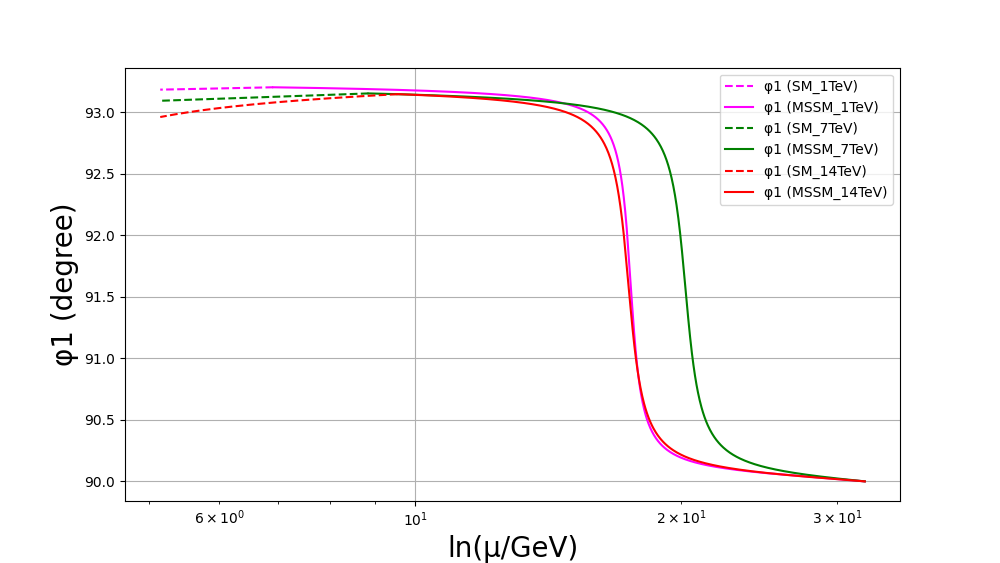}
    \caption{Evolution of $\phi_1$}
  \end{subfigure} \\[2ex]
  \begin{subfigure}[b]{0.45\textwidth}
    \centering
    \includegraphics[height=5cm]{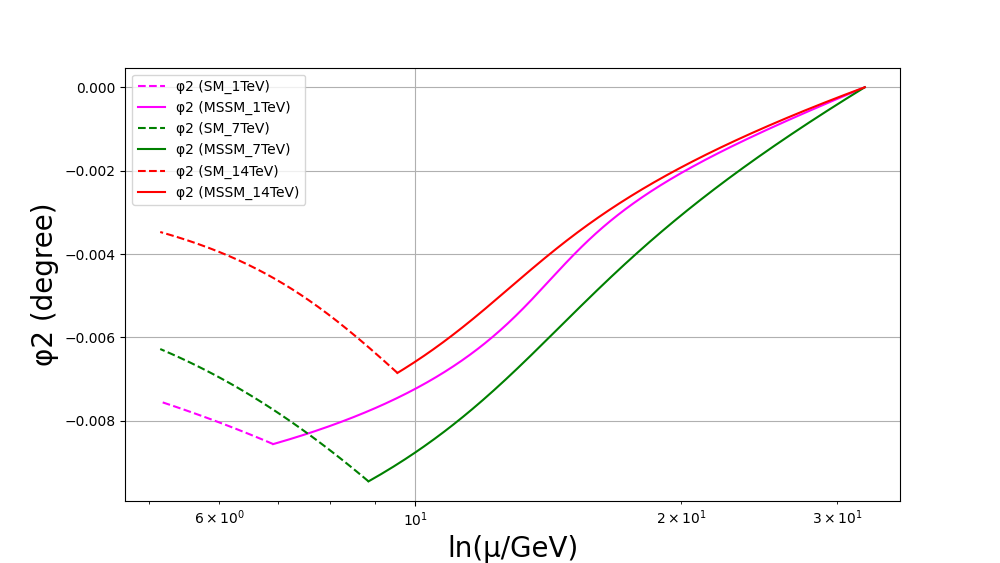}
    \caption{Evolution of $\phi_2$}
  \end{subfigure} & 
  \begin{subfigure}[b]{0.45\textwidth}
    \centering
    \includegraphics[height=5cm]{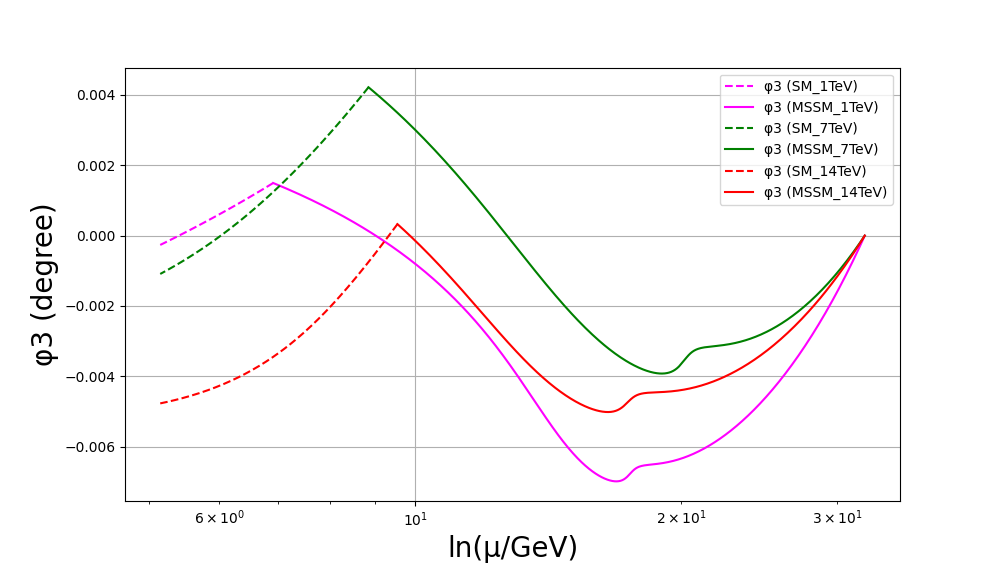}
    \caption{Evolution of $\phi_3$}
  \end{subfigure}  
  \end{tabular}
  \caption{Evolution of CP phases with energy scale for IO and case-I with three different values of $\Lambda_s$. Solid and dashed portions of each curve represent the evolution in the MSSM and SM regions respectively. Magenta , green and red colors respectively stands for $\Lambda_s=1\ TeV,\ 7\ TeV\ \text{and}\ 14\ TeV$.}
  \label{FIO12}
\end{figure}

\indent The high energy input values of $\theta_{12}$ and $\theta_{13}$ at $\Lambda_{FS}$ are taken as $34.35^\circ$ and $8.57^\circ$ respectively for the case of $\Lambda_s=1\ TeV$. Due to RG running, they evolve to $\theta_{12}=33.5174^\circ$ and $\theta_{13}=8.5781^\circ$ at $m_t$ scale. As observed in the NO scenario, the deviation of $\theta_{23}$ from its maximal value in the present scenario of IO is also very small. Its low energy value is found to be approximately $44.91^\circ$ with a tendency to lie in the first octant. It is in contrast to the observation made in NO scenario where $\theta_{23}$ tends to lie in the second octant due to RG running. Nonetheless, low energy values of all three mixing angles at the $m_t$ scale are consistent with the $3\sigma$ range of global data (Table\ref{GA}). Again in the NO scenario, RG running of CP phases were found to be relatively weak with only a small amount of deviations. In the present scenario of IO, the runnings are found to be relatively stronger for $\delta$, $\alpha$ and $\phi_1$. Low energy values of these phases are found to be $93.18^\circ$, $266.89^\circ$ and $93.18^\circ$ respectively for the case of $\Lambda_s=1\ TeV$. 
\begin{table}[!t]
\begin{center}
\begin{tabular}{c cc cc cc}
\hline
\multirow{2}{*}{Parameter}& 
\multicolumn{2}{c}{$\Lambda_s=1\ TeV$}&
\multicolumn{2}{c}{$\Lambda_s=7\ TeV$}&
\multicolumn{2}{c}{$\Lambda_s=14\ TeV$} \\
\cline{2-7}
 &\makecell{Input at \\ $\Lambda_{FS}$}  & \makecell{Output at \\$\Lambda_{EW}$} &\makecell{Input at \\ $\Lambda_{FS}$}&\makecell{Output at \\ $\Lambda_{EW}$ }&\makecell{Input at \\ $\Lambda_{FS}$}&\makecell{ Output at \\ $\Lambda_{EW}$} \\ \hline
 $m_1\ (eV)$& 0.03270 &  0.049026 & 0.03345 & 0.049296 & 0.03381& 0.049403 \\
 $m_2\ (eV)$& 0.03699 & 0.049763 & 0.03699  & 0.050089 & 0.03699  &0.050206 \\
 $m_3\ (eV)$& 0.00255 & 0.003726 & 0.00255 &  0.003686 & 0.00255  &0.003661\\
$\theta_{13} (/^\circ)$&8.8 & 8.8088 & 8.8 &  8.8068 & 8.84 & 8.8460\\
$\theta_{12} (/^\circ)$& 34.95 & 35.2992 & 34.96 & 35.2857 & 34.96 & 35.3643 \\
$\theta_{23} (/^\circ)$& 45 & 44.9459 & 45 & 44.9568 & 45 & 44.9612 \\
 $\delta  (/^\circ)$& 270 & 269.9907 & 270 & 269.9928 & 270 & 269.9918\\
 $\alpha  (/^\circ)$& 90 & 89.9995 & 90 &  89.9994 & 90 & 90.0001 \\
 $\beta  (/^\circ)$& 90 & 90.0084 & 90 &  90.0065 & 90 & 90.0060 \\
 $\phi_1 (/^\circ)$& 270 & 269.9986 & 270 & 269.9989 & 270 &269.9980 \\
 $\phi_2 (/^\circ)$& 0 & -0.0015 & 0 &-0.0011 & 0 & -0.0009 \\
 $\phi_3 (/^\circ)$& 0 &-0.0069  & 0 &-0.0055 & 0 & -0.0050\\
 $g_1$& 0.633482 & 0.461240 & 0.625793 & 0.461241 & 0.623121 &  0.461241 \\
 $g_2$& 0.702064 & 0.662409 & 0.684943 & 0.662409 & 0.679140 &0.662409 \\
 $g_3$& 0.745271 & 1.210910 & 0.721910 & 1.193682 & 0.715024 & 1.191968\\
 $y_t$& 0.763613 &  0.988794 & 0.700936 & 0.967551 & 0.685736 &  0.963090 \\
 $y_b$& 0.679651 & 0.899020 & 0.601777 & 0.860137 & 0.583075& 0.850472\\
 $y_{\tau}$& 0.779141 & 0.569774 & 0.735524 & 0.555126 & 0.725049 & 0.550269\\
 $\Delta m^2_{21}(10^{-5}eV^2)$&- & 7.28 & - & 7.88 &- & 7.99 \\
 $\Delta m^2_{32}(10^{-3}eV^2)$& -& -2.46 & - & -2.49 &- & -2.50 \\
 $\sum_i m_i(eV)$&- & 0.10251 & - & 0.10307 &- & 0.10327 \\
 \hline
\end{tabular}
\end{center}
\caption{Input values at $\Lambda_{FS}$ and corresponding low energy values at $m_t$ scale of all the parameters for three different values of $\Lambda_s=1\ TeV,\ 7\ TeV$ and $14\ TeV$ in IO and case-II.}
\label{TIO2}
\end{table} 

\begin{figure}[!t]
  \centering
  \begin{tabular}{cc}
   \begin{subfigure}[b]{0.45\textwidth}
    \centering
    \includegraphics[height=5.cm]{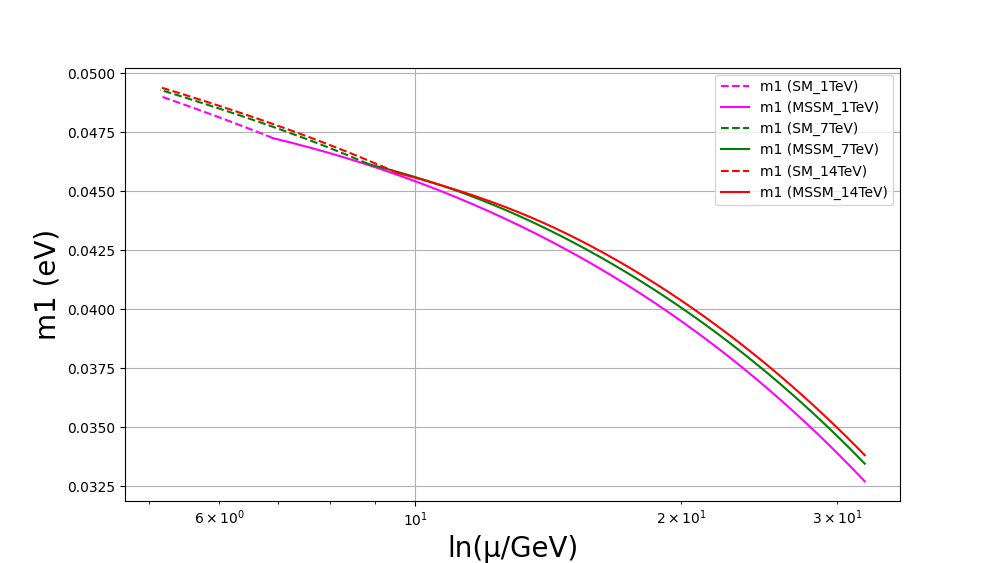}
    \caption{Evolution of $m_1$}
  \end{subfigure}&
  \begin{subfigure}[b]{0.45\textwidth}
    \centering
    \includegraphics[height=5.cm]{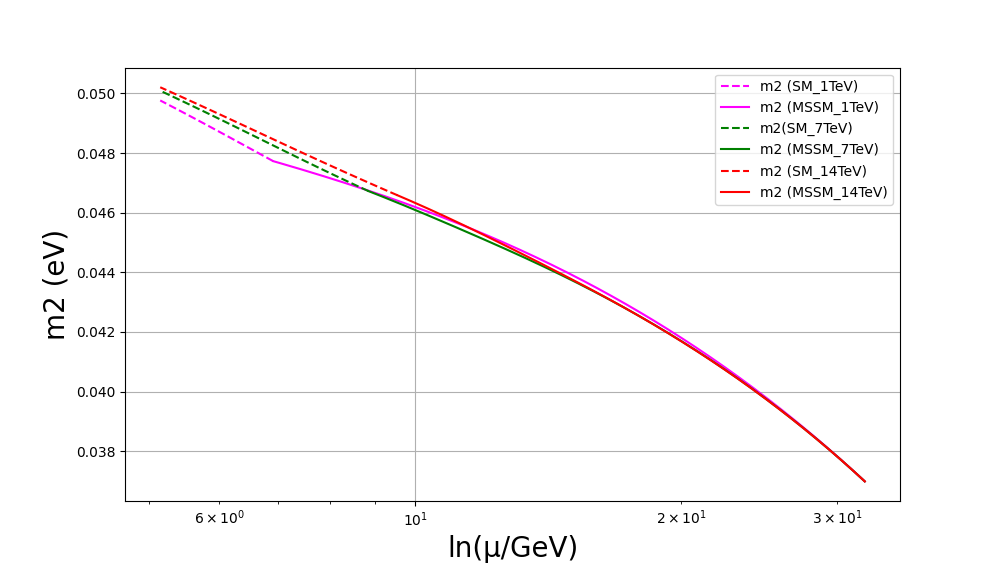}
    \caption{Evolution of $m_2$}
  \end{subfigure} \\[2ex]
  \begin{subfigure}[b]{0.45\textwidth}
    \centering
    \includegraphics[height=5cm]{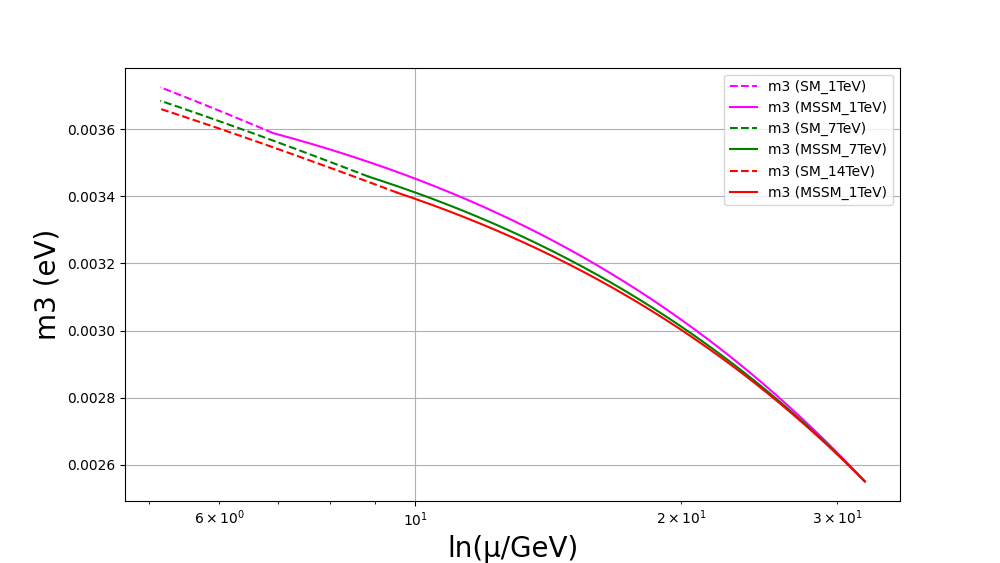}
    \caption{Evolution of $m_3$}
  \end{subfigure} &
  \begin{subfigure}[b]{0.45\textwidth}
    \centering
    \includegraphics[height=5cm]{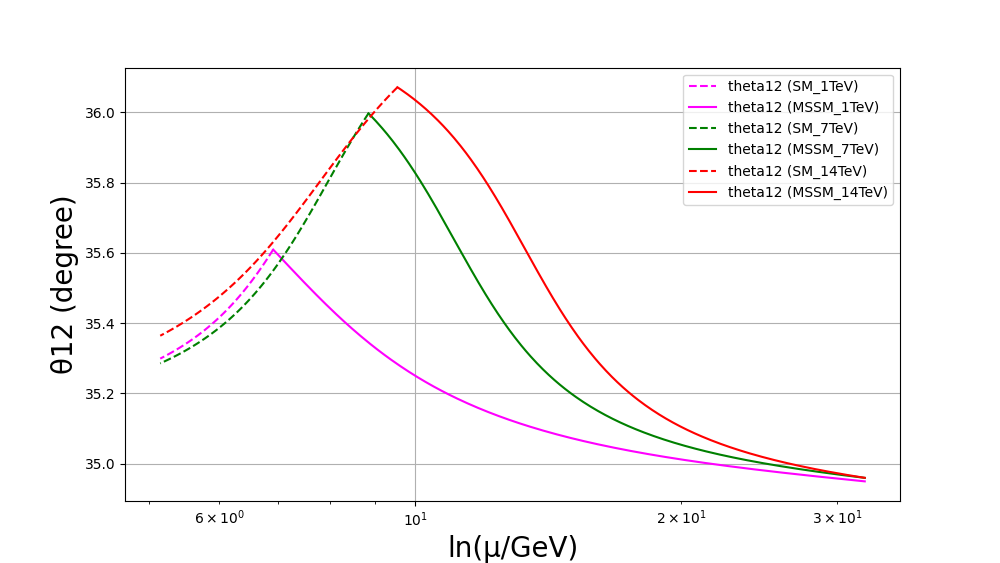}
    \caption{Evolution of $\theta_{12}$}
  \end{subfigure} \\[2ex]
  \begin{subfigure}[b]{0.45\textwidth}
    \centering
    \includegraphics[height=5cm]{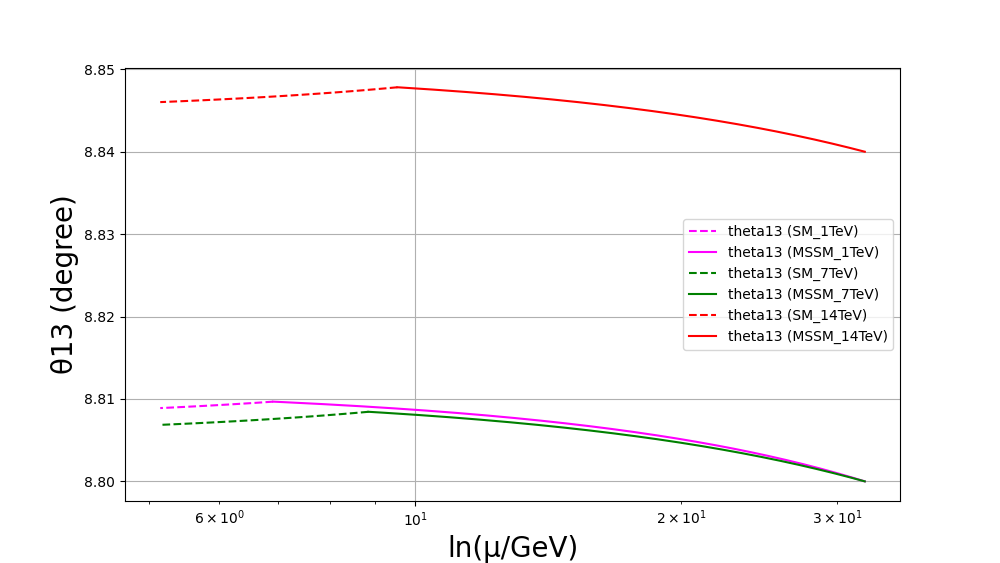}
    \caption{Evolution of $\theta_{13}$}
  \end{subfigure} &
  \begin{subfigure}[b]{0.45\textwidth}
    \centering
    \includegraphics[height=5cm]{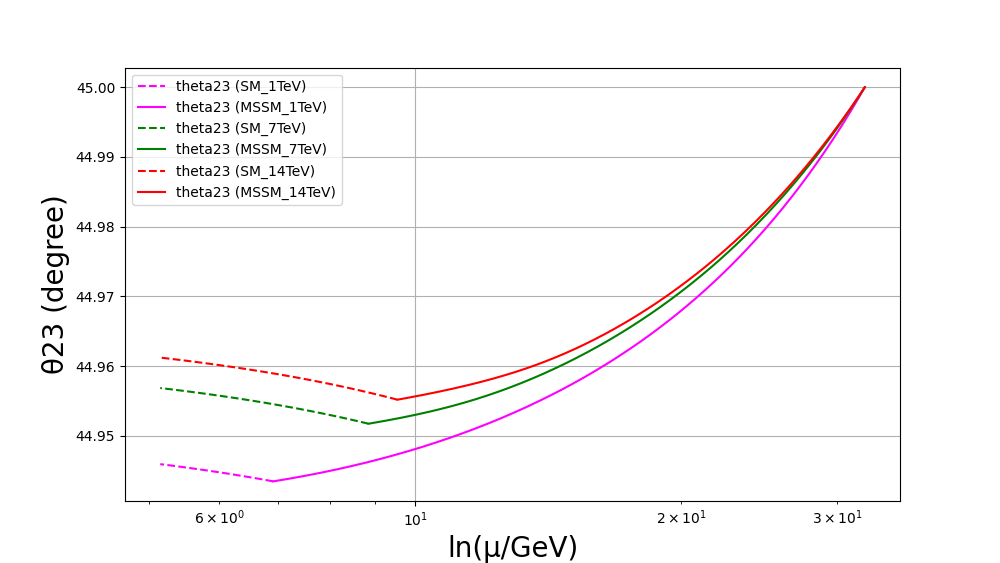}
    \caption{Evolution of $\theta_{23}$}
  \end{subfigure} \\[2ex]
    \end{tabular}
  \caption{Evolution of mass eigenvalues and mixing angles with energy scale for IO and case-II with three different values of $\Lambda_s$. Solid and dashed portions of each curve represent the evolution in the MSSM and SM regions respectively. Magenta , green and red colors respectively stands for $\Lambda_s=1\ TeV,\ 7\ TeV\ \text{and}\ 14\ TeV$.}
  \label{FIO21}
\end{figure}
\indent To investigate the impact of variation of SUSY breaking scale on the running parameters as before, we first study the low energy output values of all the parameters for the case of $\Lambda_s=7\ TeV$ and $14\ TeV$ using the same set of input values as taken for the case of $\Lambda_s=1\ TeV$. Unlike the NO scenario, in the present case, we notice a significant impact on the low energy prediction due to an increase in the SUSY breaking scale. While in the NO case, same set of input values yields low energy parameters consistent with $3\sigma$ range of global data for all the three different values of $\Lambda_s$, in the present case we find that for $\Lambda_s=7\ TeV$ and $14\ TeV$, low energy values of mass squared differences get shifted to some inconsistent value. For instance, we get $\Delta m_{21}^2=4.17\times 10^{-5}\ eV^2$ and $6.13\times 10^{-5}\ eV^2$ for $\Lambda_s=7\ TeV$ and $14\ TeV$ which lie outside the $3\sigma$ range of global data. We find that for $\Lambda_s=7\ TeV$, by adjusting the input values of $\theta_{12}$ from $34.35^\circ$ (chosen for $\Lambda_s=1\ TeV$) to $34.4^\circ$ while keeping the all other input parameter same as taken in case of $\Lambda_s=1\ TeV$, low energy value of $\Delta m_{21}^2$ can be made consistent with the $3\sigma$ range. Similarly, for $\Lambda_s=14\ TeV$, we need to adjust additionally the input values of $m_3$ along with $\theta_{12}$ in order to make the mass squared difference consistent with $3\sigma$ range. We have taken $m_3=0.00755\ eV$ and $\theta_{12}=34.37^\circ$ keeping other free parameters same as taken for $\Lambda_s=1\ TeV$. All the input values and corresponding output values of the parameters for the three different cases of SUSY breaking scale are summarised in Table \ref{TIO1}. The energy scale evolution of mass eigenvalues and mixing angles are depicted in Figs. 5(a)-(f). Similarly, the evolution of CP phases is presented in Figs. 6(a)-(f). As before, each figure contains three evolution curves of a given parameter with magenta, green and red colors corresponding to the case of $\Lambda_s=1\ TeV,\ 7\ TeV$ and $14\ TeV$ respectively. From Figs. 5(a)-(b), we see that the three curves in each figure are very much close to each other. This implies that the variation of the SUSY breaking scale does not cause significant change in the running of the mass eigenvalues. Again, Fig. 5(d) shows that the splitting between different curves are relatively distinct as compared to those observed in case of $\theta_{13}$ and $\theta_{23}$ (Figs. 5(e) and (f)). This reflects significant impact on the running effects of $\theta_{12}$ due to variation of $\Lambda_s$. The low energy output values of $\theta_{12}$ also vary with a comparatively large amount with the variation of $\Lambda_s$. Similar kind of distinct splitting and variation of low energy output values are also visible for the CP phases $\beta$, $\phi_2$ and $\phi_3$ from Figs. 6(c), (e) and (f). On the other hand variations of low energy output values $\delta$, $\alpha$ and $\phi_1$ are insignificant, as can be seen from Figs. 6(a), (b) and (d).
\begin{figure}[!t]
  \centering
  \begin{tabular}{cc}
  \begin{subfigure}[b]{.45\textwidth}
    \centering
    \includegraphics[height=5cm]{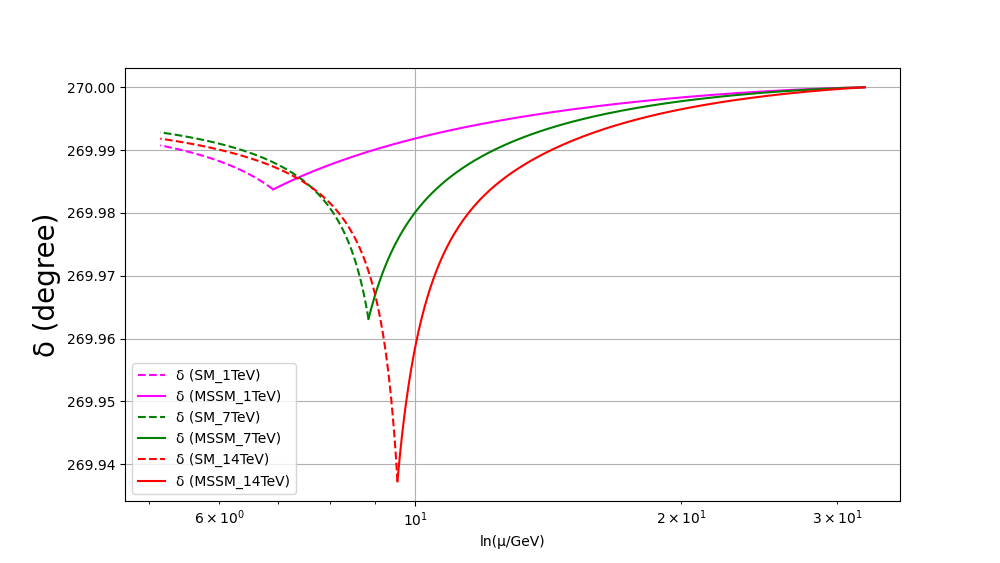}
    \caption{Evolution of $\delta$}
  \end{subfigure} &
  \begin{subfigure}[b]{0.45\textwidth}
    \centering
    \includegraphics[height=5cm]{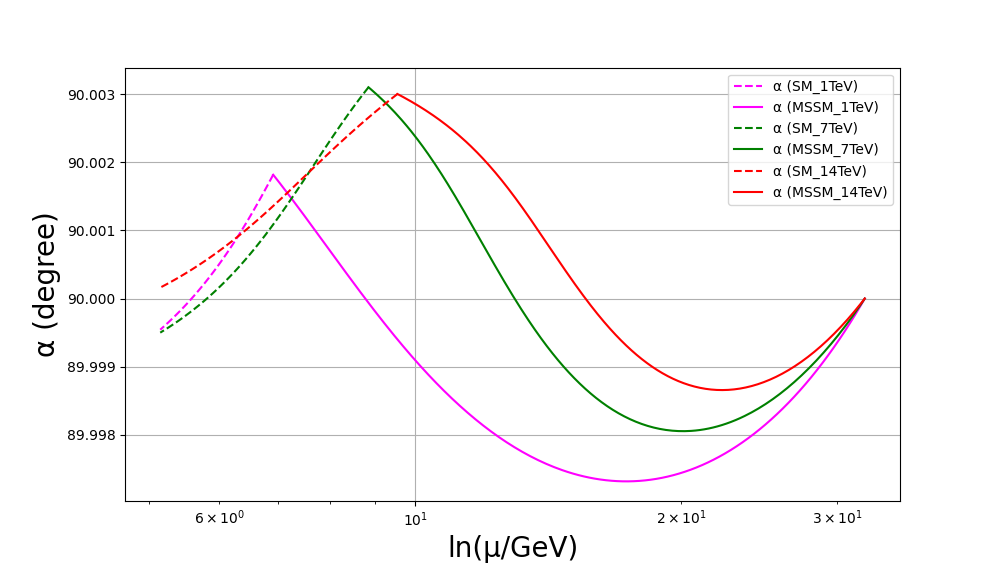}
    \caption{Evolution of $\alpha$}
  \end{subfigure} \\[2ex]
  \begin{subfigure}[b]{0.45\textwidth}
    \centering
    \includegraphics[height=5cm]{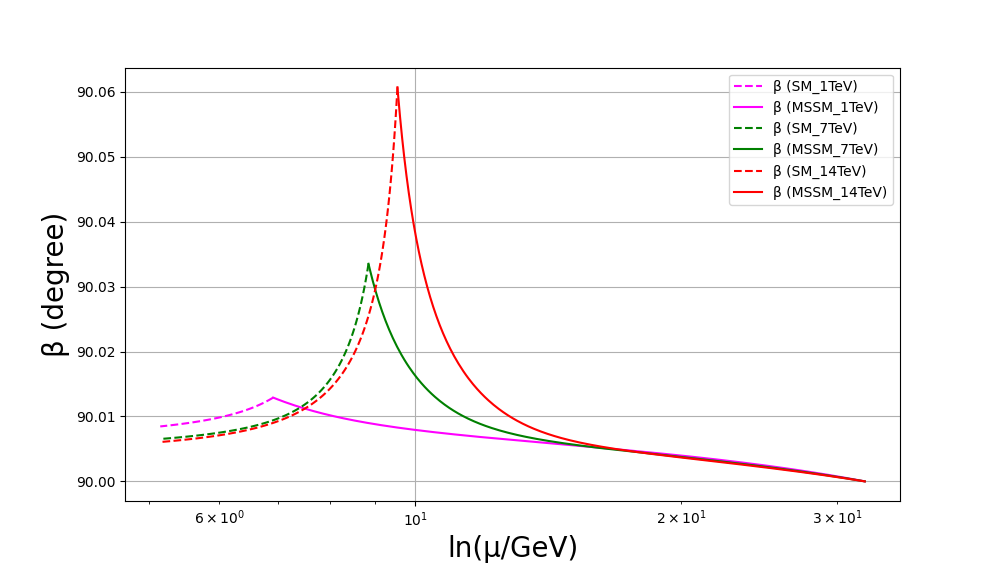}
    \caption{Evolution of $\beta$}
  \end{subfigure} &
  \begin{subfigure}[b]{0.45\textwidth}
    \centering
    \includegraphics[height=5cm]{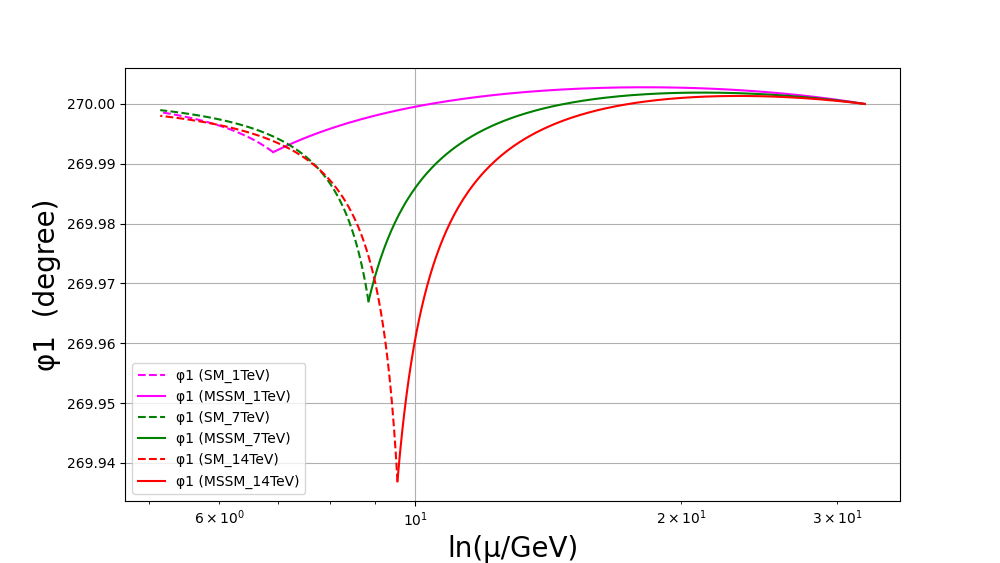}
    \caption{Evolution of $\phi_1$}
  \end{subfigure} \\[2ex]
  \begin{subfigure}[b]{0.45\textwidth}
    \centering
    \includegraphics[height=5cm]{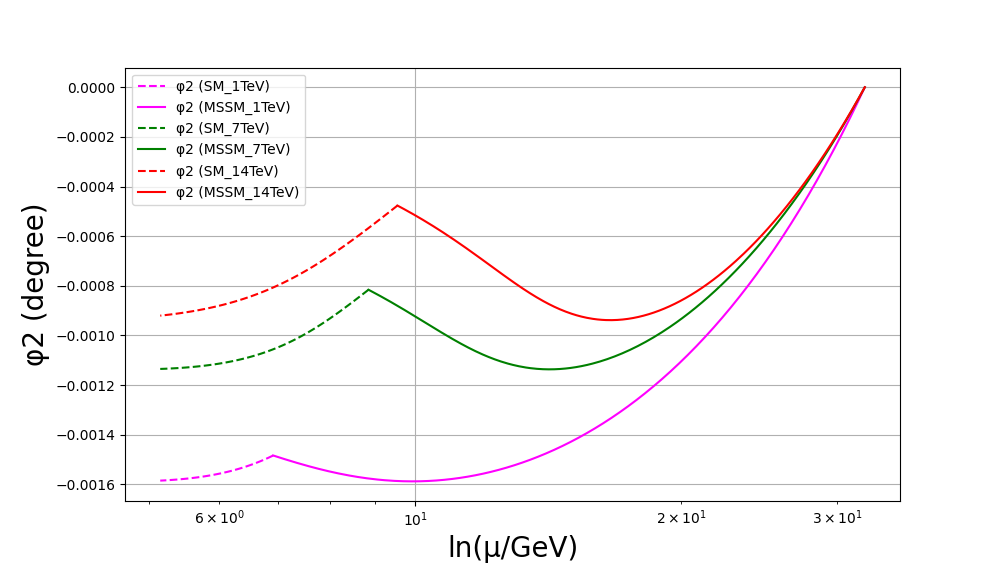}
    \caption{Evolution of $\phi_2$}
  \end{subfigure} & 
  \begin{subfigure}[b]{0.45\textwidth}
    \centering
    \includegraphics[height=5cm]{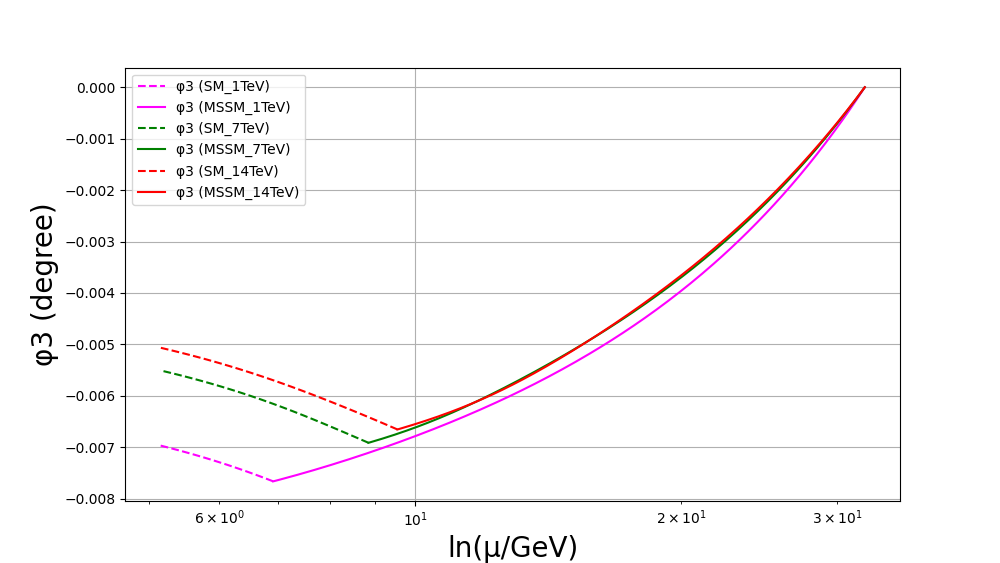}
    \caption{Evolution of $\phi_3$}
  \end{subfigure}  
  \end{tabular}
  \caption{Evolution of CP phases with energy scale for IO and case-II with three different values of $\Lambda_s$. Solid and dashed portions of each curve represent the evolution in the MSSM and SM regions respectively. Magenta , green and red colors respectively stands for $\Lambda_s=1\ TeV,\ 7\ TeV\ \text{and}\ 14\ TeV$.}
  \label{FIO22}
\end{figure}

\indent For case II in the IO scenario, high energy input values of Dirac phase and atmospheric mixing angle are $\delta=270^\circ$ and $\theta_{23}=45^\circ$ and those of other CP phases are given by Eq. (\ref{ch2}). Unlike previous cases, it is highly challanging in this case to choose an appropriate set of input values of the free parameters which can lead to consistent results under RG running. Despite selecting precise input values which yield low energy parameters consistent with $3\sigma$ range and cosmological upper bound, energy evolution curves of $\theta_{12}$ and some of the CP phases generally show some unusual distortion at certain points. After numerous trials, we could identify an optimal set of input values of the free parameters that results in low energy parameters best consistent with observational data, along with evolution curves of the parameters free from any distortions. The chosen high energy input masses are $m_3=0.00255\ eV<m_1=0.03270\ eV<m_2=0.03699\ eV$ for SUSY breaking scale $\Lambda_s=1\ TeV$. Along with these, the input values of $\theta_{12}$ and $\theta_{13}$ are taken as $34.95^\circ$ and $8.8^\circ$ respectively. Even a minor change in these chosen input values causes drastic inconsistencies in the desired results. Neveretheless with this set of input values, the low energy masses result in $\sum m_i \approx 0.10251\ eV$, which satisfy the cosmological upper bound. Further, we get the low energy values of mass squared differences given by $\Delta m^2_{21}=7.28\times 10^{-5}eV^2$ and $\Delta m^2_{32}=-2.46\times 10^{-3}eV^2$ for $\Lambda_s=1\ TeV$ which lie within the $3\sigma$ range of global data (Table \ref{GA}). Low energy prediction of $\theta_{12}\approx 35.3^\circ$ and $\theta_{13}\approx 8.81^\circ$ are also consistent with the $3\sigma$ range. Atmospheric angle $\theta_{23}$ tends to decrease with the decrease of energy scale and thereby lies in the first octant with a predicted value of $44.946^\circ$. This running behavior of $\theta_{23}$ is similar to that observed in the case I under IO scenario. Since the results from $T2K$ and $NO\nu A$ experiments and also the global analysis of oscillation data reflect a maximal value of Dirac phase near $270^\circ$ in the IO scenario, it is important to note the running of the phase $\delta$ in the present case. As per global analysis, best-fit value of $\delta$ is $285^\circ$ (without SK data)/ $274^\circ$ (with SK data). We observe that under RG running $\delta$ tends to decrease with the decrease in energy scale, with a predicted value of $269.99^\circ$. The prediction is consistent with the $3\sigma$ range. Similar to $\delta$ other CP phases also recieve a small amount of perturbations under RG running. 

\indent Similar to the previous analysis, we study the low energy output values of the parameters for $\Lambda_s = 7~TeV$ and $14~TeV$ as well. As found in case I, the high energy input values of parameters in the present case, chosen for the case of $\Lambda_s=1\ TeV$, do not result in consistent predictions when applied for the case of $\Lambda_s=7\ TeV$ and $14\ TeV$. For the case of $\Lambda_s=7\ TeV$, we adjust the value of $m_1$ to $0.03345\ eV$ from $0.03270\ eV$ (corresponding to $\Lambda_s=1\ TeV$) and that of $\theta_{12}$ from $34.95^\circ$ (corresponding to $\Lambda_s=1\ TeV$) to $34.96^\circ$. Similarly for $\Lambda_s= 14~TeV$, input values of $m_1$ is adjusted to $0.03381\ eV$ while that of $\theta_{13}$ is taken as $8.84^\circ$ instead of $8.8^\circ$ (corresponding to $\Lambda_s=1\ TeV$). Corresponding low energy predictions of all the parameters can be read off Table \ref{TIO2}, which are all consistent with the $3\sigma$ range of global data. Regarding the effects of variation of SUSY breaking scale on the running of parameters in this case, they are same for the mass eigenvalues and mixing angles $\theta_{13}$ and $\theta_{23}$ as observed in case I. However in case of $\theta_{12}$ and CP phase $\delta$, $\alpha$, $\beta$ and $\phi_1$ (Figs. 7(d) and 8(a)-(d)), we observe that the impact is prominent only in the MSSM region (splittings between evolution curves are quite distinct), but interestingly, the running in SM region become indistinguishable resulting into no significant variation in the low energy output values.  

\section{Summary and discussion}
The $\mu-\tau$ reflection symmetry plays an attractive role in explaining the observed pattern of neutrino mixing as it accommodates maximal value of $\theta_{23}$ as well as non-zero $\theta_{13}$ compatible with experimental data. It further constrains the Dirac CP phase $\delta$ to be maximal ($\pi/2$ or $3\pi/2$). As the recent results from T2K and NO$\nu$A experiments indicate the value of $\delta$, which is close to the maximal value $3\pi/2$ in the IO scenario (the same is also reflected in the global analysis of neutrino oscillation data), the studies of $\mu-\tau$ reflection symmetry become more prominent. Since experimental results generally indicate certain deviations from maximality, it becomes reasonable to study the breaking of an exact symmetry. In this work, we consider RG running effects on neutrino masses and mixing as a perturbation to the $\mu-\tau$ reflection symmetry and study the consequent deviations. We choose the seesaw scale $10^{14}\ GeV$, where light Majoran neutrino masses are generated effectively, as the flavor symmetry scale $\Lambda_{FS}$, such that $\mu-\tau$ reflection symmetry remains preserved at this scale. Accordingly atmospheric angle $\theta_{23}$ and Dirac CP phase $\delta$ assume maximal values at this scale as per the symmetry. In addition to $\delta$, we also constrain the values of Majorana phases and unphysical phases  to be maximal in order to maintain consistency of reflection symmetry with the standard parametrization of the lepton mixing matrix. Remaining two mixing angles $\theta_{12}$ and $\theta_{13}$, along with three neutrino mass eigenvalues, remain as free parameters at $\Lambda_{FS}$. As we run down from the seesaw scale, all the parameters evolve according to their RGEs and get deviated from their high scale values. To account for the deviations, we consider MSSM as the effective theory below the mass scale of the seesaw. We further assume three different choices for the SUSY breaking scale $\Lambda_s$ at $1\ TeV,\ 7\ TeV$ and $14\ TeV$ to study the running behavior seperately. We first derive the one-loop RGEs of mass eigenvalues, mixing parameters and CP phases from the one-loop RGE of the Weinberg operator $\kappa$ following a standard method as available in the literature \cite{KS, RGE2, R1, R3}. While obtaining the RGEs for the mass eigenvalues, we have considered energy-dependent vevs of the Higgs field. We then solve all the coupled RGEs numerically employing a top-down approach, taking into account the RGEs of gauge coupling, Yukawa coupling and Higgs quartic constant as well. We run all the RGEs from the seesaw scale down to the electroweak scale, which is chosen at the top quark mass scale $m_t=172\ GeV$. Due to the choice of the SUSY breaking scale, intermediate between the seesaw scale and the EW scale, the running was carried out in two definite steps. In the first step, parameters are allowed to evolve from $\Lambda_{FS}$ to $\Lambda_s$ in the MSSM framework, while in the next, they evolve from $\Lambda_s$ to $\Lambda_{EW}$ in the SM scenario. Appropriate matching conditions are applied for the gauge and Yukawa coupling constants at the intermediate scale $\Lambda_s$. 

\indent The numerical analysis is carried out for both the cases of Normal Order (NO) and Inverted Order (IO), considering a higher value of $tan\beta=58$. The analysis further differs for case I, corresponding to $\theta_{23}=\pi/4$ and $\delta=\pi/2$ and for case II, corresponding to $\theta_{23}=\pi/4$ and $\delta=3\pi/2$, under each scenario of NO and IO. Besides $\theta_{23}$ and CP phases whose values are constrained according to $\mu-\tau$ reflection symmetry at $\Lambda_{FS}$, one needs to choose the high-energy input values of the other two mixing angles and three mass eigenvalues. For each case under the NO and IO scenarios, we could find an optimal set of input values for the free parameters that can yield low energy mass squared differences ($\Delta m^2_{21}$ and $\Delta m^2_{32}$) and mixing angles consistent with $3\sigma$ range of global analysis data \cite{GlobAnal}. Low-energy mass eigenvalues simultaneously satisfy the cosmological upper bound on the sum $\sum m_i<0.12\ eV$ \cite{mbound}. We have observed that in the IO scenario, it is relatively difficult to choose the input values of the free parameters such that all the low-energy parameters become consistent with $3\sigma$ range. In case II, under the IO scenario, it is even harder to choose the free parameters. We notice that, despite selecting an appropriate set of input values of the free parameters, evolution curves of $\theta_{12}$ and some of the CP phases generally exhibit some unusual distortions at some points. Through careful investigation, we could identify an optimal set of input values that results in low energy parameters best consistent with observational data, along with evolution curves of the parameters free from any distortions.

\indent Both in the NO and IO scenarios, running behaviours of the mass eigenvalues are found to be the same, they all tend to increase with the decrease of energy scale. However, mixing angles and CP phases do not exhibit a uniform running behaviour in both scenarios. The running of $\theta_{13}$ and $\theta_{23}$ are relatively weak as compared to $\theta_{12}$ in all cases and possess a little amount of deviation with respect to the high energy input values. $\theta_{13}$ and $\theta_{23}$ suffer a deviation of the order of $0.01^\circ$ and $0.1^\circ$ respectively while $\theta_{12}$ possesses a deviation of approximately $0.5^\circ-1.0^\circ$. In NO scenario, $\theta_{23}$ prefers to lie in the second octant due to RG running with low energy predicted values $\approx 45.1^\circ$ and $\approx 45.03^\circ$ respectively in case I and case II. These predictions are in agreement with the global analysis data without SK results \cite{GlobAnal}, which predicts a best-fit value of $48.5^\circ$ for $\theta_{23}$. The running of $\theta_{23}$ in the IO scenario is opposite to that observed in the NO scenario, where it tends to lie in the first octant. Like $\theta_{13}$ and $\theta_{23}$, running of the CP phases are also observed to be weak except in case I under IO scenario. Their deviations with respect to the high energy values in generally found to be of the order of $0.1^\circ$, but in the case I under the IO scenario, $\delta$, $\alpha$ and $\phi_1$ exhibit a deviation of approximately $3^\circ$. In light of the results from $T2K$ and $NO\nu A$ experiments, as they indicate a near maximal value of Dirac CP phase centred at $270^\circ$ for IO scenario \cite{T2k1,T2k2,Nova1,Nova2}, it is of particular interest to note the impact of RG running on phase $\delta$ in the present analysis. As per the global analysis \cite{GlobAnal}, the $3\sigma$ range $\delta$ is $201^\circ-335^\circ$ with a best-fit value $285^\circ$ ($274^\circ$) without SK data (with SK data) for IO scenario. In the present analysis, the evolution of $\delta$ with energy scale in case II under the IO scenario is depicted in Fig. 8(a). The low-energy prediction of $\delta$ is found to be approximately $269.9^\circ$ against the maximal value $270^\circ$ at $\Lambda_{FS}$. The deviation is very small, nevertheless it is consistent with $3\sigma$ range of global analysis.

\indent We have also studied the impact of variation of the SUSY breaking scale on the running effect of the parameters. This can be realized pictorially by studying the splitting between different evolution curves in each figure, which correspond to the three different SUSY breaking scales viz., $\Lambda_s=1\ TeV,\ 7\ TeV$ and $14\ TeV$. We observe that the impact of the variation of $\Lambda_s$ is insignificant in the case of mass eigenvalues in both NO and IO scenarios. Their low energy values do not vary significantly with the variation of $\Lambda_s$. The impact is relatively larger in case of certain mixing angles and CP phases in specific cases. For instance, in case I under the IO scenario, $\theta_{12}$ suffers a deviation of $0.9^\circ$ (Input at $\Lambda_{FS}=34.35^\circ$, output at $\Lambda_{EW}=33.51^\circ$) for $\Lambda_s=1\ TeV$, but for $\Lambda_s=14\ TeV$, the deviation is only $0.05^\circ$ (input at $\Lambda_{FS}=34.37^\circ$, output at $\Lambda_{EW}=34.42^\circ$). Similarly, CP phase $\beta$ exhibits a deviation of $0.1^\circ$ for $\Lambda_s=1\ TeV$ while it remains almost undeviated for $\Lambda_s=14\ TeV$ (low energy output value is $270.001^\circ$ against the high energy input value $270^\circ$). Again, in case II under the IO scenario, we observe that running effects on $\theta_{12}$ and CP phases $\delta$, $\alpha$ and $\beta$ vary distinctly with variation of $\Lambda_s$ in MSSM region, but interestingly, the running effects become indistinguishable in the SM region. The resultant low energy output values remain almost the same irrespective of the variation of $\Lambda_s$.

\indent In conclusion, the $\mu-\tau$ reflection symmetry of lepton mixing remains stable under RG running effects. Numerical estimation of perturbations caused by RG running shows that low-energy oscillation parameters are consistent with the $3\sigma$ range of global analysis data and they also satisfy the cosmological upper bound on the sum of mass eigenvalues.

\end{document}